\documentclass[a4paper,11pt]{article}
\usepackage{xr-hyper}
\usepackage[hidelinks]{hyperref}

\makeatletter
\newcommand*{\addFileDependency}[1]{
  \typeout{(#1)}
  \@addtofilelist{#1}
  \IfFileExists{#1}{}{\typeout{No file #1.}}
}
\makeatother

\usepackage[utf8]{inputenc}
\usepackage[T1]{fontenc}
\usepackage{graphicx,url}
\usepackage{a4wide}
\usepackage{mathpazo}
\usepackage[font=footnotesize,labelfont=bf]{caption}
\usepackage{amsmath}
\usepackage{bm}
\usepackage{amssymb}
\usepackage{multirow}
\usepackage{color,soul}
\usepackage{xurl}
\usepackage{authblk}
\usepackage[super]{cite}

\usepackage{changepage}
\usepackage{adjustbox}
\usepackage[shortlabels]{enumitem}
\setlist[enumerate]{nosep}

\usepackage[table,xcdraw]{xcolor}
\usepackage{tabulary}
\usepackage{booktabs}

\newcommand{\figstruct}[1]{Fig.\,\ref{fig:structure-pairwise-distance}#1}
\newcommand{\unit}[1]{\,\mathrm{#1}}

\newcommand{\Vext}{V_\mathrm{ext}}

\usepackage{verbatim}

\title{Decoupling geographical constraints from human mobility}
\date{}

\author[1,2]{Louis Boucherie}
\author[1,2]{Benjamin F. Maier}
\author[1,2]{Sune Lehmann\thanks{Email: \href{mailto:sljo@dtu.dk}{sljo@dtu.dk}}}

\affil[1]{DTU Compute, Technical University of Denmark}
\affil[2]{Center for Social Data Science, University of Copenhagen}

\usepackage{lineno}

\begin{document}

% \linenumbers %line numbers

\maketitle

\begin{abstract}
Driven by access to large volumes of movement data, the study of human mobility has grown rapidly over the past decades. 
The field has shown that human mobility is scale-free, proposed models to generate scale-free moving distance distributions, and explained how the scale-free distribution arises. 
It has not, however, explicitly addressed how mobility is structured by geographical constraints. How mobility relates to the outlines of landmasses, lakes, and rivers; by the placement of buildings, roadways, and cities. 
Based on millions of moves, we show how separating the effect of geography from mobility choices, reveals a power law spanning five orders of magnitude. 
To do so, we incorporate geography via the `pair distribution function' that encapsulates the structure of locations on which mobility occurs. 
Showing how the spatial distribution of human settlements shapes human mobility, our approach bridges the gap between distance- and opportunity-based models of human mobility.
\end{abstract}

\section*{Introduction}
Human mobility is a fundamental aspect of societies, with constant flows of individuals and goods circulating through cities and landscapes. 
This flow of individuals moving from place to place underpins our social, economic, and cultural lives \cite{fujita2001spatial}. 
Therefore, understanding mobility patterns enables us to mitigate the spread of infectious diseases \cite{colizza2006role, brockmann2013hidden}, reduce pollution \cite{setton2011impact}, or organize urban planning \cite{coutrot2022entropy}. 
Despite the fact that human movement is both shaped by and shaping our geography \cite{pumain1997title, arcaute2020hierarchies},  the field of data-driven human mobility has not considered these two processes together \cite{arcaute2022recent}. 

Geography trivially constrains human mobility. 
Consider two simple examples: 
Two houses cannot occupy the same physical space, thereby imposing a minimum distance for any movement between them. 
On a larger scale, residents of an isolated island of a $10\unit{km}$ radius cannot travel distances greater than $20\unit{km}$: they would end up in the sea.

Previous work has shown that human mobility is highly structured, with power law distributions of distance \cite{brockmann2006scaling, gonzalez2008understanding} and trip frequency \cite{schlapfer2021universal}, inflows between administrative boundaries \cite{zipf1946p1p2, simini2012universal}, conservation laws \cite{alessandretti2018evidence}, and scales corresponding to structures of the built environment \cite{alessandretti2020scales}. 
Simultaneously, the structure of the built environment is well understood, from the hierarchical organization of human settlements \cite{pumain2006hierarchy} at the scales of countries \cite{arcaute2016cities, wilson1970entropy, ripley1977modelling} to street-level networks \cite{samaniego2008cities, lee2017morphology}, as well as the fractal geometries of cities \cite{batty1994fractal, frankhauser1998fractal}, and how they developed \cite{pumain1997title}.

A key concern, however, is that the work cited above arises from two literatures that have evolved separately. 
The recent surge of data-driven work on human mobility has not explicitly addressed how the observed mobility data is structured by the constraints set by geography \cite{arcaute2022recent}. 
In that sense, large-scale data-driven human mobility research \cite{barbosa2018human} and the community of geographers \cite{pumain1997title, wilson1970entropy} remain distinct fields with complementary approaches.

Within data-driven human mobility studies there are two main theories for human movement \cite{barbosa2018human}:  the first theory considers physical distance to deter mobility \cite{carrothers1956historical, wilson1967statistical}, a concept often expressed through gravity models \cite{zipf1946p1p2}.
The alternative viewpoint instead suggests that human movement does not depend on distance, but rather on the quantity of opportunities nearby \cite{stouffer1940intervening, noulas2012tale}. 

There have been scattered efforts to incorporate a notion of geography in data-driven human mobility, but this work has been limited by incomplete data, e.g. relying on coarse-grained data such as data in grids \cite{mazzoli2019field}, administrative units \cite{yan2017universal}, or artificial geographical space \cite{han2011origin}, obscuring smaller scales and reducing geography to discrete density maps \cite{Hong2019, cohen2017modeling}. 
The use of neural embeddings also achieves a similar effect \cite{gao2017identifying}, and  leads to the emergence of gravity laws \cite{murray2023unsupervised}. 

Therefore, the question of how to incorporate geography in the data-driven approaches to human mobility still lacks a comprehensive and satisfying answer.
Below, we address this gap and present a framework to understand geography and the distribution of locations (addresses, points of interest, stop-locations).  
We argue here that the set of locations where people stop capture deep aspects of the mobility constraints set by geography: the built environment and urban structure, the placement of rivers and lakes, country outlines, etc. 

Our primary analysis is based on a  dataset comprised of 36 years of registry data on 39M  migrations between addresses in Denmark, pinpointed with uncertainty of only $2\unit{m}$, and previously unused for studying human mobility. 
This resolution far exceeds the state-of-the-art $50\unit{m}$ accuracy typical of GPS data. 
Below, we extend the analysis to day-to-day mobility and point-of-interests.

Analyzing these datasets using tools from statistical physics (e.g.~the pair distribution function), and by considering locations as particles, our approach unveils mobility patterns at scales that were previously obscured.

\section*{Results}
%\subsection*{Mobility takes place on the physical structure of the built environment}
% \subsection*{Mobility is shaped by the structure of the built environment} %60 characters

\begin{figure}
\includegraphics[width=\textwidth]{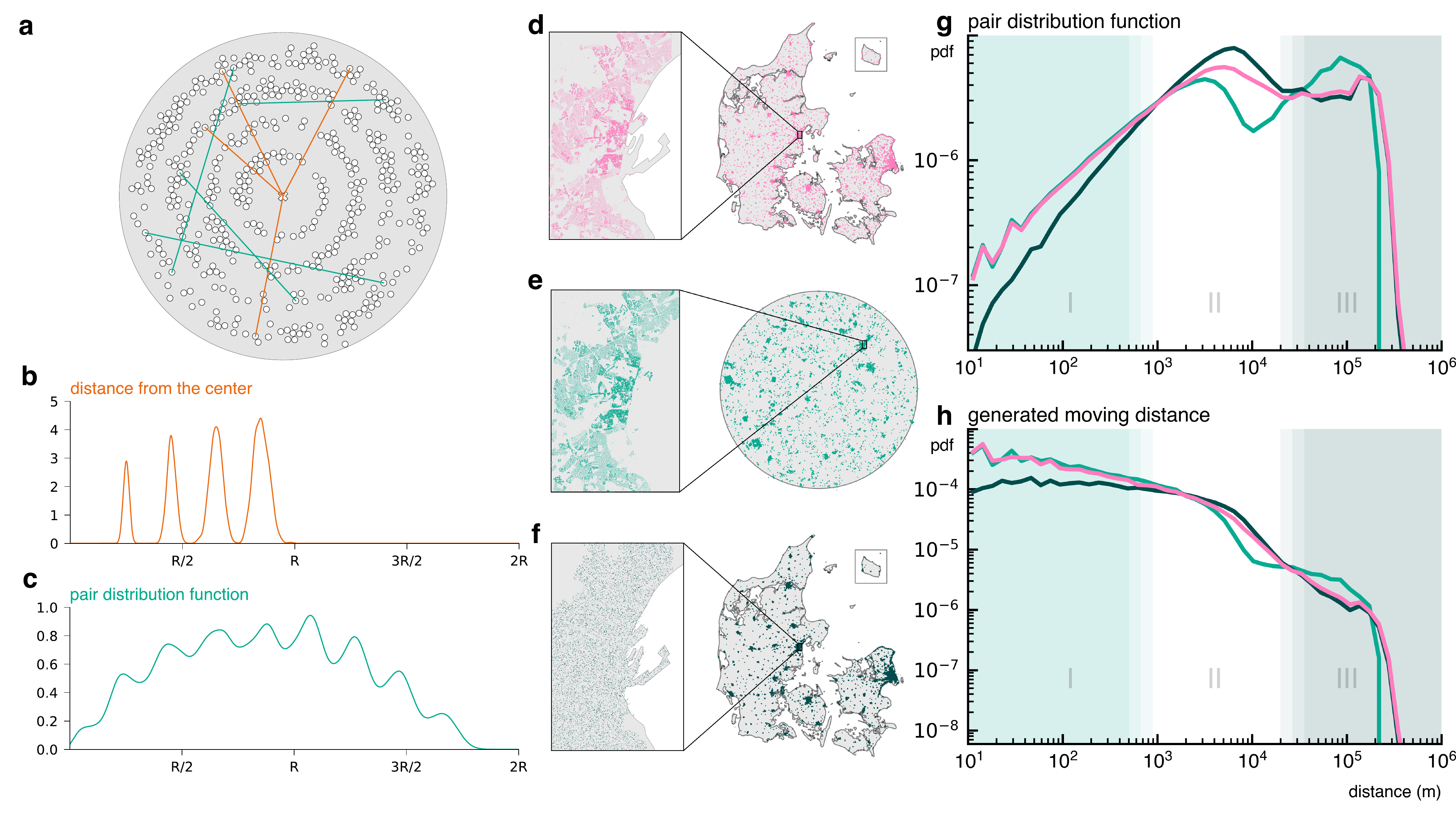}
\caption{\textbf{Illustration of geography's influence on human mobility} \textbf{(a)} Spatial arrangement of concentric circles of locations, \textbf{(b)} distance distribution of points to center of the disk (white cross). \textbf{(c)} Pair distribution of the points in the concentric circles. Examples of more shapes and distributions of locations are studied in Supplementary Section 1.1. \textbf{(d,e,f)} Three geographies: \textbf{(d)} \textit{Real Denmark} with 3.3M addresses, \textbf{(e)} \textit{Disk Denmark}, where the macro structure has cities distributed uniformly at random on a disk, but the microstructure of the cities is preserved, and \textbf{(f)} \textit{Uniform Denmark} with the macro structure of Denmark (shape of landmasses and city positions) but no local microstructure, i.e.~uniformly distributed locations within cities. \textbf{(g)} Pair distribution of the three geographies, and \textbf{(h)} the simulated observed moving distance for an intrinsic distance attractiveness of $1/r$. In \textbf{(g)} and \textbf{(h)}, I, II, III respectively represent the intra-city scale, the city scale, and the inter-city scale. }
\label{fig:1}
\end{figure}

\textbf{The concept of the pair distribution function}.
To characterize geography we study the \textit{pair distribution function} (`pair distribution' below) between locations.
The pair distribution is a powerful tool from statistical mechanics developed to understand the structural properties of materials \cite{HansenMcDonald1986}.
Here, we argue that the pair distribution is able to capture the structural properties of locations in a given geography. 
To get a sense of this distribution, consider a hypothetical spatial arrangement of concentric circles of $n$ locations (stop-locations, point of interest, addresses) on a disk of radius $R$ (Fig.\,\ref{fig:1}a). 
For an individual positioned at the center of these circles (a white $\times$), the potential moves are constrained by the position of other locations. 
For instance, in the absence of a location at distance $R/2$, a movement of that exact distance is impossible. 
To understand the potential movements from the disk's center, we compute the distribution of distances of every location to the center (Fig.\,\ref{fig:1}b) which can be interpreted as what we would observe if an individual was to move many times from the center to one of the locations on the disk, choosing their new location independently of distance. 
We can place each of the $n$ locations in the focus of such an analysis, finding a unique distance distribution for each of them, and calculate an average distance distribution. 
The result quantifies the expected number of locations found at distance $r$ from an \emph{average} location and is mathematically equivalent to the pair distribution, which simply encodes the number of location pairs found at distance $r$ (Fig.~\ref{fig:1}c).

\textbf{Observed data is a product of two separate components}.
This simple observation has a profound consequence: 
Any study of the observed movement distances $f(r)$ thus ends up analyzing a biased set of observations: a quantity that includes the geometry of space $\Omega$ \textit{and} the density $\varrho^d(x),\, x \in \Omega$, both reflected in the pair distribution $p(r)$ (see Methods: Pair Distribution Function).
We argue that, any `universal' behavior of the system, i.e.~a geometry-independent law that encapsulates the behavior of the system, must be captured entirely in a quantity that only depends on distance. We call this the intrinsic distance attractiveness.

Thus, to uncover any geometry-independent behavior of our system, we must adjust our observation $f(r)$ by the geometry encoded in $p(r)$, i.e.

\begin{align}
\label{eq:pi=f/p}
    \pi(r) \propto \frac{f(r)}{p(r)}.
\end{align}

% \textbf{Illustrating scale via three Denmarks}.
To illustrate how geography shapes the pair distribution and the observed moving distances, we consider the mobility trace simulation for three different geographies (Fig.\ref{fig:1}d,e,f and Supplementary 1.3). 
The first geography, 'Real Denmark,' uses the actual geography of Denmark, including 3.3M precise locations (addresses) (Fig.\,\ref{fig:1}d). 
The second, 'Disk Denmark,' maintains the microstructure of cities but alters city positions and landmass shapes, distributing real city centers uniformly on a disk (Fig.\,\ref{fig:1}e). 
The third, 'Uniform Denmark,' keeps the macro structure of landmasses and city positions from 'Real Denmark' but features uniformly dense cities (Fig.\,\ref{fig:1}f).

The pair distribution effectively encapsulates the nuances of each geographical layout (Fig.\,\ref{fig:1}g). 
On the local scales (less than $1\mathrm{km}$), the pair distribution of the `Real Denmark` geography aligns with the one of `Disk Denmark`, reflecting city layouts.
However, at broader scales (more than $10\mathrm{km}$), the pair distribution of 'Real Denmark' matches 'Uniform Denmark', indicating the influence of the shape of land masses and city positions.

To generate mobility traces we impose the same intrinsic distance attractiveness on each geography, and make a gravity law-inspired ansatz that it follows,

\begin{equation}
    \label{eq:ansatz}
    \pi(r)=1/r.
\end{equation}

Remarkably, despite having the same intrinsic distance attractiveness, the resulting observed mobility traces are distinct across the three Denmarks (Fig.~\ref{fig:1}h). 
This outcome is significant as it shows that even under a similar movement law, the variations in geography alone can lead to distinct mobility patterns.

\begin{figure}
\includegraphics[width=\textwidth]{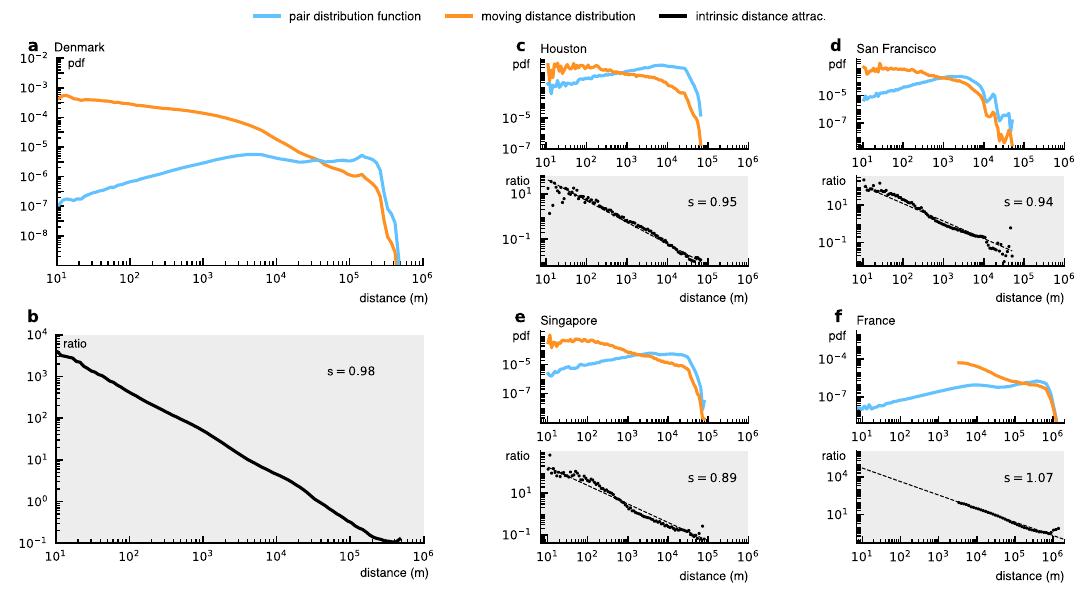}
\caption{\textbf{Renormalizing the moving distance distribution by the pair distribution uncovers a power law.}
\textbf{(a)} Empirical moving distance distribution and pair distribution in Denmark. Dividing the mobility by the pair distribution yields an intrinsic distance attractiveness \textbf{(b)} that follows a power law from 10m to 500km (maximum likelihood estimation). \textbf{(c)} Observed moving distance distribution and pair distribution for day-to-day mobility between points of interest (POI) in San Francisco, \textbf{(d)} Houston, and \textbf{(e)} Singapore. \textbf{(f)} Observed moving distance distribution and pair distribution for residential mobility in France. y-axis ranges vary across plots because the geographical areas under consideration differ in size.}
\label{fig:2}
\end{figure}

\textbf{A power law that spans five orders of magnitude.} 
We now demonstrate the impact of taking into account the pair distribution, by first analyzing an exhaustive dataset—39M residential moves between 3.3M Danish addresses over 36 years. 
The dataset has very low selection bias, as it covers all residential moves (Methods: Data).
To uncover the intrinsic distance attractiveness of residential mobility, we first eliminate the geographical component by normalizing the observed moving distance distribution by the pair distribution (Fig.~\ref{fig:2}a).
This normalization reveals that the \textit{intrinsic distance attractiveness function} (Eq.~\eqref{eq:pi=f/p}) follows a power law remarkably well:
\begin{align}
\pi(r)=\frac{f(r)}{p(r)}=\frac{1}{r^s}.
\end{align}
This power law is consistent across scales ranging from 10m to 500km, covering five orders of magnitude (Fig.~\ref{fig:2}b). 
A maximum likelihood fit provides an exponent of $s = 0.98$, which validates the ansatz in Eq.~\eqref{eq:ansatz}; the intrinsic distance attractiveness follows $\pi(r)=1/r$. 
We consider this power law to be an extension of the gravity model to the continuous domain (see Methods: Continuous Gravity Law) \cite{nijkamp1986synthesis}. 

We find that the results above are neither particular to the geography of Denmark nor to residential mobility. 
Investigating several empirical data sets, Figure~\ref{fig:2}c-e illustrates that the normalization by the pair distribution unveils power laws for day-to-day mobility across the diverse geographies of Houston, Singapore, and San Francisco.
We also find the same patterns for residential mobility in France (Fig.~\ref{fig:2}f). 
The empirical fact that our results take the same shape across countries and forms of mobility highlights the generality of the intrinsic distance attractiveness' functional shape: a power law that decreases with distance.

Next, we focus on the geographic regularities manifested through the power laws and the fluctuations in the pair distribution of Denmark (Fig.~\ref{fig:2}a).

%\section{Multi-scale urban structure shapes the distribution of pair distribution}
\textbf{The condensed matter physics of locations}.
Having illustrated the fundamental way geography is encoded in a country's location pair distribution and how it affects the dyadic process of human mobility, we now focus on a deeper analysis of its shape and origin. 
Specifically, we show how its properties naturally emerge from statistical physics arguments, and how understanding the pair distribution allows us to identify key attributes of geography in terms of shaping human mobility. 
Thus, in this section, we go beyond simply appropriating the concept of the pair distribution function for analyses of human mobility, but use the tools from condensed matter physics \cite{HansenMcDonald1986} to create simple models for the micro-, meso-, and macro-structure of the geography of an entire country.

% \textbf{Regimes of the pair distribution}.
The pair distribution between residential locations (buildings and addresses) in Denmark shows the behavior displayed in \figstruct{a}. 
In this view buildings are simply addresses that stack on top of each other (e.g.~apartment buildings).
On the micro-scale (I) within distances of $r\lesssim25\unit{m}$, the `immediate neighborhood', we observe a linear onset of neighbor density, oscillatorily modulated. 
For larger distances at the meso-scale (II), this growth assumes a scaling of approximately $p(r)\propto r^{0.67}$ between $\sim25\unit{m}$ and $\sim 200\unit{m}$ (until $\sim 4000\unit{m}$ for addresses). 
Afterward, on the macro-scale (III), this growth curbs, decays  rapidly, and eventually approaches zero as we reach the limit of the finite system. 

%\textbf{Regime I: The density of cities as an ideal gas in a potential}. %62 characters
\textbf{Regime I: City density as an ideal gas in a potential}. %54 characters
Our basis for modeling cities is a simple particle model in two dimensions -- where buildings are modeled as particles distributed uniformly at random, corresponding to an ideal gas kept in place by an external potential.
The key constraint to identify the properties of this 2D confining potential (\figstruct{b}) is the observation from the literature that population density per unit area decays exponentially with distance to its center (\figstruct{c}) \cite{clark1951urban, newling1969spatial, bertaud2003spatial}. 
A generalized Gamma distribution of shape $p(r) = (r/R^2)\exp[-(r/R)^m]/\Gamma(2/m)$ accurately describes the pair distribution of this ensemble, where variable $R$ allows us to associate every city with a radius that can be estimated from the data via a maximum-likelihood fit, without having to rely on a definition of city center (\figstruct{c} and Supplementary: City Sizes).

Interactions between locations: to understand the oscillatory behavior modulating the onset of the pair distribution, we take the condensed matter approach a step further.
We argue that there is both a repulsive and attractive effective force between locations.
The repulsive force originates from the simple fact that two buildings cannot occupy the same space. 
The effective attractive force results from the lower cost of placing buildings near one another as it reduces infrastructural cost for shared amenities \cite{Duranton2004}. 

In implementing this idea, we recognize that -- unlike gases -- cities are not constructed according to strict laws, so there is not a single, `true' description of the system. 
Thus, we illustrate the influence of such effective forces on the pair distribution by modeling the system as two distinct ensembles of repulsive-attractive forces: (i) an ensemble of Lennard-Jones (LJ) disks, a canonical model for sphere-like particles \cite{LennardJones1931}, and (ii) attractive hard disks of heterogeneous size, both in the presence of a linear external potential (Methods: Locations as interacting particles). 
Both models reproduce the oscillatory features observed at the micro scale in our high-resolution dataset (\figstruct{d}).

We can study the oscillatory structure in terms of the pair-correlation function $g(r)=p(r)/i(r)$, where $i(r)$ is the pair distribution of an ideal gas. 
This function measures the over- and under-representation, respectively, of neighbors at distance $r$ in relation to the expectation if no interaction forces were present \cite{OrnsteinZernike1914}. 
In both models and the data, there is a considerable lack of neighboring buildings for distances $r \lesssim 4\unit m$, suggesting presence of repulsive forces, later reaching a peak $g(r)>1$ at $r\approx 6\unit m$ which is typically observed for systems with attractive forces. 
This peak becomes wider when we consider heterogeneous building radii. 
The discrepancy between the data and model(s) below 5$\unit m$ (distance between black and teal/pink lines in \figstruct{d}) is due to the definition of building location in the data as the location of its front door, rather than the building center (Methods: Data).

\begin{figure}
    \vspace{-3cm}
    \centering
    \includegraphics[width=\textwidth]{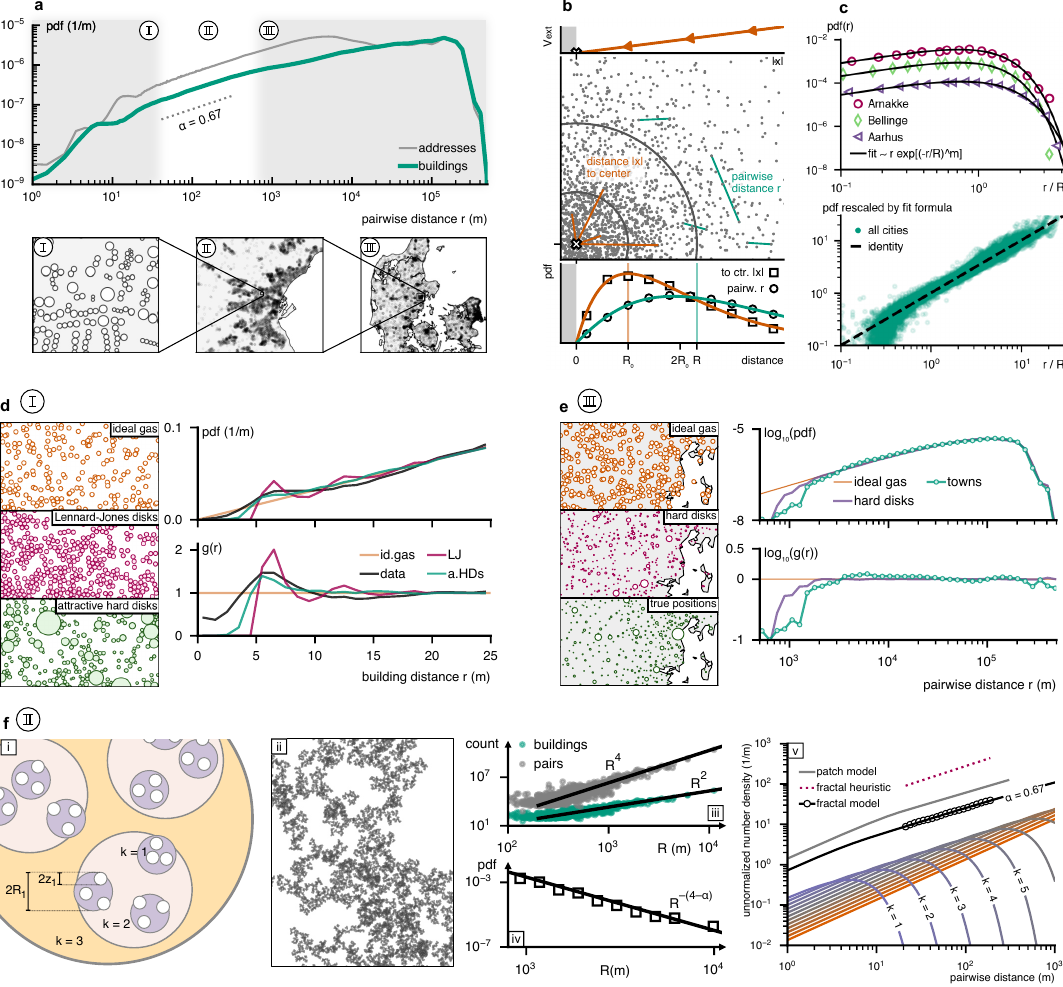}
\caption{\textbf{(a)} Pair distribution of residential locations in Denmark. (I) Immediate neighborhood-scale with a linear onset and an oscillating modulation for the immediate neighborhood, (II) city-scale showing a power-law growth with exponent $\alpha = 0.67$, and (III) country-scale with a slower growth and eventual fast decay.
    \textbf{(b)} With external potential $\Vext(|\bm x|)$ that grows linearly with distance to the city center $|\bm x|$, an ideal gas of buildings will form a city-like structure with radial building density following an Erlang distribution of scale $R_0$ (squares) (Methods: Model of locations). The corresponding pair distribution approximately assumes a generalized Gamma distribution with scale $R$ providing a definition for the radius of a city (circles).
    \textbf{(c)} Maximum-likelihood fits of the generalized pair distribution against pair distribution of buildings in three Danish towns (upper) and all Danish towns, rescaled (lower).
    \textbf{(d)} Pair distributions for four systems of building positions on the micro-scale, (i) buildings in Denmark, (ii) random positions (ideal gas), (iii) Lennard-Jones disks, and (iv) attractive hard disks of heterogeneous size. All systems display linear onset with oscillatory modulations. The pair-correlation function $g(r)$ quantifies the under- and over-presence, respectively, of neighbors at distance $r$ when comparing to the ideal gas. Repulsive and attractive forces lead to valleys and peaks in $g(r)$ (see Methods:
    Model of locations, Hard disks and Molecular Dynamics simulation)
    \textbf{(e)} Pair distribution and $g(r)$ for all towns in Denmark for (i) random positions (thin, orange lines), (ii) non-attractive hard disks (thick, violet lines), (iii) real data (dotted, teal lines). Systems (ii) and (iii) display similar behavior with slow onset, quickly approaching the ideal gas.
    %Weighting with the numbers of building pairs per pair of towns (pink and green lines) replicates the macro regime behavior (III) of panel (a) but introduces slight deviations to the model system (ii).
    \textbf{(f)} (i) Fractal model for building positions in cities, as a hierarchically nested structure of self-similar patches. On each hierarchy layer, a patch of buildings consists of $b$ sub-patches distributed non-overlapping with constant density and packing fraction $\theta$. At the lowest level, a sub-patch is a building. (ii) A sample configuration of this model with $b=4$, $z_1=3\unit{m}$, $\theta=0.8$, and $L=8$. (iii-iv) Independent-patch model. Equating the concept of ``town'' with ``patch of radius $R$ containing buildings'', the number of buildings per patch grows as $R^2$ and the number of pairs of buildings per patch grows as $R^4$ (with $R$ from the analysis in panel c). At the same time, the tail of the patch radius distribution falls as $R^{-(4-\alpha)}$ with $\alpha=0.67$. (v) For both the fractal model and the independent-patch model, we recover the pair distribution scaling observed in the data.}

    \label{fig:structure-pairwise-distance}
    %\end{adjustwidth} % delete this for final submission
\end{figure}

\textbf{Regime III: The spatial distribution of cities}.
We now use the condensed matter tool-set to understand the macro-scale (III) of the pair distribution. 
For large distances, its shape is dominated by the locations of cities. 
We compare an ideal gas of cities (i.e.~random positions uniformly distributed in the landmass of Denmark) to the empirical pair distribution of all Danish towns (\figstruct{e}). 
While the ideal-gas distribution has a linear onset, there is a considerable lack of neighboring towns for small distances $r\lesssim 1\unit{km}$, but after that, the observed town pair distribution rapidly approaches and matches the ideal-gas distribution remarkably well, demonstrated by the fast approach of $\log(g(r))\rightarrow0$.

The fact that there is not a clear peak in $g(r)$ suggests the absence of a strong attractive force. 
When we compare the observed pair distribution to that of a non-attractive hard disks-model (Methods: City pair distribution) we observe a similar behavior, indicating that the positions of towns are almost statistically indistinguishable from random locations except for the condition that towns may not overlap in territory. Next we weigh every contributing pair distance for a pair of towns by the product of their respective numbers of buildings (see SI). While the hard disks-model aligns with the empirical observations for smaller distances, for larger distances the building-location pair distribution is only recovered by the empirical weighted city-location distribution -- and not by the hard disks-model. This is in line with observations that size plays a non-negligible role in where towns are located \cite{1933zentrale}.

\textbf{Regime II: Fractal structures explain the meso-scale}.
As described above, the initial linear growth of the pair distribution quickly approaches a sub-linear scaling law of $\alpha\sim 0.67$ (\figstruct{a}).
The emergence of this scaling law is related to the non-regular shapes of cities composed of smaller `patches' of buildings that form larger clusters in a space where occupation is limited by local geography (e.g.~bodies of water, hills) or for developmental reasons (e.g.~industrial areas, parks, agricultural use).

Specifically, we adapt a model that was recently used to generate surrogate city positions \cite{Hong2019, Soneira1978}. 
Assume that a city is constructed as a hierarchically nested structure of patches containing $b$ buildings with packing fraction $\theta$ (\figstruct{f.i}, Methods: Fractal model). For a range of realistic parameters, this model yields $\alpha=0.67$ for both single realizations (\figstruct{f.ii}) and analytically. Deriving the heuristic approximation $\alpha=1-2\ln\theta/\ln(\theta/b)$, we further show that the scaling emerges independently from the choice of patch geometry (see Supplementary Section 2.7).
Our model replicates the behavior observed for the whole country and is in agreement with $\alpha=0.69\pm0.04$ for individual cities. 
Furthermore, this nested description is consistent with earlier observations that settlement size distributions are scale-free \cite{batty1994fractal, makse1995modelling}.

% \subection*{Intrinsic Distance Attractiveness function}

\begin{figure}
\includegraphics[width=0.5\textwidth]{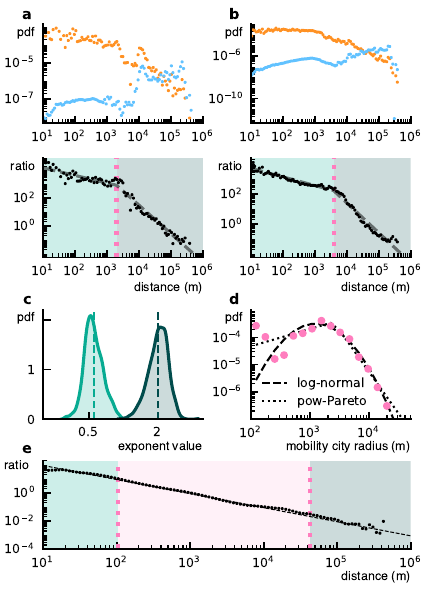}

\caption{\textbf{A piece-wise process}.
\textbf{(a-b)} Residential mobility analysis in a small \textbf{(a)} and a large \textbf{(b)} Danish town reveals a consistent piece-wise power law across all cities. For more cities, see Supplementary\,Fig.22-25, for details on the piece-wise power-law definition and maximum likelihood estimation see Supplementary Section 3.5.
\textbf{(c)} Distribution of the piece-wise power laws' exponents, for the first and second parts, respectively center around $0.6 \pm 0.20$ and $2.0 \pm 0.21$ (mean $\pm$ standard deviation).
\textbf{(d)} The transition point in the piece-wise power law aligns with a city size distribution. We fit different distributions (log-normal, power-Pareto) via maximum-likelihood (see Methods: Power-law estimation).
\textbf{(e)} We recover the intrinsic distance attractiveness from simulating the piece-wise power law (a,b) with the mean values of $0.6$ and $2$ (c), over the city-scale (d). We assume random position of cities as in \figstruct{e}. For alternative derivation of the intrinsic distance attractiveness and details see Supplementary Section 3.5.}
\label{fig:4}
\end{figure}

\textbf{From local to global power laws}. 
Considering the intrinsic distance attractiveness starting from individual Danish cities (as opposed to globally), we observe a piece-wise process (Supplementary\,Fig.22-26) \cite{cadwallader1992migration}.
Specifically, for each city, we consider the distribution of moves originating in the city of interest to the rest of the country. 
The pair distribution is limited to pairs that include at least one address in the city of interest. 
At the city scale, we find that the intrinsic distance attractiveness is a piece-wise power law (Fig.\,\ref{fig:4}a-b) with the first exponent centered around $0.60\pm 0.20$(SD) and the second exponent around $2.0 \pm 0.21$(SD) (Fig.~\ref{fig:4}c). 
The process is universal in the sense that the exponents of the piece-wise power laws are consistent across the 1400 cities of Denmark. 
The transition point between the two power laws defines a \textit{mobility city radius} (Fig.~\ref{fig:4}d) with a distribution that lies between a log-normal and a power-Pareto distribution (Methods: Power-law estimation).

\textbf{Putting the pieces together}. 
Now we can put the pieces together. We do this by exploring how the global power law (Fig.\,\ref{fig:2}b) relates to the local piece-wise picture for each city (Fig.\,\ref{fig:4}a-b).
Here, our statistical physics framework describing the components of the pair distribution can be used to complete the picture of human mobility. Finally, we can reproduce the intrinsic distance attractiveness starting from the city-level piece-wise description.
We simulate the piece-wise power law over a toy geography that reproduces the key characteristics of geography: scale of cities (Fig.~\ref{fig:4}.e), local pair distribution that follows a generalized gamma distribution (Supplementary Equation 17, \figstruct{c}), and random position of cites (\figstruct{e.III}), we recover the empirical intrinsic distance attractiveness, $\pi(r)=1/r$, validating our geographical model and the piece-wise power law (see Supplementary Section 3.5).

\section*{Discussion}
Geography trivially constrains human mobility. 
While the existing human mobility literature has extensively documented the structure of human mobility, it often remains disconnected from studies focused on the organization of the built environment. 
This led to the emergence of two distinct perspectives on human mobility \cite{barbosa2018human}:  one based on distance, similar to the gravity model  \cite{zipf1946p1p2}, and another focused on the availability of opportunities \cite{stouffer1940intervening}.
Here, we have proposed the pair distribution of locations as a key normalization variable of human mobility. 
The normalization reveals an intrinsic power law consistent across various geographies and mobility types. 
Additionally, we proposed a model inspired by statistical physics to account for the shape of the pair distribution between locations. 
A limitation of our work is that while the model replicates pair distribution characteristics for general addresses at dwelling and country scales, the analysis does not yet extend to the pair distribution between non-residential of locations, like home and work pairs.
 One could, for example, compute the pair distance function for all possible (`home', `work') location pairs. The resulting empirical `home'-`work' trip distance distribution would capture the spatial constraints on travel patterns. Then, using the approach outlined in the manuscript: dividing the commute trip distance distribution by the (`home', `work') pair distance distribution, we can reveal an intrinsic distance pattern for this type of activity. Such an extension could help us understand commuting patterns, a significant component of day-to-day mobility \cite{pappalardo2023future, clark2003does, weber1929alfred}. 
The temporal scope of the analysis is also a limitations. We consider the period from 1986 to 2020 as an aggregated timescale. Therefore, our study captures long-term average characteristics but does not resolve dynamics or evolutionary trends within this 34-year span, nor does it account for the impact of specific events or changes during these decades. Further, the fact that we do not have data from periods prior to 1986 prevents a study of historical changes of the observed mobility structures. This restricts our ability to fully understand how these patterns have evolved from earlier states of the built environment and societal organization, or to ascertain whether the observed power law is a long-standing feature or one that emerged recently.
Finally, we explained our normalization as a continuous gravity model that reconciles the two paradigms of human mobility. 
We also show that this description does not hold from the perspective of individual cities and emerges from the aggregation of scales \cite{alessandretti2020scales}. 
Moreover, our approach provides a baseline in the form of the intrinsic distance attractiveness, $\pi(r) = 1/r$, which serves as a null model for exploring mobility patterns of sub-populations. Exploring deviations from this baseline across different demographic groups—such as age, gender, or socioeconomic status—could offer valuable insights into how specific factors shape human movement. Incorporating such new demographic insights could deepen our understanding of the underlying variability in mobility patterns across different population segments. At the city level, other demographic factors could explain relocation decisions, such as housing prices, employment opportunities, and family evolution \cite{cadwallader1992migration}. 
Our framework expands the understanding of human mobility and provides a bridge connecting two rich bodies of literature.

% \section*{Word count (main text and abstract only)}
% \detailtexcount{main}

% %TC:ignore

\section*{Methods}

\subsection*{Data}

\subsubsection*{Uncertainty Quantification}
Estimates reported in the main text are reported as mean $\pm$ standard deviation.

\subsubsection*{Residential Mobility in Denmark}

The data on residential mobility within Denmark is derived from the \textit{Befolkning} database of Danmarks Statistik \cite{dst_befolkning_valg}. This database contains 39,297,646 residential moves between different addresses in Denmark from 1986 to 2020. The location data for the 2,345,453 buildings that make up the 3,251,464 addresses are also provided by Danmarks Statistik, with an accuracy better than $\pm2$ meters \cite{dataforsyningen2023adgangsadresse}. The location of an address is defined as the location of the building's entrance door, we investigate the effect of this definition of pair distribution function in Supplementary Section 2.4. These datasets are not publicly available due to Danish Data Protection regulations. However, access to the data is possible for research purposes. Data can be obtained via \textit{Statistics Denmark for Researchers} in accordance with Statistics Denmark’s Research Scheme: \url{www.dst.dk/en/TilSalg/Forskningsservice/}

\subsubsection*{Bias in residential mobility in Denmark}

In contrast to studies using digital tools such as mobile phones, which may introduce biases \cite{schlosser2021biases}, the Danish residential mobility dataset we study is inherently representative of the entire population. In fact, the dataset includes every residential move, a requirement for all residents to report, thus eliminating   potential biases associated with selective data collection methods \cite{wesolowski2013impact}. Demographic information (gender and year of birth) are reported in Extended Data Figure 2-3 \cite{jacobsen1999natural, Denmark2024Births}. Cities are defined according to the definition of Danmarks Statistik; to ensure the robustness of the results, we compare their definition with a hierarchical clustering of the addresses (HDBSCAN \cite{campello2013density} in Supplementary Section 4)

\subsubsection*{Residential mobility in France}

The location data for France is publicly available and compiled from the \textit{Base Adresse Nationale}, accessible via \url{https://adresse.data.gouv.fr/donnees-nationales}, which contains precise location data for 21,567,447 buildings. To match the housing unit data, we utilized information from \url{https://www.data.gouv.fr} (FiLoSoFi), which segments the country into a grid of 200m squares, each annotated with the count of housing units and collective housing entities. We augment the data by uniformly distributing the housing unit count amongst each collective housing entity, resulting in 34,041,910 individual addresses. The residential mobility data is publicly available and was obtained from \url{https://www.insee.fr/fr/statistiques}. It contains 40,465,288 inter-city migrations from 2016 to 2020. The distances are calculated from one city center to another. The city centers are computed as the centroid of contained addresses. As we only have access to inter-city moves, migration distances that are less than the largest city's dimensions in France were disregarded.

\subsubsection*{Day-to-Day Mobility}

The data on daily mobility was obtained from user check-ins (points of interest) on the location-centric  social network Foursquare over six-months (May 27, 2010 to November 3, 2010) as documented \cite{noulas2012tale, yang2015participatory}.
The Foursquare dataset was sourced entirely from publicly available data: check-ins were accessed via Twitter's Streaming API and venue geo-coordinates from the Foursquare website, both being public at the time of collection. User consent was implied by their public sharing of this information on these platforms, and as such, data access was governed by the terms of these public channels.
The data comprises a total of 239,788 movements across 43,395 distinct locations.
The regional breakdown of the data is, 47,996 movements occurred across 11,808 locations in Houston, 79,624 movements across 15,617 locations in Singapore, and 112,168 movements across 15,970 locations in San Francisco.

\subsubsection*{GeoBoundaries}

We obtain the country, state, and municipality shapes from the geoboundaries project \cite{runfola_geoboundaries_2020}.
Throughout the paper, we work with the EPSG:23032 projection for all points that lie within Denmark and the EPSG:27561  projection for points that lie within France. We then perform analyses on the Euclidean geometry of the projection. 

\subsection*{Pair distribution function}

In this section, we provide a formal definition of the pair distribution function. We consider points $x_i\in\Omega$  distributed in a $d$-dimensional space $\Omega\subseteq\mathbb{R}^d$ according to a density $\varrho^d(x)$, then the pair distribution $p(r)$ is uni-dimensional and determined by both the shape of the space as well as $\varrho^d(x)$.
Thus, the \textit{pair distribution} is given by,

\begin{align}
    \label{eq:definition-pairwise-distribution}
    p(r) = \int_\Omega d  x\ \int_\Omega d  y\ \varrho^d(  x) \varrho^d(  y) \delta(r - \lVert x-y\rVert).
\end{align}

with $\lVert x-y\rVert$ the distance between two points $x$ and $ y$. When we analyze flows between pairs of points in $\Omega$, we notice a distribution that depends on the distance, referred to as $f(r)$, or the \textit{observed movement distance}. The anticipated number of movements from a location $x$ to a location $y$ is determined by this distribution.

\begin{align}
    f(x,y)\mathrm{d}^2x\mathrm{d}^2y
\end{align}

with \(f(x, y)\) being a continuous number density. This number is influenced by two functions. First, the number of observations is proportional to the number of locations at \(x\) as well as the number of locations at \(y\), i.e., depending on the product of location density \(\varrho(x)\varrho(y) \mathrm{d}x \mathrm{d}y\). Second, the number of movements is proportional to the movement propensity, i.e., the tendency for a move that begins at a location at \(x\) to end at a location at \(y\), which we denote as \(\pi(x, y)\),

\begin{align}
    f(x,y)\mathrm{d}^2x\mathrm{d}^2y = \pi(x,y) \varrho^d(x) \varrho^d(y) \mathrm{d}^2x\mathrm{d}^2y.
    \label{eq:exp_mov}
\end{align}

The number of moves spanning distance $r$ within the domain $\Omega$ is

\begin{align}
    f(r) = \int_\Omega \pi(x,y) \varrho^d(x) \varrho^d(y)\delta(r - \lVert x-y\rVert) dx dy.
\end{align}

If we make the anstaz that the propensity to travel from $x$ to $y$ solely depends on the distance between the two,~i.e,
\begin{align}
    \pi(x,y) = \pi(\lVert x-y\rVert))=\pi(r)
    \label{eq:pi_r}
\end{align}
as $\pi(r)$ is now independent of location $x$ and $y$, it can be factored out of the integral, we obtain
\begin{align}
    f(r) \propto \pi(r)p(r)
\end{align}

with $p(r)$ the pair distribution function as in Eq.~\eqref{eq:definition-pairwise-distribution}, and we recover the main text's Eq.~\eqref{eq:pi=f/p} (see  Supplementary Section 1.2 for more details). 
The ansatz of Eq.~\eqref{eq:pi_r} is the key point to reconcile the distance-based and opportunity-based perspective on human mobility. 
The distance term is ~$\pi(\lVert x-y\rVert)$, the opportunity term is the expected number of pairs of locations between location $x$ and location $y$,   $\varrho^d(x) \varrho^d(y) \mathrm{d}^2x\mathrm{d}^2y$ (Supplementary Section 1.2, 3.3) \cite{liang2013unraveling, maier_generalization_2019}. 

Moreover, the pair distribution function characterizes the geometry of the space as it represents the joint probability of finding two locations at particular positions in the system, (Supplementary Section 1.1 and Supplementary Section 1.5). 
The computation of the pair distribution function scales as $\mathcal{O}(n^2)$, computing it for 
a large number of locations is challenging (see Supplementary Section 2.5)

\subsection*{Locations as interacting particles}

Our basis for modeling cities is a simple particle model in two dimensions---where buildings are modeled as particles uniformly distributed at random, corresponding to an ideal gas kept in place by an external potential. Consider the city center, e.g.\ the central business district, as the primary point of interest due to its proximity to amenities. This attractiveness implies a higher cost of being located further away from the center, represented by the distance $| \bm x|$.

At the same time, two simple mechanisms will prevent buildings from accumulating in the exact center of the city. First, buildings have a certain average radius $z$, so they cannot be too close (at distance $<2z$). There is an advantage to buildings not being too far apart, as they can share local amenities.  Rather than explicitly modeling this phenomenon, we assume that there is an inherent temperature, $T$, in the system. This temperature determines the distribution of house locations, which is influenced by interactions and external potential.

From a statistical physics perspective, we describe the system with a simple external potential that linearly increases with distance.
\begin{align}
    V^\mathrm{ext}(\bm x) = \gamma|\bm x|
\end{align}
where we assume that the origin of the coordinate system is located at the city center. In order to model repulsion and attraction between houses, we further assume a Lennard-Jones interaction potential

\begin{align}
    V^{\mathrm{int}}(\bm x_i,\bm x_j) = \varepsilon \left[
                                \left(\frac{2z}{|\bm x_j - \bm x_i|}\right)^{12}
                            -   2\left(\frac{2z}{|\bm x_j - \bm x_i|}\right)^6
                              \right],
\end{align}

which is commonly used to model intermolecular attraction and repulsion \cite{LennardJones1931,HansenMcDonald1986}.

In total, a system with these properties evolves according to the Hamiltonian

\begin{align}
    \mathcal H(\{\bm x, \bm p\}) = \frac12\sum_{i=1}^N \bm p_i^2 + \gamma\sum_{i=0}^N |\bm x_i| + \frac 12 \sum_{i=1}^N\sum_{j\neq i}^N V^{\mathrm{int}}(\bm x_i, \bm x_j)
\end{align}

with two-dimensional momenta $\bm p_i$ and locations $\bm x_i$. In a canonical-ensemble formulation of the system, i.e. at constant inverse temperature $\beta$, the probability of finding a configuration $\{\bm  x, \bm p\}$ in volume-element $d^2\bm x d^2\bm p$, is given by

\begin{align}
    \varrho[\{\bm  x, \bm p\}] d^2\bm x d^2\bm p &=
        \exp[-\beta \mathcal H(\{\bm x, \bm p\})] d^2\bm x d^2\bm p.
\end{align}

We will initially consider an ideal gas first where $V^\mathrm{int}=0$, which will enable us to say something about the density of particles around the center of city, i.e.~we want to find the particle dwelling probability $p(r)dr$ with $r$ being the particle's distance to the center. For the sake of simplicity, we set $N=1$ without loss of generality. This is justified by the ergodicity of the system, which implies that the trajectory of a single particle will eventually follow the density of the entire distribution. This can be expressed as taking the $N$th root of the $N$-particle density. Integrating over the momenta yields
\begin{align}
    \varrho[\bm x] d^2\bm x  &=
        \exp\left(-\beta \gamma |\bm x|\right) d^2\bm x% \\
\end{align}
With a change of variables to polar coordinates, we find
\begin{align}
    \varrho[r, \phi] r dr d\phi = r \exp\left[-\beta \gamma r\right] dr d\phi,
\end{align}
i.e. 
\begin{align}
    p(r) dr = r \exp\left[-\beta \gamma r\right] dr.
\end{align}
This implies that in the absence of interactions, the distribution of houses around the city center should follow a $k=2$ Erlang distribution with a scale parameter of $\lambda=\beta \gamma=1/R_0$ where $R_0$ is half the city radius. The role of the inverse temperature term, $\beta$, is explained in Supplementary Section 2.1.4.

To gain insight into the radial particle density in the context of strong repulsion, we return to the Lennard-Jones perspective. In the limit of $\varepsilon/T\gg1$, particles will have a strong tendency to be found in their respective potential minimum, i.e. at distance $2z$ from each other. Effectively, we can think of them as hard disks of radius $z$ that have a tendency to form clusters. If the temperature is low, the effective radius of the city will be small, we expect a crystal to form in the center (see Supplementary Section 2.1.4).

\subsubsection*{Molecular Dynamics simulation}

We are interested in finding configurations $\mathcal C$ that accurately depict the canonical ensemble with number of particles $N$, an average-constant temperature $T$, and a constant but irrelevant volume. We assume that the particles are constrained to a radially symmetric external potential, with the total volume of the system containing the particles being irrelevant.  Finally, the total potential must be

\begin{align}
    V = \sum_{i=1}^N \left[ V^{\mathrm{ext}}(\bm x_i) + \frac{1}{2}\sum_{j\neq i}V^{\mathrm{int}}(\bm x_i, \bm x_j)\right].
\end{align}

To this end, we integrate the equations determined by the system's Hamiltonian numerically using the velocity-Verlet algorithm \cite{verlet1967computer}. Furthermore, we rescale particle velocities according to the stochastic Berendsen thermostat with relaxation time $\tau$ \cite{berendsen1984molecular}. The details and parameters of the molecular dynamics simulation are available in Supplementary Section 2.1.5.

\subsubsection*{Hard disks}

We generate a configuration of hard disks with a non-specified attractive force (i.e.~we do not explicitly integrate the equations of motion for a hard-sphere interaction potential with an additional attractive force). To do so, we first generate $N=10^4$ random positions according to a radial Erlang distribution of scale parameter $R=675$ (i.e.~an ideal gas). Subsequently, we draw a random radius for each position from a heterogeneous distribution with power law tail. We first draw values $\hat z_i$,

\begin{align}
    p(\hat z) = \frac{a^2-1}{2a}\times\begin{cases}
            \hat z^a, & \hat z \leq 1, \\
            1/\hat z^{a}, & \hat z > 1,%
    \end{cases}
\end{align}

and then assign disk radius $z_i = 3 \hat z_i/\left<\hat z\right>$. We choose $a=4$ to obtain a heterogeneous distribution with non-finite variance in disk area. Afterwards, we run the collision algorithm outlined in Supplementary Section 2.2.

The modelling of interaction between the house in the external potential is further developed in the Supplementary Section 2.3.

\subsection*{Non-overlapping disks}
To emulate the position of cities, we randomly distribute $N$ hard disks of radii $R_i$ in shape $\Omega$. We start with largest disk of radius $R_\mathrm{max}$ and iterate over all disks, ordered decreasingly in size. For every disk $j$, we generate a random position $\bm x \in \Omega$ until the condition $|\bm x - \bm x_i| > R_j+R_i,\, \forall j<i$ is met, i.e.\ drawing new random positions until there are no overlaps with other, already placed disks. Then, assign $\bm x_j\leftarrow \bm x$ and continue with the next disk.

\subsection*{City pair distribution and radius}

The pair distribution of each city is best represented by a generalized Gamma distribution with linear onset
\begin{align}
    \label{eq:gen_pdd}
    p(r,R,m) = \frac{r}{R^2\,\Gamma(2/m)} \exp\left[-\left(\frac r R\right)^m\right],
\end{align}
as depicted in ~\figstruct{c}. Here, $r$ is the distance between two locations, $R$ is the scale parameter corresponding to the city radius and $m>0$ is a shape parameter controlling the decay of the tail. Assuming that population density decays exponentially with distance to the city center and that consequently the pair distribution of buildings assumes a generalized pair distribution of the form of Eq.~\eqref{eq:gen_pdd}, we perform maximum-likelihood fits to infer parameters $R$ and $m$ for each city in Denmark with more than 30 buildings, we generally find strong correspondence to the empirical distribution, as illustrated by the correlation of $r/R$ and the inverse of Eq.~\eqref{eq:gen_pdd},
\begin{align}
    (r/R)^*=-(\log[p(r)R^2\Gamma(2/m)/r])^{1/m},
\end{align}
albeit with small deviations at smaller distances (cf.~\figstruct{c}) that can be associated with the sub-linear scaling behavior observed on the meso-scale of \figstruct{a}. Remarkably, measuring the `radius' $R$ of a city in terms of its pair distribution's scaling parameter comes with the advantage of not having to rely on any definition of `city center'.

Moreover, no assumptions about urban growth are necessary; instead, the observed behavior emerges from the simple ansatz that being placed at distance $|\bm x|$ from the center of a town comes at a cost $\propto |\bm x|$, which can be overcome for a multitude of reasons encoded in temperature $T$.

\subsubsection*{Fractal Model}
The fractal model we adopt was originally devised to explain the positions of galaxies and was recently used to generate surrogate city positions \cite{Hong2019, Soneira1978}. We assume that a city is constructed as a hierarchically nested structure of patches containing buildings (cf.~\figstruct{f.i} and Supplementary Section 2.7).
At layer $k = 1$, a patch of radius $R_1$ contains $b$ buildings of radius $z_1$ placed in a non-overlapping manner with packing fraction $\theta$. 
At layer $k=2$, a patch of radius $R_2$ contains $b$ building-patches of radius $z_2 = R_1$, again placed without overlap and packing fraction $\theta=\theta_k=bz_k^2/R_k^2$. 
Continue in that fashion until reaching layer $k=L$. Due to the constant packing fraction $\theta < 1$, larger areas of absent buildings will form at each hierarchy layer, mirroring similar observations in real urban structures. 
Note that demanding $\theta$ to be constant across hierarchy layers also implies a predetermined patch radius of $R_k=(\theta/b)^{k/2}z_1$. A sample configuration with $b=4$, $z_1=4\unit{m}$ and $\theta=0.8$ can be seen in \figstruct{f.ii}.

To evaluate the overall pair distribution of such a fractal structure, we assume that for a patch of hierarchy layer $k$, the $b^{k-1}$ buildings that are contained in each of its $b$ sub-patches are all located at their respective centers. Then, the pair distance number density (pdnd) of an average such patch will approximately assume Eq.~\eqref{eq:gen_pdd}, weighted with a total of $b^{2k-2}b(b-1)/2$ building pairs (between subpatches). Moreover, at every hierarchy layer $k$ there will be $b^{L-k}$ such patches. In total, each hierarchy layer therefore contributes a pdnd of
\begin{equation}
    \label{eq:number-density-k}
    n_k(r) \propto b^k p(r, R_k, m),
\end{equation}
implying that the observed pdnd of the whole structure is proportional to
\begin{equation}
    \label{eq:number-density-all}
    n(r) =  \sum_{k=1}^L n_k(r) \propto \sum_{k=1}^L \frac{b^{2k} r}{\theta^k z_1^2} \exp\left[-\left(\frac{r}{z_1}\right)^{m}\left(\frac{b}{\theta}\right)^{mk/2}\right].
\end{equation}
In the limit of $L\rightarrow\infty$, we use Laplace's method to find

\begin{equation}
    \label{eq:number-density-approx}
    \log n(r) \propto \underbrace{\left({1-\frac{2\log\theta}{\log(\theta/b)}}\right)}_{=\alpha} \log r,
\end{equation}
i.e.~a sub-linear scaling law emerges from considering the pair-weighted sum of single-scale pair distributions with linear onset. The pair distance number density $n(r)\propto r^\alpha$ with $\alpha = 1-2\log\theta/\log(\theta/b)$. Notably, the result is independent of the parameter $m$ that controls the tail of the single-scale pair distribution, suggesting that the exact shape of the composite pair distributions are irrelevant, which we further demonstrate in the Supplementary Section 2.1 by discussing another functional form of the pair distribution.

\subsubsection*{Patch Model}

We consider a \emph{multiverse} consisting of an infinite amount of universes indexed $i$, each of which is inhabited by a patch of size $R_i$ with $N_i$ buildings inside, contributing to the multiverse with a pair distribution number density of $n(r,R_i) \approx N_i^2 p(r,R_i,m)$ (each universe contributes a number of building pairs amounting to $N_i^2$). We also assume that there's a constant population density $\rho_0\propto N/R^2$, demanding that $N\propto R^2$.

For each of these universes (or rather, each of these patches), we calculate the number of building pairs at distance $r$ and assume that patch sizes $R$ are distributed according to some---at this time still unknown---distribution $f(R)$. Then, the joint distribution of house pairs at distance $r$ is given by
\begin{align}
    n(r) &\propto \int\limits_0^\infty dR\ n(r,R) f(R)\\
     &\propto \int\limits_0^\infty dR\ f(R) R^4 \frac{r}{R^2}\exp\left(-\left(\frac {r}{R}\right)^m\right)
     .
\end{align}
To simplify the integral, we change variables to $\beta = 1/R$, such that $\mathrm dR = -\beta^{2}\mathrm d\beta$ and 
\begin{align}
    n(r) &\propto \int\limits_0^\infty d\beta \ f(\beta^{-1}) \frac{r}{\beta^4}\exp\left(-(\beta r)^m\right).
\end{align}
This integral yields a solution if $f=\beta^{4-\alpha}$ with $\alpha < 1$. Let's assume that this is the case, which means we have to solve the integral
\begin{align}
    n(r) = C \int_0^\infty\beta^{-\alpha} r \exp\left(-(r\beta)^m\right) d\beta 
\end{align}
where $C$ is a normalization constant. We recognize the integral of the gamma function (see Supplementary Section 2.6) and therefore find,
\begin{align}
    n(r) \propto r^\alpha C\,\Gamma\left(\frac{1-\alpha}{m}\right).
\end{align}

That is, if the radius of patches in the multiverse would be distributed as $f(R) \propto R^{-(4-\alpha)}$, the initial growth of the joint pair distribution would follow a sub-linear power law. Notably, the scaling exponent does not depend on the tail parameter $m$. 

Let's see what this would mean in terms of the area of the patches. The area would scale as $A\propto R^2$ (or $R\propto\sqrt A$), i.e. the distribution of the area would follow
\begin{align}
    \tilde f(A) &= \frac{\mathrm dR}{\mathrm d A} f(R=\sqrt{A})\\
                &\propto \frac{1}{A^{(5-\alpha)/2}}.
\end{align}
In the data, we observe $\alpha=0.67$. That would mean that the area of the patches would have to be distributed according to a power law with exponent $\mu=-(5-\alpha)/2\approx -2.17$, which is well within what has been found empirically for cities \cite{makse1995modelling}. Furthermore, we find $f(R)\propto1/R^{3.29}$ from our city radius inference analysis (cf.~Supplementary Figure~11) which leads to $\mu=-2.15$, showing that the results of these two separate analyses are consistent.

\subsection*{A `continuous' gravity law}

In its original form, the gravity law for human migration states that the probability of moving between two cities is proportional to the product of their population and inversely proportional to the distance separating them \cite{zipf1946p1p2},
\begin{align}
\label{eq:gravity}
 P_{1 \to 2} \propto \frac{N_1 N_2}{d_{1 \to 2}}.
\end{align}
This relationship requires administrative units to define cities and their respective populations, thereby rendering it inherently discrete and subject to arbitrary choices regarding administrative boundaries. 
Yet, constructing a product of origin$\times$destination populations closely aligns with the idea of normalizing by the pair distribution. 
To illustrate, consider two cities of populations $N_1,\, N_2$, with radii $r_1$ and $r_2$ separated by a distance $d$. If we assume a uniform location density within these cities (as illustrated in Fig.\ref{fig:2}c), the density of locations is given by
\begin{align}
    p^d( x) =\frac{N_1}{\pi r_1^2} \delta_{\Omega_1}(x) + \frac{N_2}{\pi r_2^2} \delta_{\Omega_2}(x),
\end{align}
with $\Omega_j$ the domain of city $j$. This leads to the pair distribution (using Eq.\,\ref{eq:definition-pairwise-distribution}) :
\begin{align}
    p(r) = N_1 N_2  \int_\Omega\int_\Omega d  x d  y \frac{1}{\pi r_1^2} \frac{1}{\pi r_2^2}
           \delta_{\Omega_i}(x)\delta_{\Omega_j}(y)\delta(r-R(  x,   y)).
\end{align}
In the limit case where $d$ significantly exceeds both $r_1$ and $r_2$, when the city radii are negligible compared to the distance between cities, the pair distribution around $d$ tends towards a Dirac distribution of height equal to the product of the number of addresses in each city, $p(r)\xrightarrow[d\gg r_{1},r_{2}]{} N_1 N_2 \delta_d(r)$, therefore (Extended Data Fig.4),
\begin{align}
    f(r) = p(r)\pi(r) \xrightarrow{} \frac{N_1 N_2 \delta_d(r)}{r}.
    \label{eq:continuous-gravity}
\end{align}
Equation~\eqref{eq:continuous-gravity} illustrates how in the context of the the gravity model for migrations, the `mass-product' $N_iN_j$ emerges.  
Given that the number of addresses in a city is proportional to its population, we recover the discrete gravity law. 
We refer to our finding as `continuous' since it operates at the granularity of individual addresses, the finest possible scale (see Supplementary Section 3.1.1). A comparison between the gravity model, the radiation model, and the pair distribution framework is shown in Extended Data Figure~5. 

We can also interpret the intrinsic distance attractiveness $\pi(r) = 1/r$ as an emergent property of random utility theory, where individuals appreciate distance with a logarithmic scale. This interpretation is consistent with the logarithmic utility function described in Supplementary Section 3.2.

\subsection*{Power-Law Estimation}
The parameters for power-law distributions are determined using maximum-likelihood estimation, as outlined in \cite{clauset2009power}. For continuous distributions, the probability density function (pdf) for a power-law distribution is given by
\begin{align}
\label{eq:pdf-powerlaw}
    p(x) = C x^{-\alpha}
\end{align}
where $x>x_{\text{min}}>0$ and $C$ is a normalization constant. The parameter $\alpha$ is typically greater than 1, as the probability density function described in equation~\eqref{eq:pdf-powerlaw} does not integrate over $x \in [x_{\text{min}},+\infty )$ for $\alpha\leq1$. However, in our analysis, the intrinsic distance attractiveness has an exponent $\alpha\sim1$. Moreover, the maximum-likelihood estimation method from \cite{clauset2009power} is known to be biased when $\alpha < 1.5$, a limitation that has been confirmed by \cite{hanel2017fitting}. Consequently, to estimate power laws with $\alpha < 1.5$, we employ the maximum-likelihood estimator developed in \cite{hanel2017fitting,bauke2007parameter}, with a maximum boundary for the power law, so that the pdf in equation \ref{eq:pdf-powerlaw} is integrable over $x \in [x_{\text{min}},x_{\text{max}}]$ for any $\alpha>0$. It is reasonable to assume that power-law distributions are bounded within the context of this study, given that geographical limitations inherently bound the movement distances in our data set. We define $x_{\text{max}}$ as the diameter of each domain $\Omega$, which corresponds to the largest distance between two points (Extended Data Table~1). We assess the power-law parameter estimates through goodness-of-fit tests employing the Kolmogorov–Smirnov (KS) statistic and likelihood ratios. The Kolmogorov–Smirnov tests are presented with p-values in Extended Data Table\,1. Likelihood ratios provide evidence on whether the power-law distribution is a superior fit compared to other distributions. The methodology for estimating piece-wise power laws for each Danish city (Fig.\,\ref{fig:2}f-g) is based on a maximum likelihood estimator derived in \cite{maier2023maximum}, the details of the fit are available  Supplementary Section 3.4 and Supplementary \,5.
In the analysis of the \textit{Mobility City Radius} the power Pareto distribution \cite{maier2023maximum} is identified as the optimal model with a slightly better fit, measured by an Akaike Information Criterion (AIC) score of 24046.03. This is compared to the log-normal distribution, which has an AIC of 24161.17. A comparison with additional distributions is available in Extended Data Table 2.

The mesoscale power law of the pair distribution function (\figstruct{a}) and the simulation of the fractal and patch model (\figstruct{f}v) are also estimated with a maximum likelihood estimator.

\newpage
\section*{Data availability}

The data for Denmark used in this study are not publicly available due to Danish Data Protection regulations. However, access to the data is possible for research purposes. Data can be obtained via \textit{Statistics Denmark for Researchers} in accordance with Statistics Denmark’s Research Scheme: \url{https://www.dst.dk/en/TilSalg/Forskningsservice/}.
The location data for France are publicly available at \url{https://adresse.data.gouv.fr/donnees-nationales} and \url{https://www.data.gouv.fr}. The residential mobility data were obtained from \url{https://www.insee.fr/fr/statistiques}.
The Foursquare data are publicly available at \url{https://sites.google.com/site/yangdingqi/home/foursquare-dataset}.
The source data used to generate the figures are available via the Zenodo repository: \url{https://zenodo.org/records/14329837} \cite{boucherie2024decomposing}.

\section*{Code availability}

The geography model and estimation code are available at \href{https://github.com/benmaier/ljhouses}{github.com/benmaier/ljhouses}. Code for figures, data, and statistical analysis is at \href{https://github.com/LCB0B/role-of-geo}{github.com/LCB0B/role-of-geo}. The code is also available on the Zenodo repository \url{https://zenodo.org/records/14329837} \cite{boucherie2024decomposing}.

\section*{Acknowledgement}
The authors thank J.~Dzubiella and YY.~Ahn for helpful comments in the early development of this study, as well as L.~Alessandretti and S.~De Sojo for providing insightful comments on the manuscript. The work was supported in part by the Villum Foundation Grant Nation-Scale Social Networks  [grant number: 00034288] (SL)  and the Danish Council for Independent Research [grant number: 0136-00315B](SL). The funders had no role in study design, data collection and analysis, decision to publish or preparation of the manuscript.

\section*{Author contributions}
L.B., B.M. and S.L. designed the study and the model. L.B. and B.M. performed the analyses and implemented the model. L.B., B.M. and S.L. analyzed the results and wrote the paper.

\section*{Competing interests}
The authors declare no competing interests.
%TC:endignore

%TC:endignore

% \section*{Figure Captions}

\setcounter{figure}{0}
\setcounter{table}{0}
\captionsetup[figure]{name={Extended Data Figure}}
\captionsetup[table]{name={Extended Data Table}}

\newpage
\section*{Extended Figures and Tables}
\begin{figure}[h]
\includegraphics[width=\textwidth]{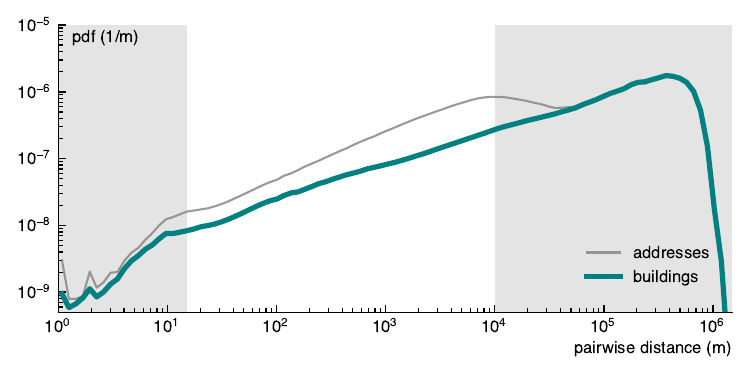}
\caption{Pair distribution of buildings and addresses in France shows three scales: (I) Immediate neighborhood-scale with linear onset and oscillating modulation, (II) city-scale with power-law growth ($\alpha = 0.67$, for addresses), and (III) country-scale with slower growth and rapid decay (see SI.\,1.6)}
\label{fig:pairwise_fr}
\end{figure}

\newpage

\begin{figure}[h]
\includegraphics[width=\textwidth]{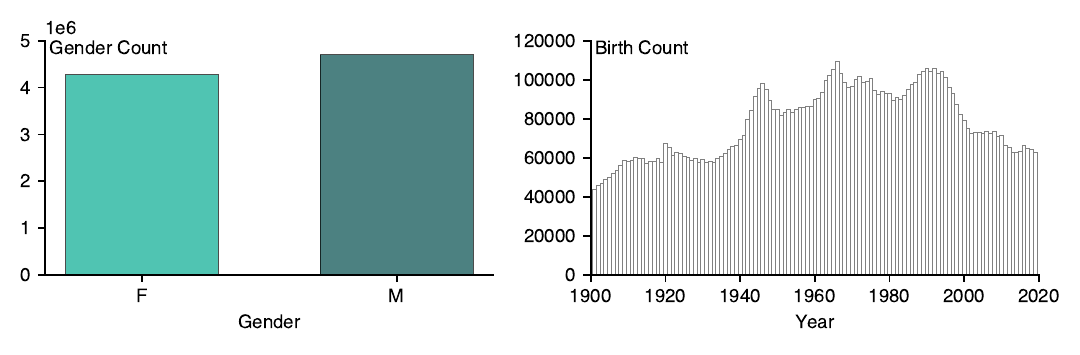}
\caption{Demographic characteristics of the Danish Residential Mobility dataset. The gender imbalance is the first panel due to the $1.07$ male/female ratio rate at birth \cite{jacobsen1999natural}. Indeed we study the total population, not the currently alive population, that is balanced for gender due higher death rate for men. Therefore although there is an imbalance in gender, the dataset is not biased. The second panel shows the birth year of every individual in the dataset, it closely follows the birth rate of Denmark, and the differences are due to migrations \cite{Denmark2024Births}. }
\label{fig:demographic}
\end{figure}

\newpage

\begin{figure}[h!]
    \centering
    \includegraphics[width=\textwidth]{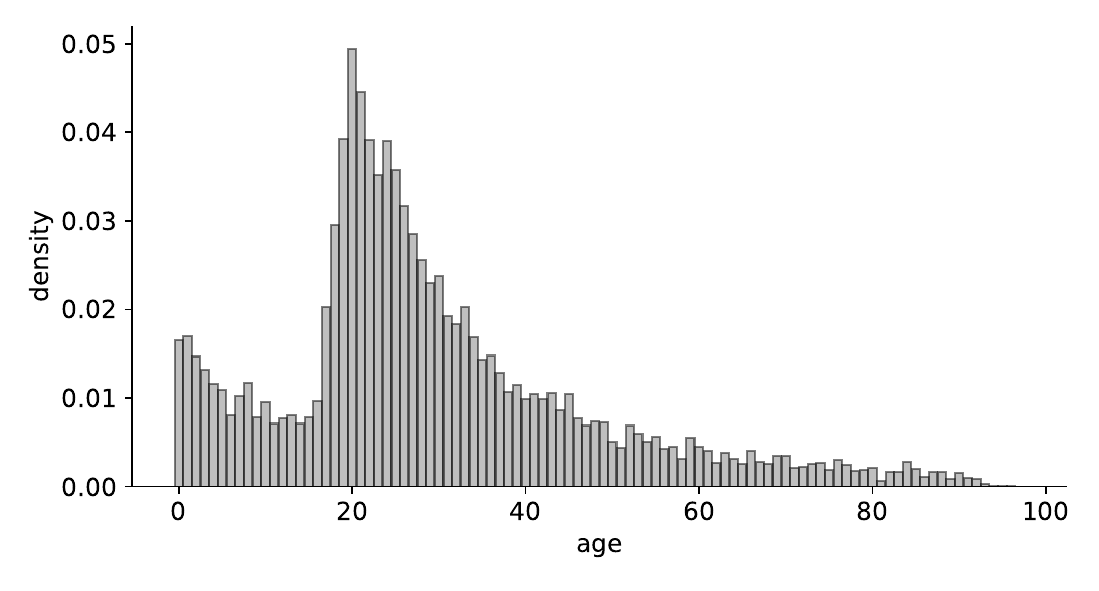}
    \caption{Distribution of Age at the time of moving in the Danish residential mobility data (see Methods: Data). This figures has to be put in persepective with the Figure \ref{fig:demographic}b}
    \label{fig:age}
\end{figure}

\newpage

\begin{figure}[h]
\centering
\includegraphics[width=0.5\textwidth]{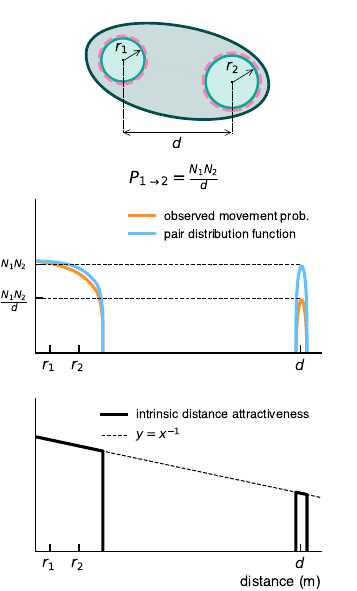}
\caption{\textbf{(a)} This framework can be represented as a \textit{continuous gravity law}, if we take a toy model of two distant cities, \textbf{(b)} the pair distribution at the distance $d$ between the two cities collapses into a Dirac distribution, with the height equal to the product of cities population, $N_1 N_2$.
\textbf{(x)} The toy \textit{intrinsic distance cost} of the simulation model. For a comparison with a continuous radiation model, see SI.\,3.1.2.}
\label{fig:continuous_gravity_law}
\end{figure}

\newpage

\begin{figure}[h]
\includegraphics[width=\textwidth]{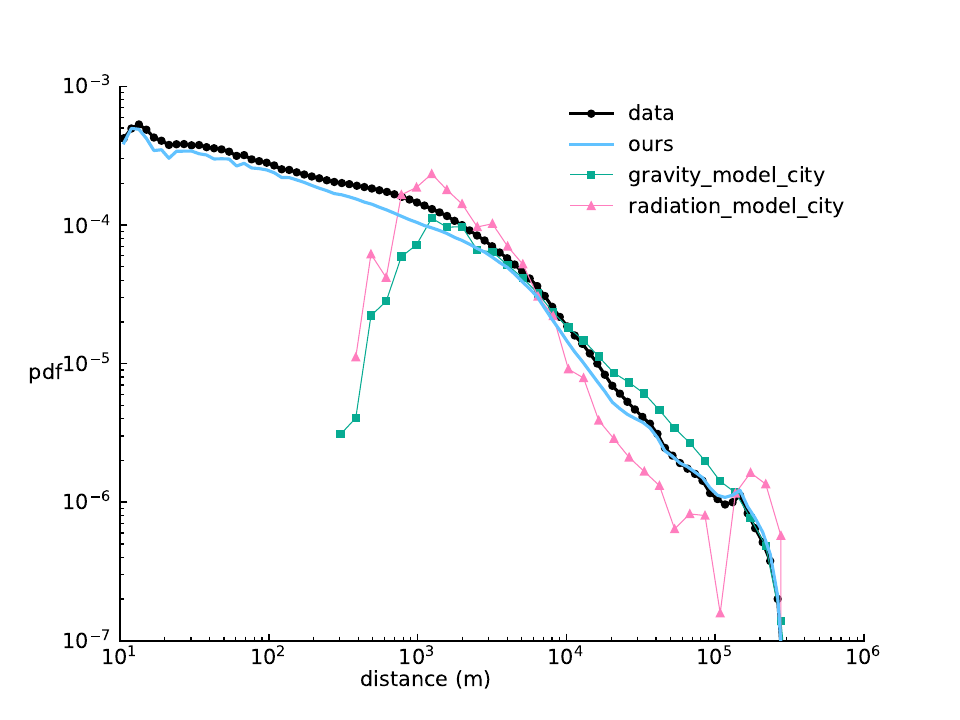}
\caption{Comparison between our framework, the gravity model and the radiation model for cities.}
\label{fig:comparison_grav_rad}
\end{figure}

\newpage

\begin{table}[h]
\centering
\caption{\textbf{Power law estimation}}
\begin{tabular}{@{}llrrrrrr@{}}
\toprule
\toprule
Type & Dataset  & $n$ &$x_{\text{min}}$  & $x_{\text{max}}$ & Exponent & KS & p\\

\midrule
\midrule

\multirow{1}{*}{Residential Mobility} & Denmark & 39M & 10 & 475032 & \textbf{0.98} & 0.01 & \textbf{0.99}\\
                                    & France & 40M & 3000 & 1134667 &  \textbf{1.07} & 0.01& \textbf{0.98}  \\ 
\midrule
\multirow{1}{*}{Day-To-Day Mobility} &

                                San Francisco & 112168 & 10 & 55639   & \textbf{0.94}  & 0.07 & \textbf{0.96} \\
                                
                                &Houston  & 47996 & 43 & 71209 & \textbf{0.95} & 0.03& \textbf{0.98}\\
                                
                                &Singapore  & 15167 & 10 &79674&   \textbf{0.89} & 0.12 & \textbf{0.67} \\
\midrule
\multirow{1}{*}{Geography} & City Radius    & 1367 &  2878 & N/A & $3.05$   & $0.58$& 0.11  \\
                              
\bottomrule

\end{tabular}
\label{tab:power-law}
\end{table}
\footnotesize
The table presents information about the estimation of the exponent of the power law. The estimation of the power law is based on a maximum-likelihood estimator as in \cite{hanel2017fitting}. The values of $x_{\text{min}}$ and $x_{\text{max}}$ are reported in meters. The table reports also the p-values associated with the Kolmogorov-Smirnov test for the power law exponent estimators \cite{clauset2009power}. According to \cite{clauset2009power}, the suitability of a power law model is considered statistically implausible if the p-value is less than or equal to 0.1. This threshold indicates a less than or equal to $10\%$ probability that the observed deviation of the data from the model predictions is due to random variation alone. Therefore the power-law model is statistically significant if $p >0.1$ (in bold).

\newpage
\begin{table}[h]
\centering
\caption{\textbf{Log-likelihood Ratio between power law and other heavy-tailed distributions}}
\begin{tabular}{@{}llrrrr@{}}
\toprule
\toprule
Type & Dataset   & Log-Normal & Exponential & Weibull \\
\midrule
\midrule

\multirow{1}{*}{Residential Mobility} & Denmark & $128697$  & $19209$  & $89676$  \\
                         & France    & $5042$  & $411$  & $18535$    \\
\midrule
\multirow{1}{*}{Day-to-day Mobility}    & San Francisco  & $1091$ & $91$ & $2001 $  \\
                         & Houston  & $446$ & $35$ & $2946$    \\
                         & Singapore & $729$ & $104$ & $3651$     \\
\midrule

\multirow{1}{*}{Geography} & Mobility City Radius     &$-0.96$   &  $1.91$   & $1.18 $  \\
                       
\bottomrule
\end{tabular}
\label{tab:log-likelihood}
\end{table}

\footnotesize
The table presents the log-likelihood ratio $\mathcal{R}$ (ref. \cite{clauset2009power}) comparing the power law (cf. Table\,\ref{tab:power-law}) to other heavy-tailed distributions (one per column). When $\mathcal{R}$ is positive, the power law distribution has a higher likelihood compared with the alternative. When $\mathcal{R}$ is negative, the other distribution has a higher likelihood compared with the power law.

%word count ignore:
%TC:ignore

\setcounter{figure}{0}
\setcounter{table}{0}

\captionsetup[figure]{name={Supplementary Figure}}
\captionsetup[table]{name={Supplementary Table}}

\newpage
\vspace{5cm}
\section*{Supplemental Information}
\tableofcontents 
\listoffigures
\newpage

\section{The concept of pair distribution function distribution }

\subsection{Pair distribution examples}
\label{SI:pdd-example}
\begin{figure}
    \centering
    \includegraphics[width=\textwidth]{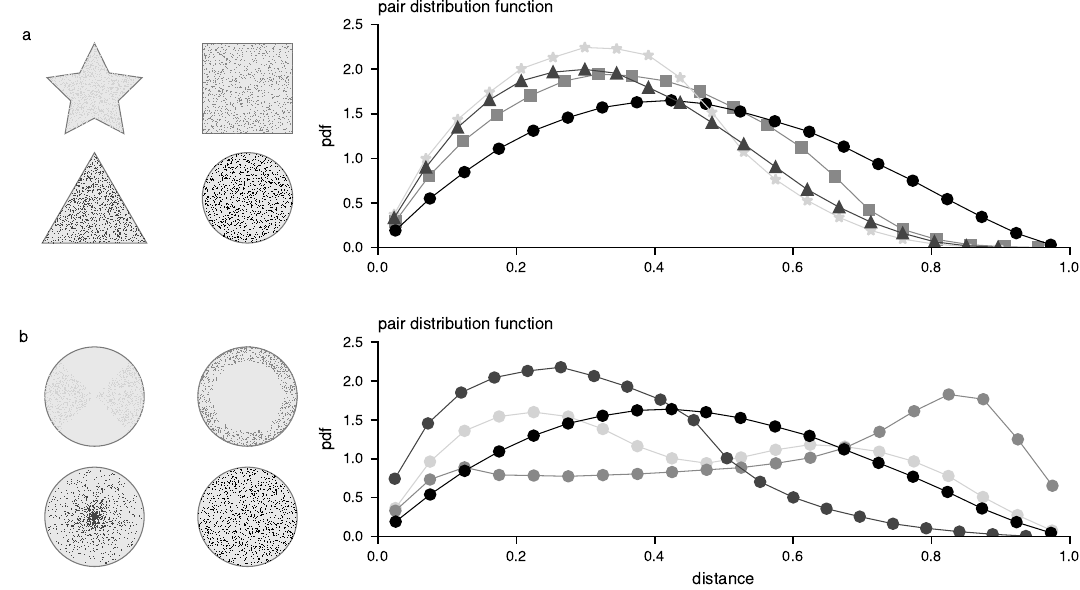}
    \caption{Example of pair distribution function for different toy geographies and density of points. 
    \textbf{(a)} Four different shapes of landmasses and their respective pair distribution functions (same maker shape and greyscale).
    \textbf{(b)} For the circle shape, four different toy layouts of the built environment and their respective pair distribution function (same greyscale).
    }
    \label{fig:pdd_examples}
\end{figure}

This section presents additional examples of toy geographies that vary both the shape of the landmasses and the density distribution of points. These examples intend to demonstrate that the pair distribution distribution can encode the geography of the built environment. Figure \ref{fig:pdd_examples}a, illustrates how the pair distribution function captures the shape of uniformly distributed points. Figure \ref{fig:pdd_examples}b illustrates the impact of spatial point distribution on a fixed shape.

\subsection{The pair distribution function emerges as a normalization for human mobility}
\label{SI:pdd-normalisation}
Whenever points $ x_i\in\Omega$ are distributed in a $d$-dimensional subspace $\Omega\subseteq\mathbb R^d$ according to some density $p^d(  x)$, their pair distribution function $p(r)$ is one-dimensional and determined by both the geometry of the space as well as $p^d(  x)$, given that there exists a (pseudo-) metric $r\equiv R(  x,  y)$ that determines a distance between two points $  x$ and $  y$. This pair distribution is given by
\begin{align}
    \label{eq:definition-pairwise-distribution-SI}
    p(r) = \int_\Omega d  x\ \int_\Omega d  y\ p^d(  x) p^d(  y) \delta(r - R(  x,  y)).
\end{align}

When observations are made between two distinct points (i.e. when a quantity is counted) in the domain of interest, a distance-dependent distribution of the observed entities is observed. This distribution is hereafter referred to as the function $f(r)$. These observations can be quantified as the number of packages sent from location A to location B at distance r, the number of people moving from A to B at distance r, or the number of power line connections that connect substations at distance r in a power grid. The observed distribution function $f(r)$ is the result of two factors: the number of pairs of points that exist in the domain of interest, $\Omega$, with associated point density $p^d(x)$, and the one-dimensional probability $\pi(r)$ with which pairs of distance $r$ are manifested in the real world to be observed. In short, the number $f(r)$ of observations of distance $r$, is determined by the number of possible pairs $p(r)$ at distance $r$ and the probability $\pi(r)$ that they would exist at this distance, i.e.
\begin{align}
    \label{eq:observation-by-prob-and-pairwise_SI}
    f(r) \propto \pi(r)p(r),
\end{align}
bar a normalization constant.

When we measure $f(r)$, we always measure with it the geometry of the subspace $\Omega$ as well as the density $p^d(  x)$, manifested in the pair distribution function $p(r)$. The ``universal'', i.e. geometry-independent law that encapsulates the behavior of the system we want to study, is encoded in $\pi(r)$. To properly deduce this geometry-independent behavior of our system, we need to adjust our observation $f(r)$ by the geometry encoded in $p(r)$, i.e.
\begin{align}
    \pi(r) \propto \frac{f(r)}{p(r)}.
\end{align}

A pertinent question is which reference topology (geometry and point distribution) would result in an observation $f$ that is directly proportional to the behavior $\pi$. This question holds practical significance, as considering behavior benefits greatly from a conceptual framework for the distribution of points within a space. Indeed, our goal is to understand the mechanisms that link two points within this space. Without defining these points or the space itself, this endeavor becomes somewhat pointless.

\begin{figure}
    \centering
    \includegraphics[width=\textwidth]{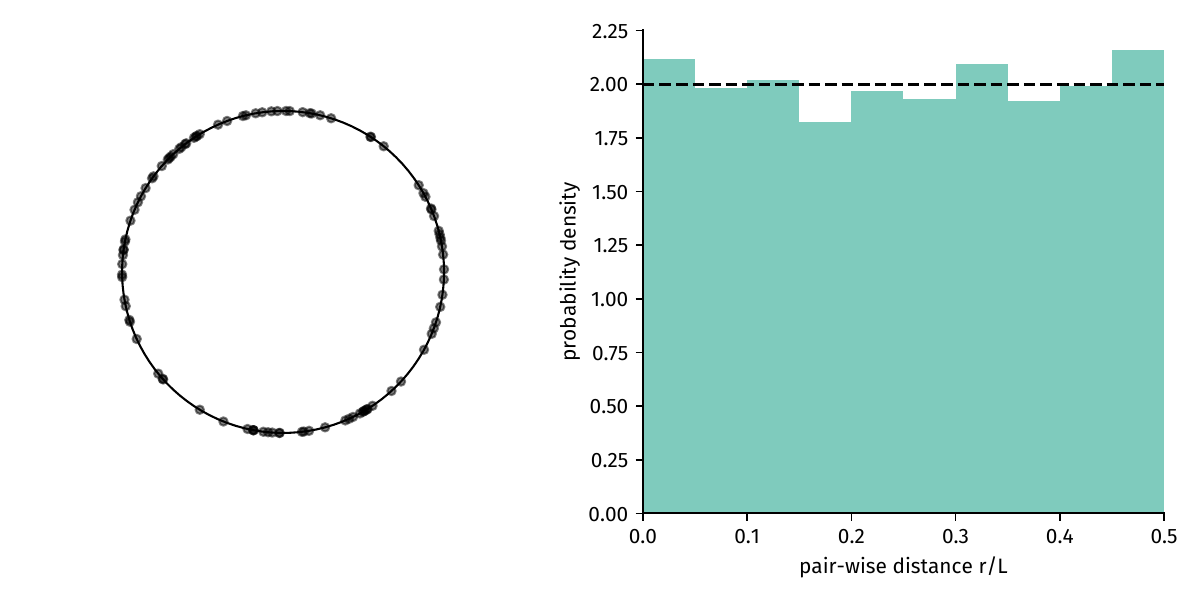}
    \caption{Given a ring with uniformly distributed points, the pair distance distribution of points will be constant.}
    \label{fig:reference-topology}
\end{figure}

In our formalism, this implies that we are seeking a topology where,
\begin{align}
    f(r) \propto \pi(r),
\end{align}
which implies
\begin{align}
    p(r) = \mathrm{const.}
\end{align}
Looking at Eq.~\eqref{eq:definition-pairwise-distribution-SI}, we have to find $\Omega$ and $p^d(  x)$ such that $p(r)=\mathrm{const}.$ While there might be a multitude of solutions, the simplest one is a one-dimensional ring, which can be conceptualized as a box of length $L$ and domain $x\in[0,L)$ with periodic boundary conditions, i.e. an associated distance
\begin{align}
    R(x,y)=\begin{cases}
        |x-y|, & 0\leq |x-y| \leq L/2,\\
        L-|x-y| & L/2 \leq |x-y| < L
    \end{cases} 
\end{align}
and uniformly distributed points, i.e.
\begin{align}
    p^{d=1}(x) = \begin{cases}
        L^{-1}, & \ 0\leq x< L,\\
        0,      & \ \mathrm{otherwise}.
    \end{cases}
\end{align}
then the pair-wise distance distribution evaluates to, 
\begin{align}
    p(r) &= L^{-2} \int_0^L dx \left(\int_{x-L/2}^x dy \delta(r - (x-y)) + \int_{x}^{x+L/2} dy \delta(r - (y-x))\right) \\
    &= \begin{cases}
            L^{-2} \int_0^L dx \left(1 + 1\right), &\  0 \leq r \leq L/2,\\
            0, &\ \mathrm{otherwise}
        \end{cases}\\
    &= \begin{cases}
            (L/2)^{-1} &\  0 \leq r \leq L/2,\\
            0, &\ \mathrm{otherwise},
        \end{cases}
\end{align}
for an illustration see Fig.~\ref{fig:reference-topology}.
Hence, the observed distance distribution $f(r)$ occurring between pairs of locations at distance $r$ on this topology will be proportional to the geometry-independent behavioral part $\pi(r)$. This technique has been used in \cite{wiedermann_spatial_2016,maier_generalization_2019,maier_modular_2019,maier_spreading_2020}.

\subsection{Three Denmark}

\label{SI:three_denmarks}

\subsection*{Illustrating scale via three Denmarks}
To illustrate how geography shapes the pair distribution and the observed moving distances, we consider in the main text the mobility trace simulation for three different geographies (Fig.\ref{fig:1}d,e,f). this section details the construction of each of the three Denmarks.

\subsubsection*{'Real Denmark'}

The first geography, 'Real Denmark,' uses the actual geography of Denmark, including 3.3M precise locations (addresses) (Fig.\,\ref{fig:1}d).  The precison of the location of adresses is +/- 2 meters (see Methods Data Section and \cite{dataforsyningen2023adgangsadresse}) and the location of an address is defined as the location of the door (see SI \ref{sec:shuffled_door}).

\subsubsection*{'Disk Denmark'}

The second geography, 'Disk Denmark,' maintains the microstructure of cities but alters city positions and landmass shapes, distributing real city centers uniformly on a disk (Fig.\,\ref{fig:1}e). 

First, the radius of the disk is chosen such that the area of the disk is the same as the area of Denmark. The area of Denmark is approximately 43,000 km\(^2\) \cite{runfola_geoboundaries_2020}. Therefore, the radius of the disk \(R\) is calculated by equating the area of a circle to the area of Denmark:

\begin{align*}
    \text{Area of Denmark} &= \text{Area of the disk}, \\
    43,000 \, \text{km}^2 &= \pi R^2 \\
        R &\approx 117 \, \text{km}.
\end{align*}

Thus, the radius of 'Disk Denmark' is 117 km. Hence, the limit at a bit more than 200km for the 'disk Denmark' in Figure \ref{fig:1}g,h.

Secondly, the position of the cities is chosen uniformly at random starting for the largest city to the smallest and ensuring they that they do not overlap as in \ref{sec:collision}.

\subsubsection*{'Uniform Denmark'}

The third geography, 'Uniform Denmark,' keeps the macro structure of landmasses and city positions from 'Real Denmark' but features uniformly dense cities (Fig.\,\ref{fig:1}f).

For this toy-geography, we use the shapes of the cities of Denmark provided by \cite{dataforsyningen2023adgangsadresse}. Defining the outlines of a city is a complex problem, and the results can vary depending on the method used \cite{arcaute2015constructing}. We tested the robustness of the official definition by applying hierarchical clustering on the precise location of the addresses (see SI \ref{sec:HDBSCAN}).

The positions of the addresses are then drawn uniformly at random within each city shape, keeping the empirical number of addresses in each city and ensuring there is no overlap with a collision detection algorithm.
\\

\subsection{Spatial descriptive statistics: Ripley's K and L}
\label{SI:ripley}

Ripley's K-function is used to describe the spatial distribution of point patterns \cite{ripley1977modelling}. It is defined as,

\begin{equation}
    K(r) = \frac{1}{\lambda} \mathbb{E}\left[N(r)\right]
\end{equation}

where \( K(r) \) is the expected number of points within distance \( r \) of a randomly selected point, \( \lambda \) is the intensity (the number of points per unit area), and \( N(r) \) is the number of other points within distance \( r \) of a typical point. The empirical estimate of the K-function from observed data is given by:

\begin{equation}
    \hat{K}(r) = \frac{A}{n^2} \sum_{i=1}^{n} \sum_{\substack{j=1 \\ j \neq i}}^{n} \delta(r_{ij} \leq r)
    \label{eq:ripley-k}
\end{equation}

where \( A \) is the area of the observation window, \( n \) is the number of observed points, \( r_{ij} \) is the Euclidean distance between points \( i \) and \( j \), and \( \mathbb{I}(r_{ij} \leq r) \) is an indicator function that takes the value 1 if the distance between points \( i \) and \( j \) is less than or equal to \( r \), and 0 otherwise.

The \( L(r) \)-function is a transformation of Ripley's K-function that normalize the expected values under complete spatial randomness (CSR) \cite{ripley1977modelling}. It is defined as:

\begin{equation}
    L(r) = \sqrt{\frac{K(r)}{\pi}} - r
\end{equation}

Under CSR, the expected value of the K-function is \( K_{\text{CSR}}(r) = \pi d^2 \), making the L-function tend to zero if the points are distributed uniformly:

\begin{equation}
    L(r) = 0 \quad \text{if points are distributed randomly.}
\end{equation}

\subsubsection*{Links between Ripley's function and the pair distribution function}

Interestingly, Ripley's K and L function are directly related to the pair distribution function. Indeed, the summation term $\sum_{\substack{j=1 \\ j \neq i}}^{n} \delta(r_{ij} \leq r)$ from equation (\ref{eq:ripley-k}) is exactly the cumulative distribution function (CDF) of the discrete pair distribution function. Indeed the discrete equivalent of equation (\ref{eq:definition-pairwise-distribution}) is,

\begin{align}
    p_{\text{discrete}}(r) = \frac{2}{n(n-1)} \sum_{i=1}^{n} \sum_{\substack{j=1 \\ j \neq i}}^{n} \delta(r_{ij} = r)
\end{align}

Thus, we can express Ripley's K function as,

\begin{align}
    K(r) = C \int_{0}^r p(s)ds
\end{align}

where the constant $C$ approaches $A/2$ as $n$ becomes large, which is typically the case. In conclusion, Ripley's K and L function are a primitive function of the pair distribution function.

\begin{figure}[h!]
    \centering
    \includegraphics[width=\textwidth]{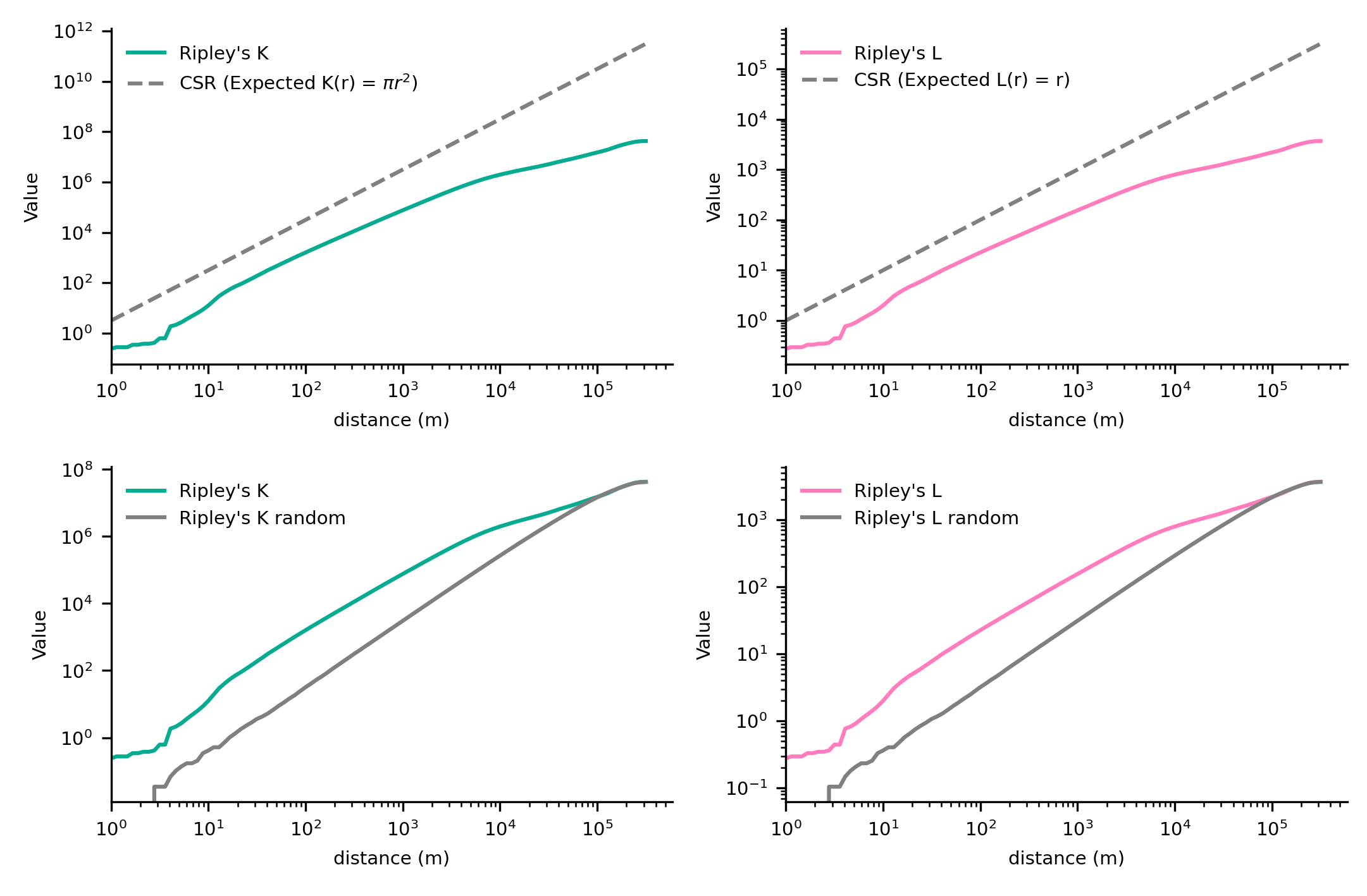}
    \caption{Ripley's K and L function for addresses in Denmark. We compare with the complete spatial randomness (first row, CSR) and the simulation of uniformly distributed addresses in the shape of Denmark (second row, random). When $K(r) > K_{\text{CSR}}(r)$, it indicates clustering, meaning more points are 
found within a given distance r than expected under complete spatial randomness. Conversely, $K(r) < K_{\text{CSR}}(r)$ suggests dispersion, where fewer points are found within distance $r$ than expected, indicating a more regular or evenly spaced distribution, while $K(r) = K_{\text{CSR}}(r)$
implies a uniformly random distribution with no significant clustering or dispersion.}
    \label{fig:mr-ripley}
\end{figure}

Figure \ref{fig:mr-ripley} shows the Ripley's K and L function for the addresses in Denmark. 
We observe that the spatial structure is broken up at same scales at the different Denmark models (Fig \ref{fig:1} and Fig \ref{fig:2}). 
First, we see a jump 6 meters (correspond to the first oscillation of Figure \ref{fig:structure-pairwise-distance}d), the one at 12 meters is harder to spot due to the cumulative nature of the figure. 
Secondly, from ~30m to 10km,the intra-city regime, the slope is 5/3, which correspond to the 2/3 regime we can see on Figure \ref{fig:1}h and Figure \ref{fig:structure-pairwise-distance}f, there is a +1 because of the Ripley's function is the anti-derivative of $p$.
Finally above 10km, the inter-city regime, the Ripley's K function converges to its uniformly random counterpart, meaning that at larges scales the position of the cities is indistinguishable from uniformly random as demonstrated in Figure \ref{fig:structure-pairwise-distance}e.

\subsection{Conditions for the pair distribution function to identify a set of points uniquely}
\label{SI:pdd-unique}
According to \cite{boutin2004reconstructing}, in general, point configurations can be uniquely determined by their pair distribution distributions, up to a rigid transformation. However. there are counterexamples, for example when two distances are equal. The counterexample starts with four points (triangles are uniquely identifiable). This counterexample can be extended to any number of points, as one can add infinitely many points on the dotted line of the two examples and their respective pair distribution would be the same.

In the context of geographic analysis, for a set of 2D points representing addresses, some distances between points will inevitably be repeated. If we consider $1 \times 10^{6}$ points located in a $1000\unit{km} \times 1000\unit{km}$ square and whose coordinates are known with a $1 \unit{m}$ precision, by a simple combinatorial argument, the number of pair distances is $1 \times 10^{12}.$ Although the maximum distance in the square is $\sqrt{2} \times 1000 \unit{km}$, or $1.414 \times 10^{6} \unit{m}$, which means that the set of possible distances contains $1.414 \times 10^{6} $ different values (due to the limited precision), but there are  $1 \times 10^{12}$ instances, so some distances must be equal, and the pair distribution function does not uniquely identify a set of 2d points. 

However, in the case of geography, due to the regularities and scaling law of the pair distribution, one can adopt a coarse-grained view of the problem. Instead of considering each point individually, we can first examine the pair distribution function between urban centers or other areas and iteratively reconstruct the geography (set of 2d points) in iterative. First, each urban center position should be a unique configuration, then independently on the local geography around each city.  This should lead to a unique configuration (up to isometries that are of the second order for the pair distribution function). The coarse-grained construction would be similar to the quad-tree partitioning or the HDSCAN clustering (see section~\ref{sec:HDBSCAN}).

\subsection{Pair distribution function for France}
\label{sec:pdd-fr}
\begin{figure}
    \centering
    \includegraphics[width=\textwidth]{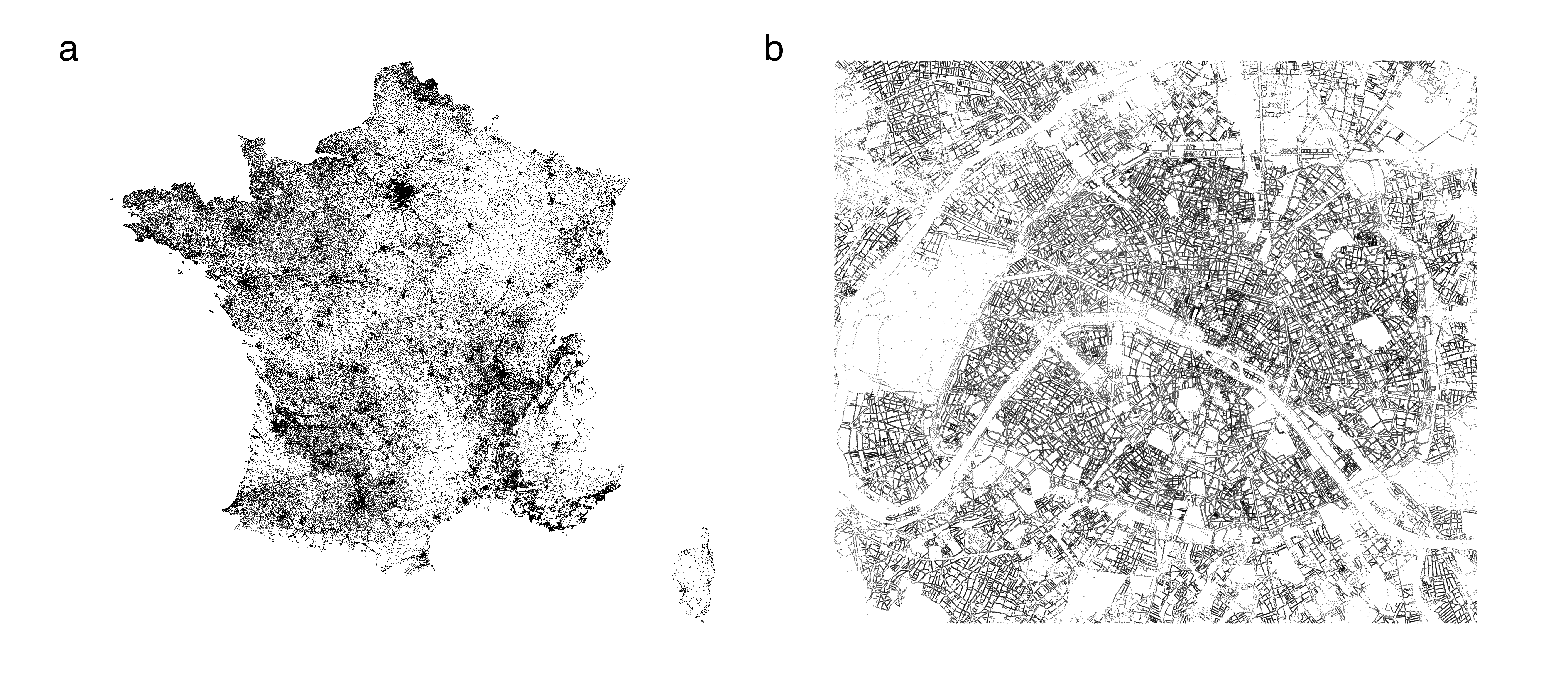}
    \caption{Map of 34,041,910 individual addresses in France \textbf{(a)}, and  zoom on the Paris area \textbf{(b)}.  }
    \label{fig:map-fr}
\end{figure}

\begin{figure}
    \centering
    \includegraphics[width=\textwidth]{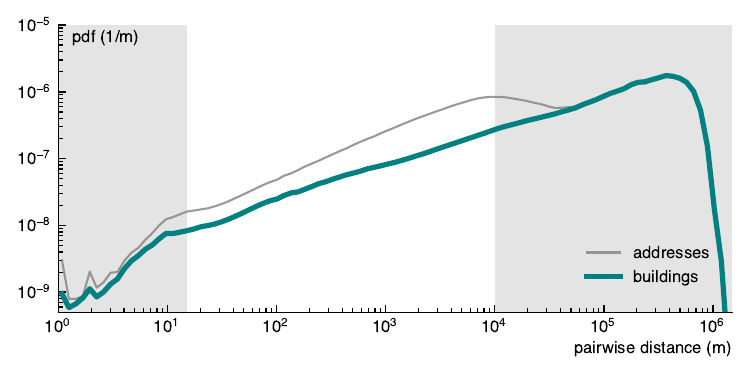}
    \caption{pair distribution of buildings and addresses in France. Neighborhood-scale with a linear onset and an oscillating modulation for the immediate neighborhood, then city-scale showing a power-law growth with exponent $\alpha = 0.67$, and finally country-scale with a slower growth and eventual fast decay.}
    \label{fig:pdd-fr}
\end{figure}

The pair distribution function between residential buildings in France shows similar patterns as the one of Denmark, as shown in Fig.\ref{fig:pdd-fr}. On the micro-scale (i.e. ~within distances of $r \lesssim 25\unit{m}$, the ``neighborhood''). We observe a linear onset of neighborhood density, oscillatory modulated. For larger distances (mesoscale), this growth assumes a scaling of approximately $p(r)\propto r^{0.67}$ between $25\unit{m}$ and $200\unit{m}$, which is even more pronounced in the distribution of address distances (up to $10000\unit{m})$. 
The mesoscale has a larger amplitude for France than for Denmark ($10\unit{km}$ vs $4\unit{km}$)due to larger urban areas (Paris $2853\unit{km}^2$ vs $526\unit{km}^2$ for Copenhagen). After that, at the macro scale, this growth slows down, decays rapidly, and finally approaches zero as we reach the limit of the finite system. 

\section{Models for pair distribution function}
\subsection{Pair distribution functions}
\label{sec:handling-generalized-pairw}
\subsubsection{Generalized pair distribution function}

In the main text, we defined the generalized pair distribution function model as
\begin{align}
        \label{eq:pdd-gen}
        p_m(r,R) = C\frac{r}{R^2} \exp\left(-\left(\frac r R\right)^m\right).
\end{align}
It has a linear onset and a tail that falls as a stretched exponential and is a special case of the generalized Gamma distribution. We find $C$ as follows. Begin with the Gamma function
\begin{align}
    \Gamma(z) = \int\limits_0^\infty t^{z-1}\exp\left(-t\right)\mathrm d t 
\end{align}
and substitute $t = (r/R)^m$ such that $\mathrm d t = (m/R) (r/R)^{m-1} \mathrm d r$ and
\begin{align}
    \Gamma(z) =\frac m R \int\limits_0^\infty (r/R)^{mz-m+m-1}\exp\left[-(r/R)^m\right]\mathrm d r.
\end{align}
Now we demand $mz-1 = 1$, i.e. $z = 2/m$, to find
\begin{align}
    \Gamma(2/m) = m{C^{-1}} \int\limits_0^\infty C\frac{r}{R^2}\exp\left[-(r/R)^m\right]\mathrm d r.
\end{align}
Due to the normalization condition, we have 
\begin{align}
    C= \frac m {\Gamma(2/m)}.
\end{align}
To obtain the first moment, we demand $mz-1=2$ such that $z=3/m$ so we find
\begin{align}
    \Gamma(3/m) &= \frac{mC^{-1}} R \int\limits_0^\infty C\frac{r^2}{R^2}\exp\left[-(r/R)^m\right]\mathrm d r\\
                &= \frac{\Gamma(2/m)} {mR} \left<r\right>\\
         \left<r\right> &= \frac{mR\Gamma(3/m)}{\Gamma(2/m)}.        
\end{align}
We can fit this distribution to data by using the per-sample log-likelihood
\begin{align}
    {\mathcal L} = \frac{1}{n}\ln L &=
    \ln {m} -\ln \Gamma(2/m) + 2\ln w 
     +\left<\ln r\right> - w^m \left< r^m \right>.
\end{align}
where $w = 1/R$ and we have an observational set $\{r_i\}$ of empirical pairwise distances with sample size $n$. From $\partial \mathcal L/\partial w=0$ we find that the inverse city scale that maximizes the likelihood is given by
\begin{align}
    \hat w(m) = \left(\frac{2}{m\left<r^m\right>}\right)^{1/m}.
\end{align}
With $\hat w(m)$, we can find the zero of
\begin{align}
\frac{\partial \mathcal L}{\partial m} = \frac{1}{m} + \frac{2}{m^2}\psi(2/m) - \Big<[\hat w(m) r]^m \ln[\hat w(m) r]\Big>
\end{align}
numerically, which gives $\hat m$. Here, $\psi(z)=\Gamma'(z)/\Gamma(z)$ is the digamma function.

\subsubsection{Circle}

The pair distribution of a uniform distribution of random points within a disk of radius $R$ is given by
\begin{align}
    p(r,R) = \begin{cases}
\frac{4r}{\pi R^2}\arccos\left(\frac{r}{2R}\right)
    -\frac{2r^2}{\pi R^3}\sqrt{1-\frac{r^2}{4R^2}},&\qquad r \leq 2R,\\
        0&\qquad \mathrm{otherwise,}
    \end{cases}
    \label{eq:circle-pairwise}
\end{align}
see \cite{mathworld}.

\subsubsection{Parabola}

Looking at Eq.~\eqref{eq:circle-pairwise}, we see that this distribution looks somewhat close to a parabola with zeros at $r=0$ and $r=2R$. A parabolic pair distribution  with such properties is given as
\begin{align}
\label{eq:pdd-parabola}
    p(r,R) = \begin{cases}
        -\frac 3 {4R} \left(\frac r R \right)^2 + \frac {3r} {2R^2}&\qquad r \leq 2R,\\
        0&\qquad \mathrm{otherwise.}
    \end{cases}
\end{align}
Note that this is the Beta distribution with $\alpha=2$ and $\beta=2$ for random variable $x=r/R$.

\subsubsection{Building locations by external and interaction potentials}
\label{sec:pdd-interactions}
Consider the location of a city as the literal center of interest, for example, the ``central business district''. Since it might be attractive for individuals to reach this center as quickly as possible (because amenities will be close to the center), we assume that there is an increased cost of living at a distance $|\bm x|$ from the center. For example, consider having to commute to a job within the central business district, which has a cost that increases with $|\bm x|$.
At the same time, two simple mechanisms will prevent buildings from accumulating in the exact center of the city. First, buildings have a certain average radius $z$, so they cannot be too close together (at a distance $<2z$). There is an advantage to buildings not being too far apart, as they can share local amenities. Instead of modeling this explicitly, we simply assume that there is an inherent temperature $T$ in the system, according to which house locations are distributed following an interaction and an external potential.

From a statistical physics point of view, we describe the system with the simplest external potential, which increases linearly with distance
\begin{align}
    V^\mathrm{ext}(\bm x) = \gamma|\bm x|
\end{align}

where we assume that the origin of the coordinate system is in the center of the city. To model repulsion and attraction between houses, we also assume a Lennard-Jones interaction potential
\begin{align}
    V^{\mathrm{int}}(\bm x_i,\bm x_j) = \varepsilon \left[
                                \left(\frac{2z}{|\bm x_j - \bm x_i|}\right)^{12}
                            -   2\left(\frac{2z}{|\bm x_j - \bm x_i|}\right)^6
                              \right],
\end{align}
which is commonly used to model simultaneous attraction and repulsion between molecules in chemical solutions \cite{LennardJones1931, HansenMcDonald1986}.

In total, a system with these properties evolves according to the Hamiltonian
\begin{align}
    \mathcal H(\{\bm x, \bm p\}) = \frac12\sum_{i=1}^N \bm p_i^2 + \gamma\sum_{i=0}^N |\bm x_i| + \frac 12 \sum_{i=1}^N\sum_{j\neq i}^N V^{\mathrm{int}}(\bm x_i, \bm x_j)
\end{align}
with two-dimensional momenta $\bm p_i$ and locations $\bm x_i$. In a canonical-ensemble formulation of the system, i.e. at constant inverse temperature $\beta$, the probability of finding a configuration $\{\bm  x, \bm p\}$ in volume-element $d^2\bm x d^2\bm p$, is given by
\begin{align}
    \varrho[\{\bm  x, \bm p\}] d^2\bm x d^2\bm p &=
        \exp[-\beta \mathcal H(\{\bm x, \bm p\})] d^2\bm x d^2\bm p.% \\
\end{align}

For now, we restrict ourselves to an ideal gas with $V^\mathrm{int}=0$, which will allow us to say something about the density of particles around the center of the city, i.e., we want to find the probability of a particle being present $p(r)dr$. Without loss of generality, we set $N=1$, because due to ergodicity the trajectory of a particle will eventually follow the density of the whole distribution (think of it as taking the $N$th root of the $N$ particle density).  Integrating over the momenta yields
\begin{align}
    \varrho[\bm x] d^2\bm x  &=
        \exp\left(-\beta \gamma |\bm x|\right) d^2\bm x% \\
\end{align}
Changing the variables to polar coordinates, where $r$ is the distance of the particle from the center, we find
\begin{align}
    \varrho[r, \phi] r dr d\phi = r \exp\left[-\beta \gamma r\right] dr d\phi,
\end{align}
i.e. 
\begin{align}
    p(r) dr = r \exp\left[-\beta \gamma r\right] dr.
\end{align}

This means that if there are no interactions, the distribution of houses around the city center should follow an Erlang distribution with scale parameter $\lambda=\beta \gamma=1/R_0$, where $R_0$ is half the city radius. Note that we have postulated that all particles have the same mass $m=1$, so $\gamma r$ has the dimension of energy. The definition of the inverse temperature $\beta = 1/T$ implies that the temperature also has an energy dimension.

Relying on the arguments of the kinetic theory of gases, we can relate the temperature to the momenta of a particle with the identity
\begin{align}
    K = N_f N \frac T 2
\end{align}
where $K=(1/2)\sum_i p_i^2$ is the kinetic energy and $N_f$ is the degree of freedom of each particle, i.e. for single-atom particles in two dimensions, $N_f=2$ (two translational degrees of freedom, no rotations, no oscillations). Note that this relates to the root-mean-square velocity of a single particle as
\begin{align}
     v_0\equiv\sqrt{\left<v^2\right>} = \sqrt{ T/2}.
\end{align}

In this sense, the instantaneous temperature plays the role of a particle's ability to overcome the potential energy. If $V^\mathrm{ext}(r)=\gamma r$ represents a \emph{cost} of being at distance $r$ from the center, the temperature gives a measure of how well particles in the system can overcome that cost. If the temperature is low, particles cannot overcome this cost and the density in the city center will be high. If the temperature is high, particles can overcome the cost easily because of the larger amount of kinetic energy available in the system.

Increasing complexity by going back to the Lennard-Jones perspective, we want to obtain an intuition about how the radial particle density changes when they strongly repel each other. In the limit of $\varepsilon/T\gg1$, particles will have a strong tendency to be found in their respective potential minimum, i.e. at distance $2z$ from each other. Effectively, we can think of them as hard disks of radius $z$ with a tendency to form clusters. If the temperature is low, the effective radius of the city will be small (remember that the Erlang shape parameter is $\lambda = \gamma/T = 1/R_0$). In this case, it may happen that the number of particles that we would expect to lie within radius $r$ from the center (according to the ideal gas) will be greater than the number of hard disks that can fit within a circle of radius $r$. When this happens, we expect a crystal to form at the center.

The maximum number of disks that can fit within a circle of size $r$ can be approximated by
\begin{align}
    N_\mathrm{disks}(r) = \theta A_\mathrm{circle}(r)/A_\mathrm{disk}. 
\end{align}
Here, $A_\mathrm{circle}(r)$ is the area of a circle of radius $r$, $A_\mathrm{disk}$ is the area of a circle of radius $z$, and $\theta$ is an optimal packing fraction, where we can approximately assume $\theta\approx0.9$ for optimal hexagonal packing. Then,
\begin{align}
    N_\mathrm{disks}(r) = \theta r^2/z^2.
\end{align}
From the ideal gas distribution, we expect to find 
\begin{align}
    N_\mathrm{id}(r) &= N P(r) \\
                     &= N \left(1 - \mathrm e^{-\gamma r/T}\left(1+\frac{\gamma r}{T}\right)\right)
\end{align}
within radius $r$ (where $P(r)$ is the Erlang cumulative density function).

Following the aforementioned argumentation, we expect a nucleation effect when $N_\mathrm{id}(r)>N_\mathrm{disks}(r)$ for $gr/T\ll1$. Linearizing the exponential factor, that happens when
\begin{align}
    N_\mathrm{id}(r) &> N_\mathrm{disks} \\
    N \left(1 - \left(1-\frac{\gamma r}{T}\right)\left(1+\frac{\gamma r}{T}\right)\right)  &>  \theta \frac{r^2}{z^2}\\
    N\left(1-\left(1-\frac{(\gamma r)^2}{T^2})\right)\right) &> \theta \frac{r^2}{z^2} \\
    \frac{N(\gamma r)^2}{T^2} &> \theta \frac{r^2}{z^2} \\
    \frac{N \gamma^2 z^2}{\theta T^2} &> 1,
\end{align}
or in terms of the reference velocity $v_0$ and the Lennard-Jones distance $d$,
\begin{align}
    \frac{N \gamma^2 d^2}{8\theta v_0^2} &> 1.
\end{align}
The radius of this cluster will approximately be given by the solution to the equation $N_\mathrm{id}(r)=N_\mathrm{disks}(r)$, which can be obtained numerically.

\subsubsection{Molecular Dynamics simulation}
\label{sec:pdd-md-simulation}
We are interested in finding configurations $\mathcal C$  that accurately represent the canonical ensemble with number of particles $N$, constant but irrelevant volume \footnote{because we force the particles to be confined within a radially symmetric external potential, the total volume containing the particles does not matter.}, and average-constant temperature $T$ with total potential energy of 
\begin{align}
    V = \sum_{i=1}^N \left[ V^{\mathrm{ext}}(\bm x_i) + \frac{1}{2}\sum_{j\neq i}V^{\mathrm{int}}(\bm x_i, \bm x_j)\right].
\end{align}
To this end, we integrate the equations determined by the system's Hamiltonian numerically using the velocity-Verlet algorithm \cite{verlet1967computer}. We also rescale the particle velocities according to the stochastic Berendsen thermostat \cite{berendsen1984molecular} with relaxation time $\tau$. To initiate the system in a state of sufficiently low potential energy, we assign initial particle positions according to the corresponding ideal gas ensemble. Then we run the collision algorithm described in Sec.~ with collision strength $u=1$. Moreover, we set the temperature by defining the root-mean-square initial velocity $v_0$ per particle and assigning a random velocity vector drawn from a two-dimensional Gaussian distribution with standard deviation $v_0$.

\begin{figure}
    \centering
    \includegraphics[width=11cm]{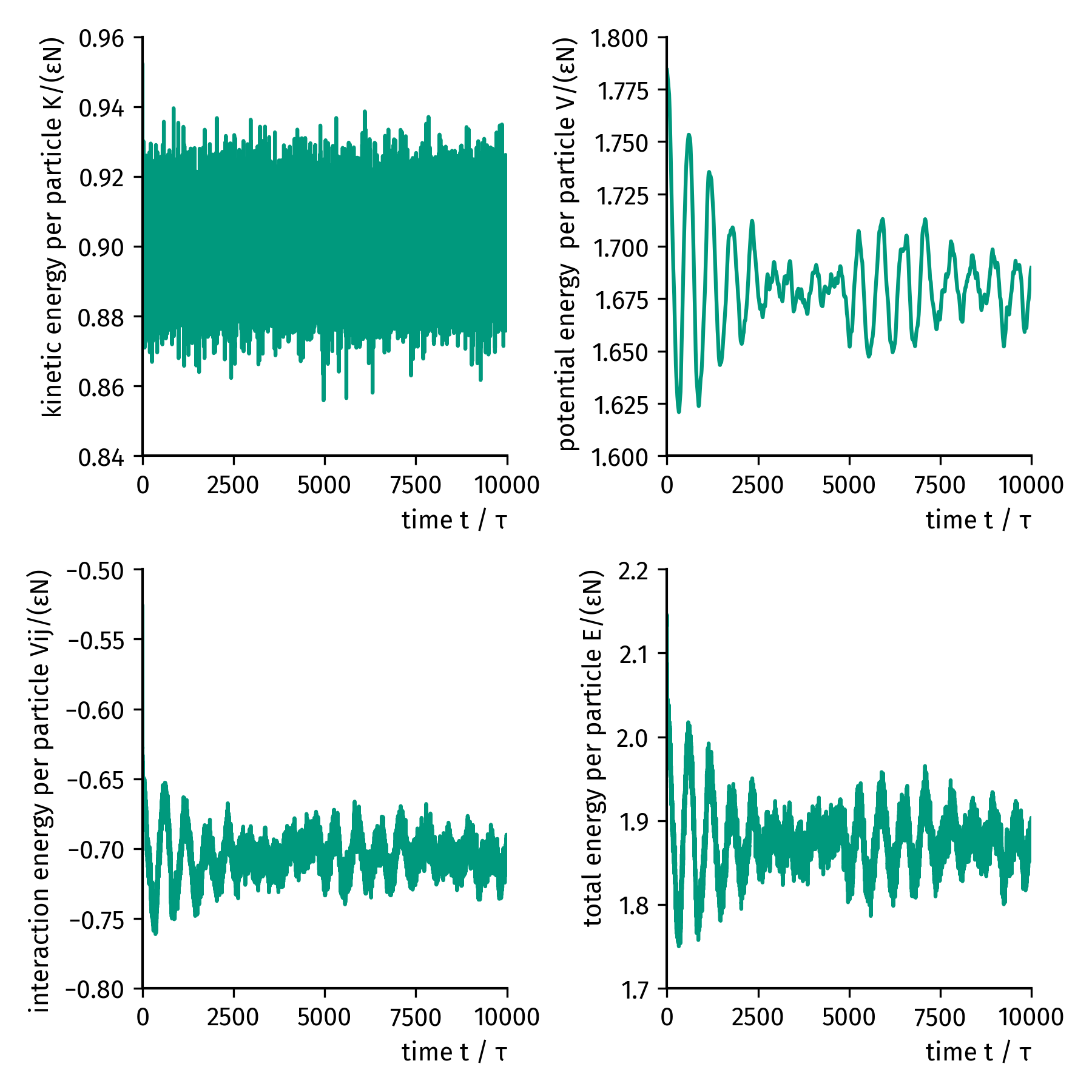}
    \caption{Energy evolution of a single MD simulation of a canonical ensemble of LJ particles in a linear external potential, thermalized by a stochastic Berendsen thermostat. Energies are displayed per particle in units of the LJ potential depth $\varepsilon$. Time is units of the stochastic Berendsen relaxation time $\tau$.}
    \label{fig:energy-time-series}
\end{figure}

\begin{figure}
    \centering
    \includegraphics[width=\textwidth]{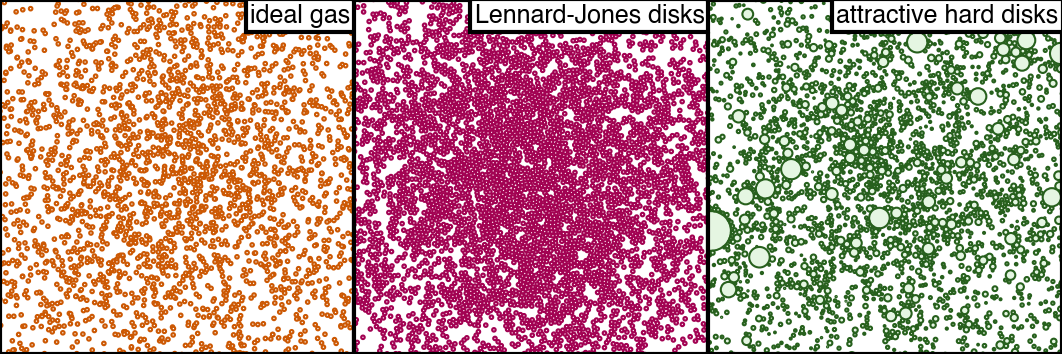}
    \caption{
    Three models for positioning buildings within an external potential of shape $V(\bm x) = \gamma |\bm x|$. (Left) An ideal gas, forming with radial density $p(|\bm x|) = (|\bm x|/R_0^2)\exp(-|\bm x|/R_0)$. (Middle) A configuration of Lennard-Jones disks. (Right) A configuration of heterogeneously sized hard disks with abstract attractive force.
    }
    \label{fig:naive-positioning-of-buildings-in-patch}
\end{figure}

As parameters, we choose $v_0=6$, $\gamma=0.08$, $N=10^4$, $\Delta t = 0.01$, $\varepsilon=20$, $z=3$, $\tau=100\Delta t$. To speed up the numerical integration, we only consider pairs of particles that lie within distance $r\leq 6z$, found by constructing and querying a k-d-tree for each time step. When calculating the interaction energies, we therefore shift the potential $V_\mathrm{LJ}$ so that $V_\mathrm{LJ}(r=6z)=0$. We integrate the equations of motion until $t=10^{4}\tau$. The energy time series for a single run can be seen in Fig.~\ref{fig:energy-time-series}. The final configuration of this run is shown in Fig.~\ref{fig:naive-positioning-of-buildings-in-patch}.

\subsubsection{Emulating attractive hard disks}

We generate a configuration of hard disks with an unspecified attractive force (i.e.~we do not explicitly integrate the equations of motion for a hard-sphere interaction potential with an additional attractive force). To do this, we first generate $N=10^4$ random positions according to a radial Erlang distribution with scale parameter $R=675$ (i.e.~an ideal gas). Then, we draw a random radius for each position from a heterogeneous distribution with power-law tail. We first draw values $\hat z_i$
\begin{align}
    p(\hat z) = \frac{a^2-1}{2a}\times\begin{cases}
            \hat z^a, & \hat z \leq 1 \\
            \hat 1/z^{a}, & \hat z > 1%
    \end{cases}
\end{align}
and then assign disk radius $z_i = 3 \hat z_i/\left<\hat z\right>$. We choose $a=4$ to obtain a heterogeneous distribution with non-finite variance in disk area. Afterward, we run the collision algorithm outlined in Sec.\ref{sec:collision} with collision strength $u=1$. This leads the initially overlapping disks to take positions where their boundaries touch, i.e. an unlikely configuration to be found in the absence of an attractive force that would cause the disks to lie right at each other's boundaries.

An example configuration of this method is displayed in Fig.~\ref{fig:naive-positioning-of-buildings-in-patch}.

\begin{figure}
    \centering
    \includegraphics[width=\textwidth]{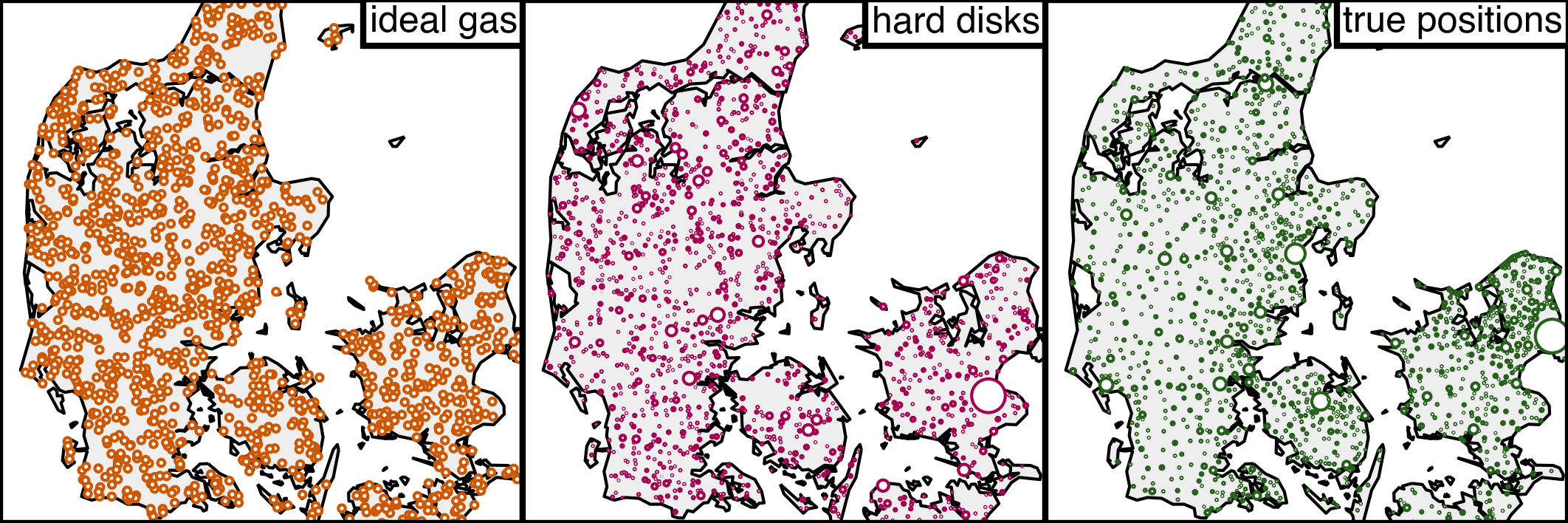}
    \caption{Three models for city positioning within Denmark. (Left) An ideal gas, i.e.~positions chosen uniform at random within the administrative boundaries. (Middle) Hard disks are placed in a non-overlapping, random manner (see Sec.~\ref{sec:hard-disks-random}). Radiuses are equal to the inferred $R$ values from the pair distribution  MLE of Eq.~\ref{eq:pdd-gen} to Denmark's towns with more than 30 buildings. (Right) Cities placed at the centroid of their respective administrative boundaries with radius equal to $R$ inferred from the respective city's empirical pair distribution .}
    \label{fig:city-positioning-in-denmark}
\end{figure}

\subsection{Collision-resolving algorithm}
\label{sec:collision}
We have $N$ disks with initial positions $\bm x_i$, radii $R_i$, and diameters $D_i=2R_i$, respectively. With the distance vector $\bm r_{ij}=\bm x_i - \bm x_j$ and distance $r_{ij}=|\bm r_{ij}|$, let
\begin{align}
	\mathcal E=
	\left\{(i,j) : (i<j) \wedge (r_{ij} < \max(D_i, D_j))
	            \wedge (r_{ij} < R_i + R_j)
	\right\}
\end{align}
be the set of pairs of disks that overlap. This set can be found by iterating over all disks $i$, finding all neighbors within distance $D_i$, for instance using a k-d-tree. With this definition, let
\begin{align}
	\mathcal J_i = \{j: (i,j) \in \mathcal E \vee (j,i) \in \mathcal E\}
\end{align}
be the set of disks that overlap with disk $i$. Then, 
\begin{align}
	\tilde {\Delta}_i = u \sum_{j\in \mathcal J_i} \frac{\bm r_{ij}}{r_{ij}}
    \big(R_i+R_j-r_{ij}\big)
 \frac{m_j}{m_i+m_j}.
\end{align}
is the demanded initial displacement, with the displacement rate $0<u\leq1$ (imagine two overlapping disks---with $u=1$, the collision would be resolved after one update). The masses $m_j$ control the strength of the displacement. Suppose that disk $i$ has s small mass and a colliding disk $j$ has a great mass. The colliding disk $j$ should move less. Hence, the influence on disk $i$ from disk $j$ should be proportional to $m_j$. Assuming homogeneous density of all disks, we can set
\begin{align}
    m_i = R_i^2.
\end{align}

We also want to avoid that disks move too far per one single update to avoid large jumps. Therefore, we move each disk by the final displacement vector
\begin{align}
  \Delta_i = {\tilde \Delta_i}\times \min(1,R_i/|\tilde \Delta_i|).
\end{align}
i.e. the disk shouldn't move more than its radius per update.

We update the whole ensemble of disks step by step until either
\begin{align}
    %\min\limits_{(i,j)\in\mathcal E} r_{ij} > r_\mathrm{crit}.
    |\mathcal E| = 0
\end{align}
or
\begin{align}
    \max\limits_i|\Delta_i| < \epsilon.
\end{align}
With a default value of $\epsilon=10^{-10}\unit m$.

If all disks are of equal radius $R$, we instead choose the second stop condition as 
\begin{align}
    \min\limits_{(i,j)\in\mathcal E} r_{ij} > (1-\epsilon) 2R
\end{align}
and $\epsilon=10^{-3}$.
	
\subsection{Random non-overlap positioning algorithm of disks}
\label{sec:hard-disks-random}
We want to randomly distribute $N$ hard disks of radii $R_i$ in shape $\Omega$. We start with the largest disk of radius $R_\mathrm{max}$ and iterate over all disks in decreasing order of size. For each disk $j$, we generate a random position $\bm x \in \Omega$ until the condition $|\bm x - \bm x_i| > R_j+R_i\ \forall j<i$ is satisfied, i.e.\ drawing new random positions until there are no overlaps with other, already placed disks. Then, assign $\bm x_j\leftarrow \bm x$ and continue with the next disk.

\subsection{Influence of building location definition on pair distribution }
\label{sec:shuffled_door}
\begin{figure}
    \centering
    \includegraphics[width=9cm]{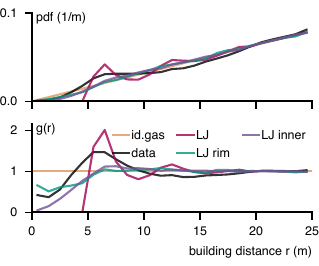}
    \caption{How the LJ pair distribution  and pair-correlation function changes when building location is redefined to (i) randomly within the disk of radius $z$ and (ii) randomly on the rim of disk of radius $z$.}
    \label{fig:shuffled-door}
\end{figure}

In the dataset, the location of a building is defined as the location of the building's entrance door. We want to check how the Lennard-Jones ensemble's pair distribution and the pair-correlation function $g(r)$ change when the location of a building is not associated with its center. To this end, we take the configuration of Lennard-Jones disks as shown in Fig.~2d in the main text and redefine a disk's location to be (i) randomly within the disk and (ii) randomly on the rim of the disk. The resulting pair distribution and $g(r)$ are shown in Fig.~\ref{fig:shuffled-door}. We see that the sharp peak almost disappears for both. At the same time, the onset of $g(r)$ becomes less abrupt, approaching a shape similar to that observed in the data.

\subsection{Measuring the pair distribution function for large dataset}
\label{SI:pdd-compute}

Calculating the pair distribution function for large datasets requires a substantial amount of memory. The number of different pairs among $N$ points is $N(N-1)/2$. Consequently, for the dataset of 34 million address coordinates within France (Fig.\ref{fig:map-fr}), we need to store $10^{15}$ distance values in memory, which is equivalent to 4,000 terabytes of data using 32-bit floats. Such a large amount of data storage is practically unfeasible. To overcome this computational hurdle, we proceed in two steps using a k-d-tree.

\subsubsection{Small distances}

Let $\mathcal D$ be the two-dimensional, non-contiguous shape containing every building (or address, respectively) in Denmark and $\delta$ the set of buildings (addresses) $i$ with locations $\bm x_i\in \mathcal D$.  For each predefined subset of this form $\Omega\subseteq\mathcal D$ (for example, the official boundaries of a city), we iterate over all building (address) positions $\omega = \{i \in D: x_i\in\Omega\}$ located within $\Omega$ and compute the distances to all buildings (addresses) $j\in \delta \wedge j\neq i$ with $0<|x_i-x_j|<r_{\max}=200\unit m$. That is, for each predefined subset of $\mathcal D$ (e.g. city), we find all distances of every one of its buildings with respect to \emph{all} buildings (addresses) that lie within radius $r_{\max}$, not just to those that also lie within its boundaries. We compute this histogram with a resolution (i.e.~bin width) of $1\unit m$. For this task, we use a k-d-tree on all locations of $\delta$.

For each city $i$ in Denmark, let $\Omega_i$ be the shape that is defined by its administrative boundaries. Note that none of the $\Omega_i$ overlap. We define as
\begin{align}
    \overline{\Omega} = \mathcal D \Big\backslash \left(\bigcup_{i} \Omega_i \right)
\end{align}
the shape that includes land that is not associated with a city. Thus, iteration over all $\Omega_i$ and $\overline{\Omega}$ allows us to analyze the small-scale structure of every city, every building (address) that is not located in a city, and - by combining all these histograms - of every building (address) in Denmark.

\begin{figure}
    \centering
    \includegraphics[width=.49\textwidth]{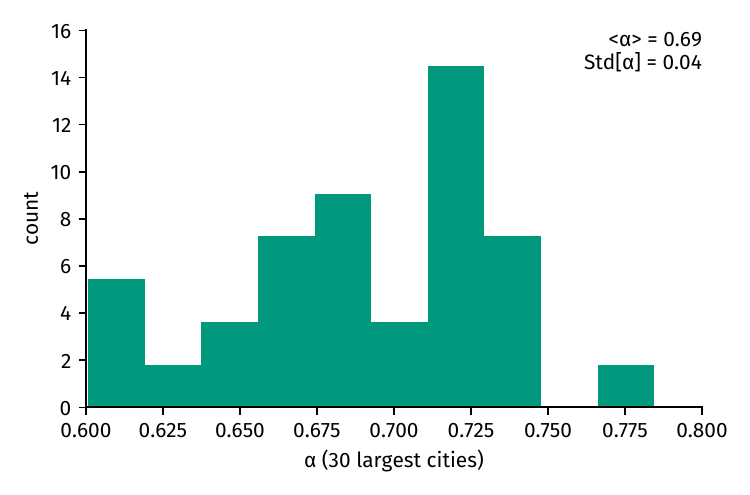}
    \includegraphics[width=.49\textwidth]{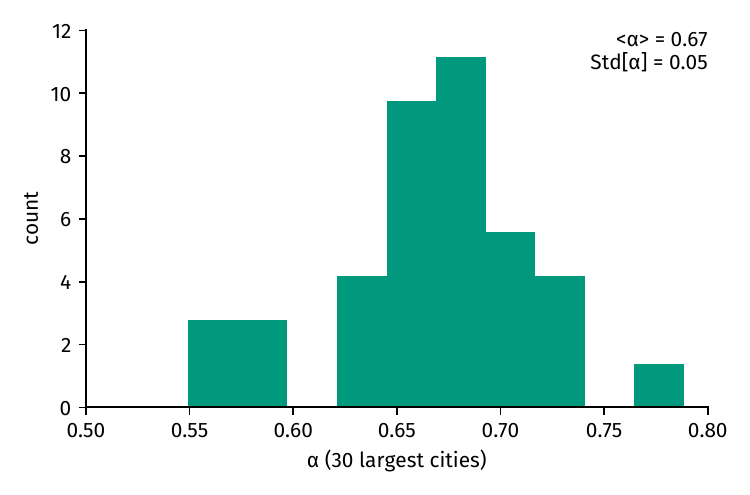}
    \caption{Inferred values of the sub-linear scaling exponent $\alpha$ in $p(r)\propto r^\alpha$ for $r\in(25\unit m, 200\unit m)$ for the 30 largest Danish cities. (Left) buildings. (Right) addresses.}
    \label{fig:inferred-alphas}
\end{figure}

\begin{figure}
    \centering
    \includegraphics[width=\textwidth]{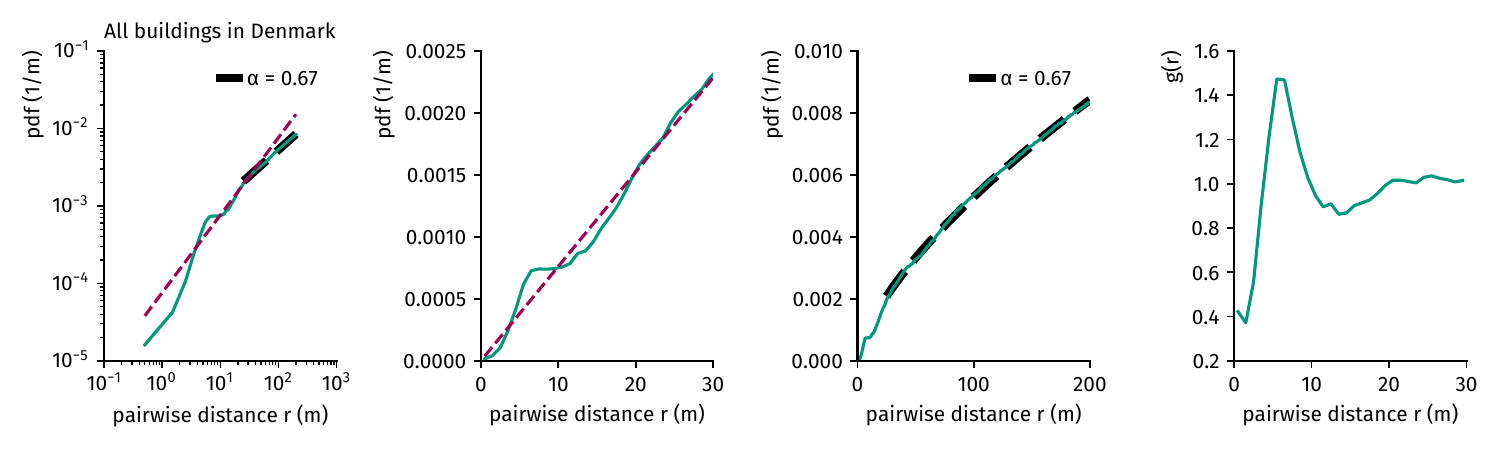}
    \includegraphics[width=\textwidth]{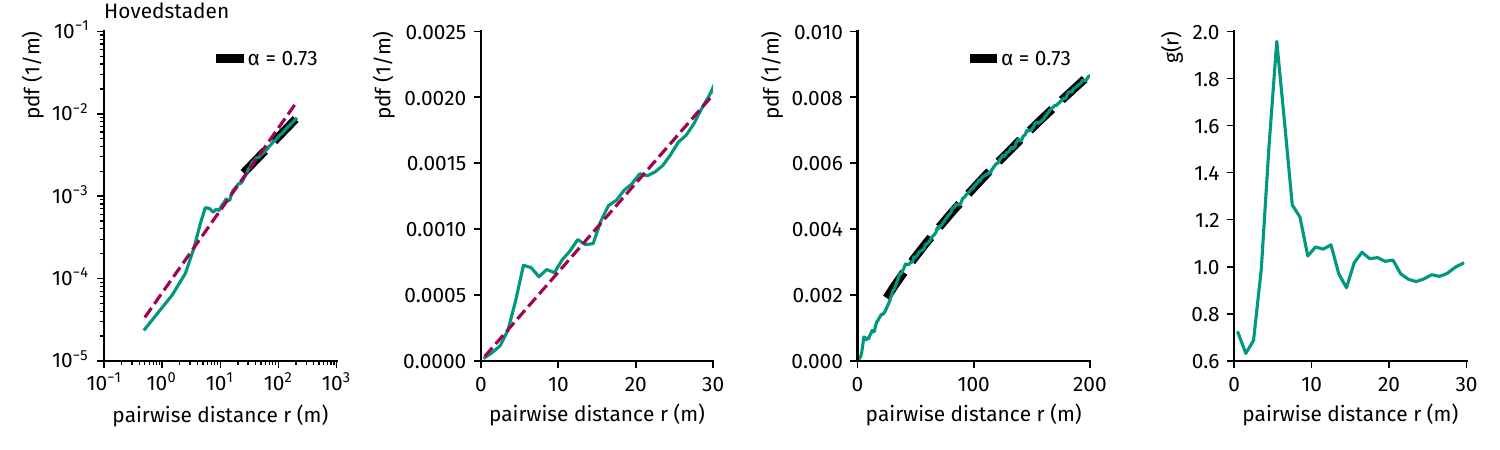}
    \includegraphics[width=\textwidth]{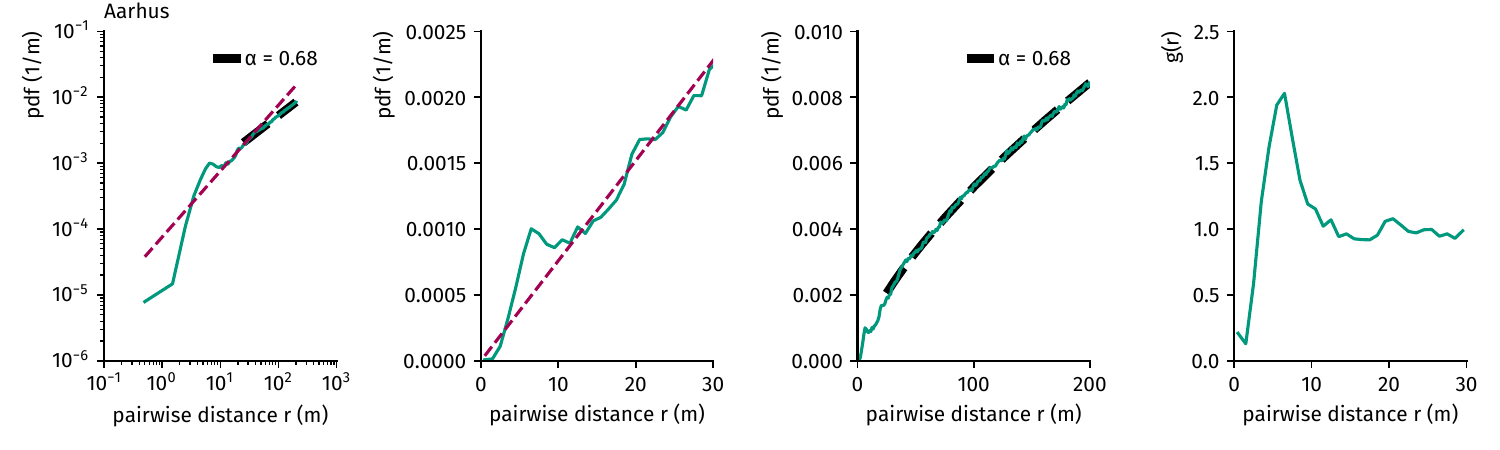}
    \includegraphics[width=\textwidth]{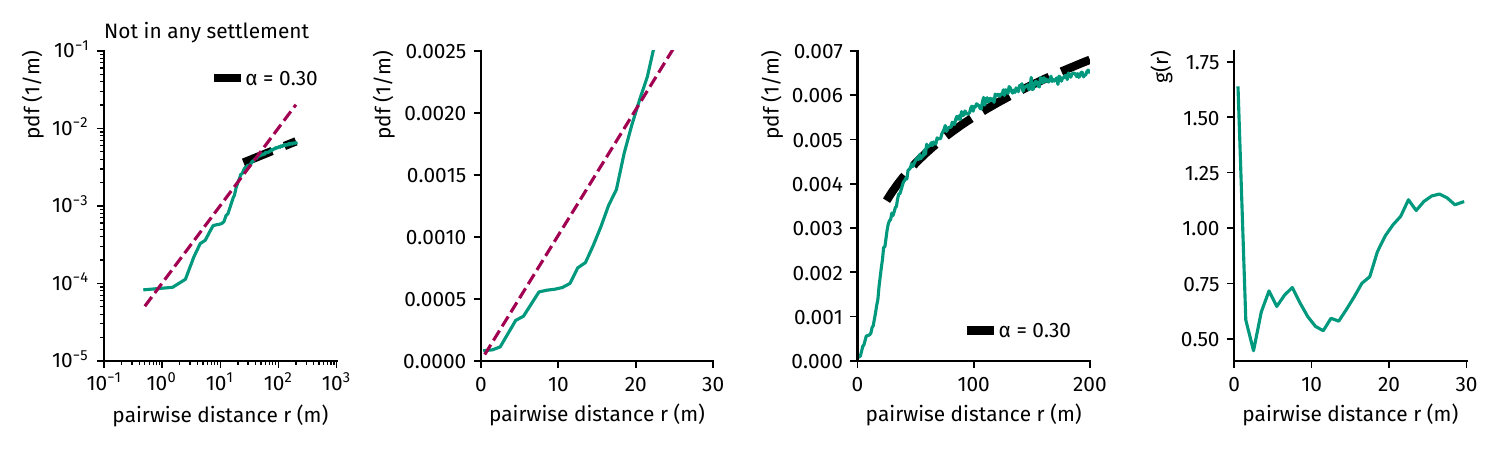}
    \caption{Building pair distribution functions and pair-correlation functions $g(r)$ with $r<200\unit m$ for (top) every building of Denmark, (second from top) Hovedstadens (capital region), (second from bottom) Aarhus, and (bottom) every building not located with the administrative boundaries of any city/town.}
    \label{fig:small-pdd-g-r-of-buildings}
\end{figure}

\begin{figure}
    \centering
    \includegraphics[width=\textwidth]{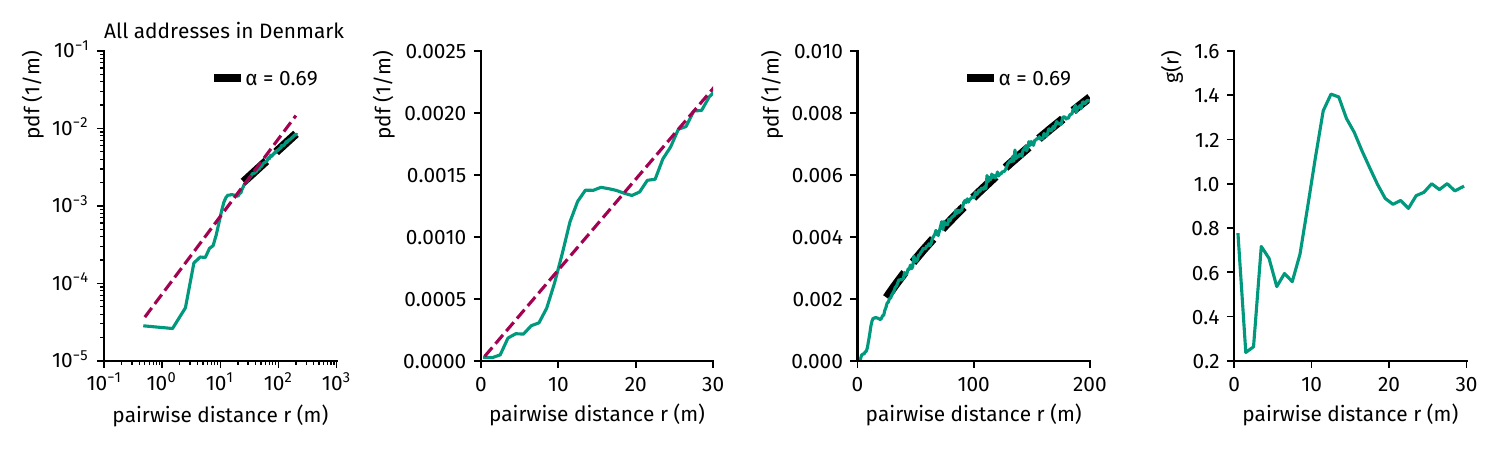}
    \includegraphics[width=\textwidth]{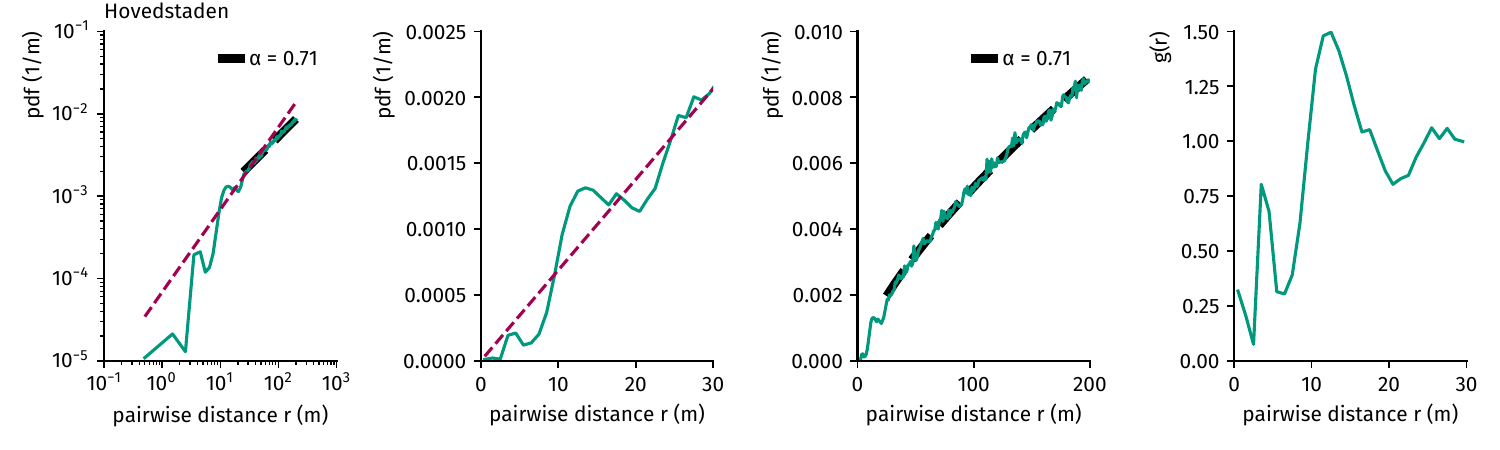}
    \includegraphics[width=\textwidth]{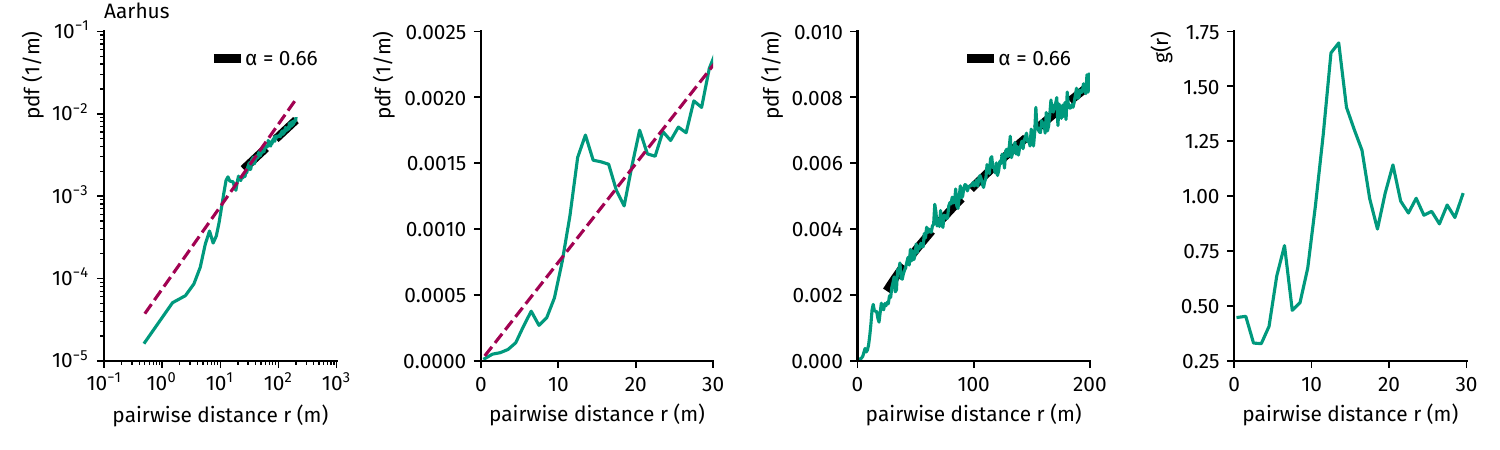}
    \includegraphics[width=\textwidth]{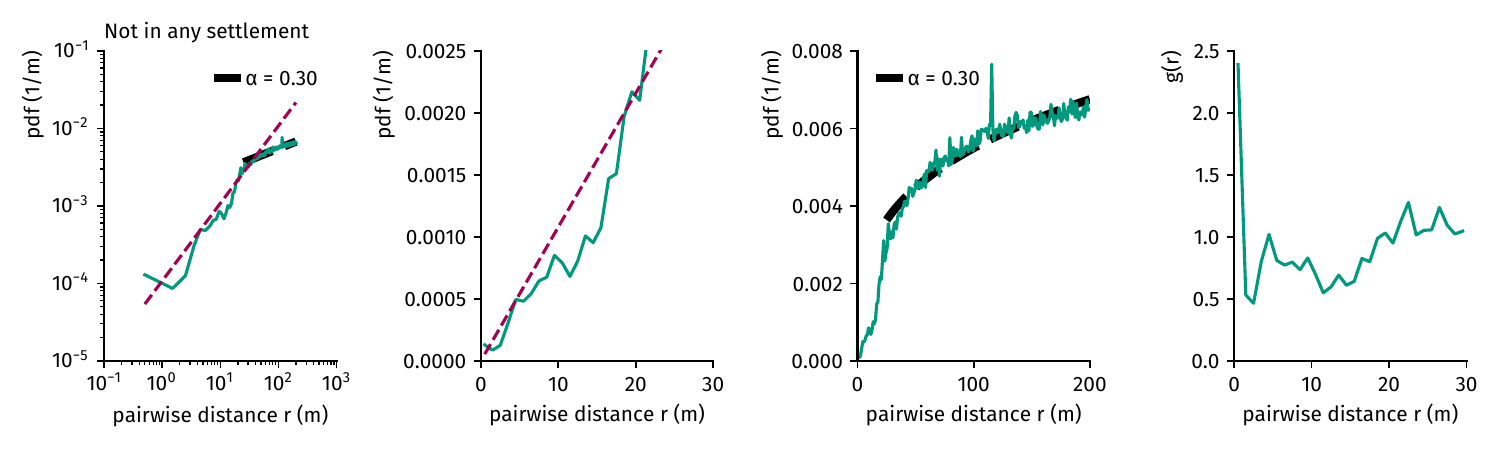}
    \caption{Address pair distribution functions and pair-correlation functions $g(r)$ with $r<200\unit m$ for (top) every building of Denmark, (second from top) Hovedstadens (capital region), (second from bottom) Aarhus, and (bottom) every address not located with the administrative boundaries of any city/town.}
    \label{fig:small-pdd-g-r-of-addresses}
\end{figure}

We inferred the mesoscale scaling parameter $\alpha$ by fitting $p(r)=C r^\alpha$ against the respective empirical pair distribution with 1m resolution in the range $r\in(25\unit m,200\unit m)$, using least squares. We find $\left<\alpha\right>=0.69$ and $\mathrm{Std}[\alpha] = 0.04$ for the building pair distribution s of the 30 largest cities as well as $\left<\alpha\right>=0.67$, and $\mathrm{Std}[\alpha] = 0.05$ for addresses, cf.~Fig.~\ref{fig:inferred-alphas}. Here, ``largest'' refers to the number of registered residential buildings with locations within the administrative boundaries of the city. For all buildings in Denmark, we find $\alpha=0.67$ and for all addresses, we have $\alpha=0.69$. Example analyses can be seen in Fig.~\ref{fig:small-pdd-g-r-of-buildings} for buildings and in Fig.~\ref{fig:small-pdd-g-r-of-addresses} for addresses. Note that the local environment around buildings and addresses that are not within a city/town boundary grows much slower with $\alpha=0.30$ for both.

\subsubsection{Larger distances}

To compute the pair distribution for larger distances, we proceed as follows. For each shape $\Omega_i$ (and $\mathcal D$, respectively), we find the set of its building (address) locations $\omega_i$ (and $\delta$, respectively). From this set we sample $n=\min(|\omega_i|,3\times10^4)$ unique locations (without replacement). Then, we find the pairwise distances of all pairs of these sampled locations and bin them to find histograms.

Note that to obtain the country-wide pair distribution displayed in Fig.~2a in the main text, we combine the respective pair distributions from the small-distance and large-distance analyses by requiring that they take the same value at $r=184.5\unit m$.

\begin{figure}
    \centering
    \includegraphics[width=\textwidth]{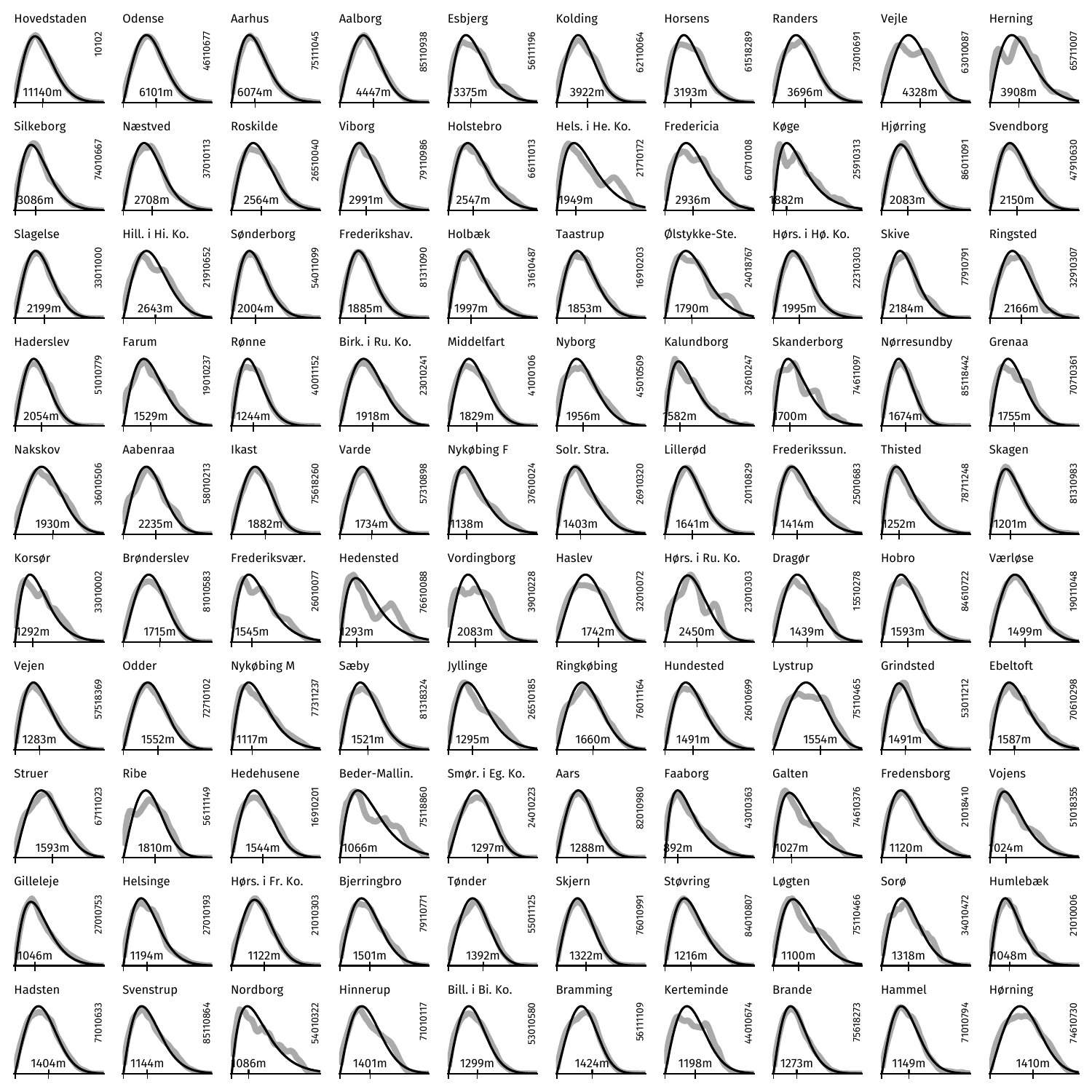}
    \caption{Empirical building pair distribution functions for the largest 100 cities in Denmark and fits of $p(r,R,m)=(r/R^2)\exp(-(r/R)^m)/\Gamma(2/m)$. We mark the inferred radius $R$.}
    \label{fig:large-pdds-buildings}
\end{figure}

\begin{figure}
    \centering
    \includegraphics[width=\textwidth]{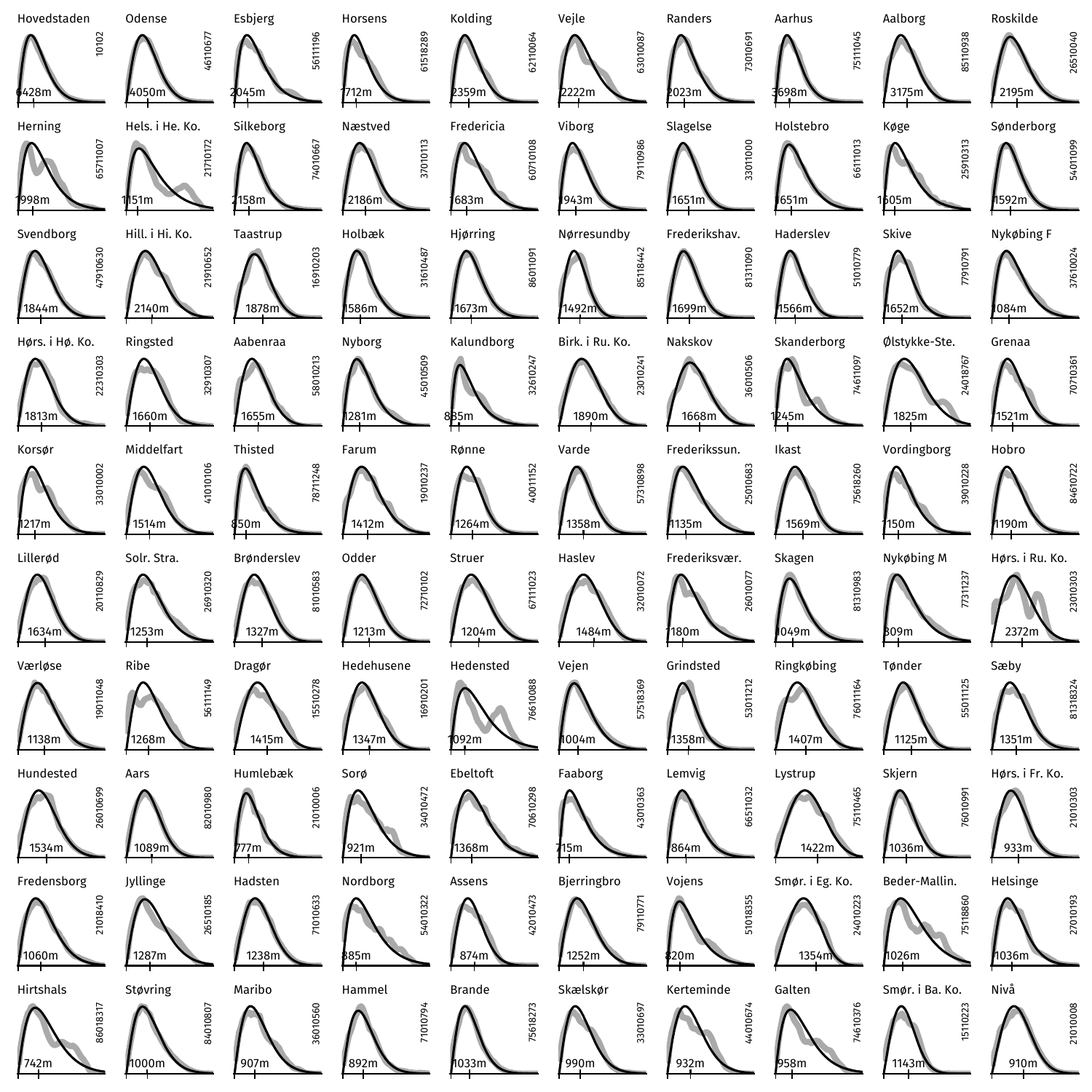}
    \caption{Empirical address pair distribution functions for the largest 100 cities in Denmark and fits of $p(r,R,m)=(r/R^2)\exp(-(r/R)^m)/\Gamma(2/m)$. We mark the inferred radius $R$.}
    \label{fig:large-pdds-addresses}
\end{figure}

\begin{figure}
    \centering
    \includegraphics{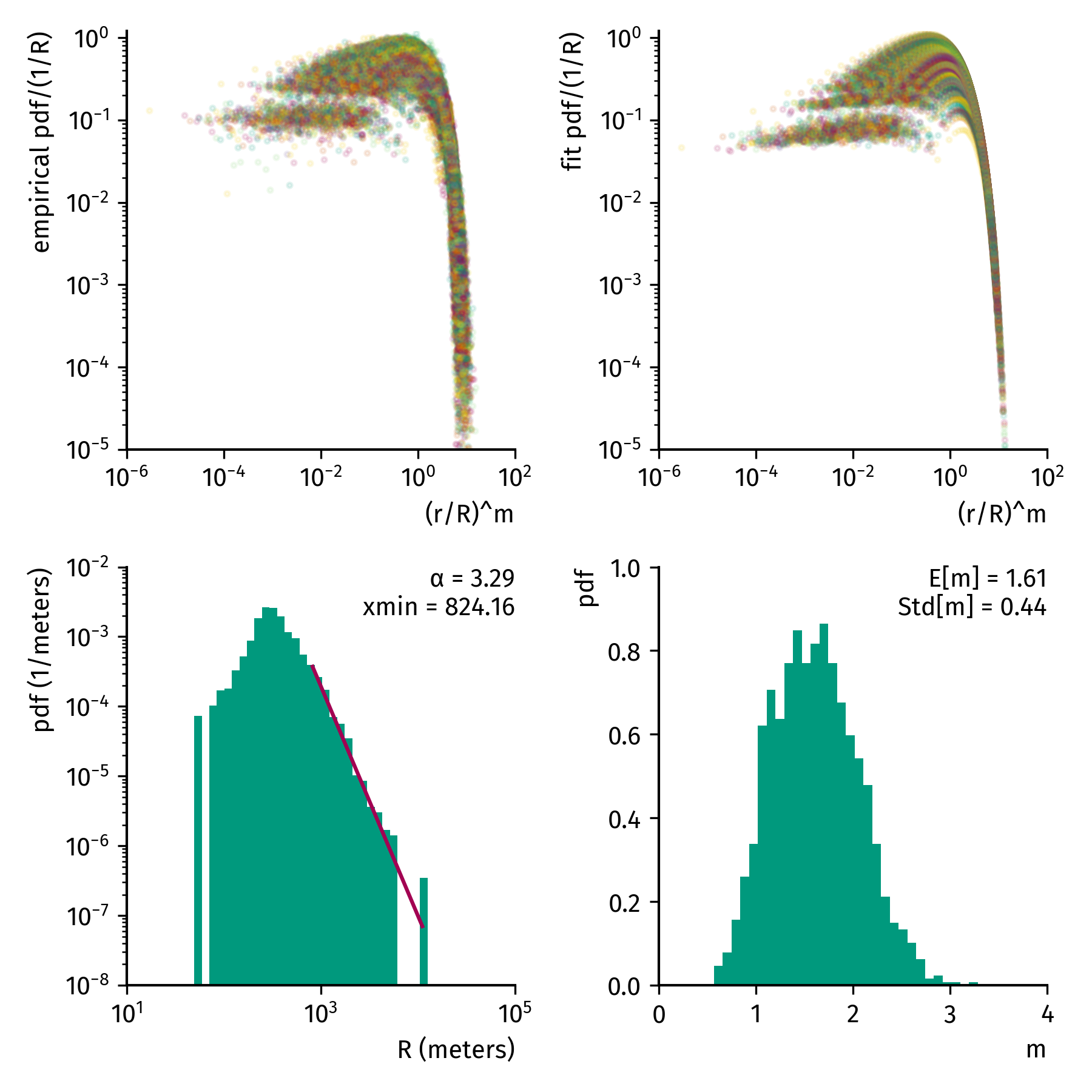}
    \caption{(Upper left) empirical pair distributions for all Danish towns with more than 30 buildings, rescaled by $R$ and $m$ obtained from (upper right) MLE fits of $p(r,R,m)=(r/R^2)\exp(-(r/R)^m)/\Gamma(2/m)$ to the empirical data. (Lower left) Distribution of inferred $R$ values and power-law fit to its tail, giving $p(R)\propto R^{-\alpha}$ with $\alpha=3.29$ and $x_{\min}=824\unit{m}$.
    (Lower right) Distribution of inferred $m$ values with $\left<m\right> = 1.61$ and $\mathrm{Std}[m]=0.44$.}
    \label{fig:cities-gpdd-fits-all-in-once}
\end{figure}

We show the pair distributions for buildings and addresses for the 100 largest Danish cities in Fig.~\ref{fig:large-pdds-buildings} and \ref{fig:large-pdds-addresses}. In Fig.~\ref{fig:cities-gpdd-fits-all-in-once} we show the empirical and fit pair distributions of all Danish cities with more than 30 buildings as well as the distributions of the inferred values of $R$ and $m$, respectively. The tail of the radius distribution $R$ scales as $p(R)\propto 1/R^{3.29}$, with $x_{\min} = 824\unit m$, inferred by the MLE technique in refs.~\cite{clauset_power-law_2009,alstott_powerlaw_2014}. The tail decay $(r/R)^m$ has values of $m$ with $\left<m\right> = 1.61$ and $\mathrm{Std}[m]=0.44$.

\subsubsection{Between cities}

To compute the pair distribution and $g(r)$ between cities, we define the `city center` as the centroid of a city's (multi-) polygon, i.e.~its geometric center: the center of mass of its shape.

\begin{figure}
    \centering
    \includegraphics[width=10cm]{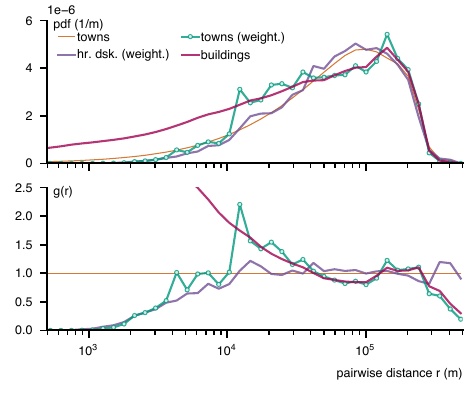}
    \caption{building-pair-weighted pair distribution function of cities, both for real city positions as well as a hard disk model without attractive force. Overlaid are the building pair distribution and the ideal gas pair distribution.}
    \label{fig:g_of_r_macro}
\end{figure}

Weighting each inter-city distance $r_{ij}$ by the number of pairs of buildings $m_im_j$ it contains, we find a pair distribution that approximates the empirical building pair distribution (see Fig.~\ref{fig:g_of_r_macro}). This pair distribution has a clear peak in $g(r)$, suggesting an abstract attractive force between cities of a certain size. However, the onset of $g(r)$ is consistent with the weighted pair distribution of the hard-disk model configuration.

\subsection{Pair distribution from independent patches of heterogeneous size}
\label{sec:patch-model}
We consider a \emph{multiverse} consisting of an infinite amount of universes indexed by $i$, each of which is inhabited by a patch of size $R_i$ with $N_i$ buildings inside, contributing to the multiverse with a pair distribution number density of $n(r,R_i) \approx N_i^2 p(r,R_i,m)$ (each universe contributes a number of building pairs amounting to $N_i^2$). Furthermore, we assume that there is a constant population density, which is proportional to the number of individuals per unit area, and that this density is inversely proportional to the square of the radius, $\rho_0\propto N/R^2$. This implies that the number of individuals in a given area is proportional to the area itself, $N\propto R^2$. 

For each of these universes (or rather, each of these patches), we calculate the number of building pairs at distance $r$ and assume that the patch sizes $R$ are distributed according to some distribution $f(R)$, which is currently unspecified. The joint distribution of house pairs at distance $r$ is then given by
\begin{align}
    n(r) &\propto \int\limits_0^\infty dR\ n(r,R) f(R)\\
     &\propto \int\limits_0^\infty dR\ f(R) R^4 \frac{r}{R^2}\exp\left(-\left(\frac {r}{R}\right)^m\right)
     .
\end{align}
To simplify the integral, we change variables to $\beta = 1/R$, such that $\mathrm dR = -\beta^{2}\mathrm d\beta$ and 
\begin{align}
    n(r) &\propto \int\limits_0^\infty d\beta \ f(\beta^{-1}) \frac{r}{\beta^4}\exp\left(-(\beta r)^m\right).
\end{align}
 The integral in question has a solution in the case where $f = \beta^{4 - \alpha}$ with $\alpha < 1$. We assume that this is the case, the integral now becomes,
\begin{align}
    n(r) = C \int_0^\infty\beta^{-\alpha} r \exp\left(-(r\beta)^m\right) d\beta 
\end{align}
where $C$ is a normalization constant.

We begin with the Gamma function
\begin{align}
    \Gamma(z) =\int_0^\infty  t^{z-1} \exp(-t)dt.
\end{align}
This integral converges for $z>0$ if $z\in \mathbf R$. Substituting $t=(r\beta)^m$ yields $\mathrm d t = m r^m\beta ^{m-1}\mathrm d\beta$ and therefore
\begin{align}
    \Gamma(z) &=\int_0^\infty (r\beta)^{mz-m} \exp\left(-(r\beta)^m\right) m (r\beta)^m \beta^{-1}d\beta\\
              &=mr^{mz-1}\int_0^\infty \beta^{mz-1} r\exp\left(-(r\beta)^m\right) d\beta\\
              &=mr^{-\alpha} C^{-1}C\int_0^\infty \beta^{-\alpha} r\exp\left(-(r\beta)^m\right) d\beta.
\end{align}
Here, we introduce $z = (1-\alpha)/m$, which, due to the lower bound of $z > 0$, leads to an upper bound of $\alpha < 1$. We recognize the integral $n(r)$ on the right and therefore find
\begin{align}
    n(r) \propto r^\alpha C\,\Gamma\left(\frac{1-\alpha}{m}\right).
\end{align}

If the radius of patches in the multiverse were distributed according to the function $f(R) \propto R^{-(4-\alpha)}$, the initial growth of the joint pairwise distribution would follow a sublinear power law. Notably, the scaling exponent would not depend on the tail parameter $m$. 

\begin{figure}
    \centering
    \includegraphics[width=\textwidth]{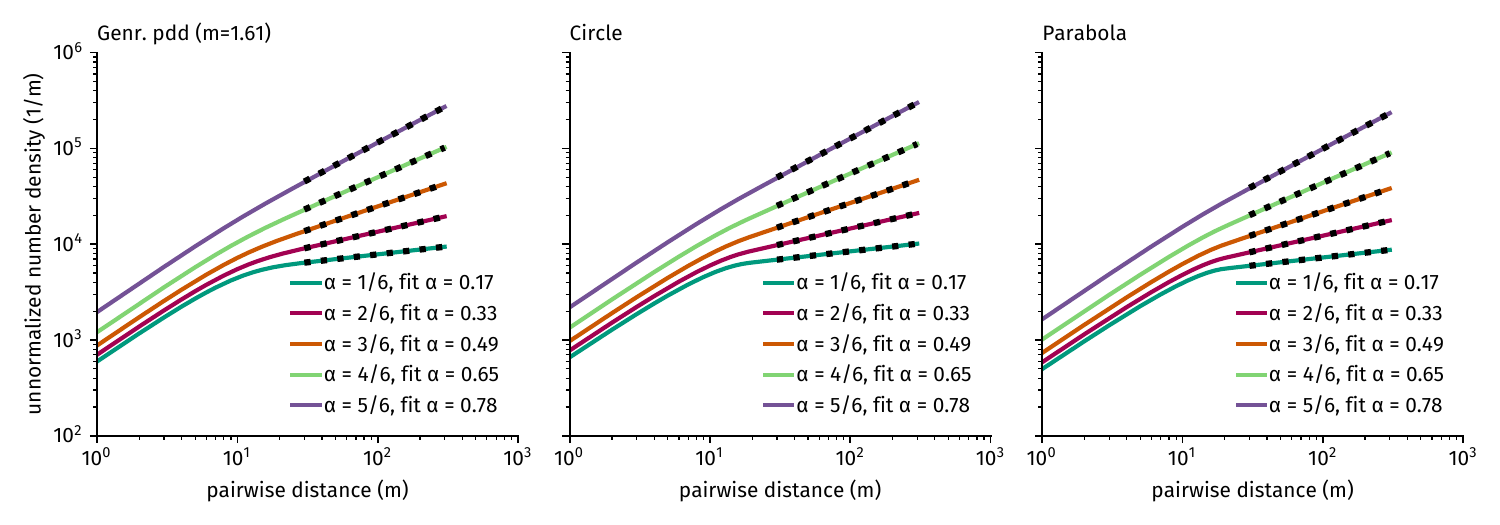}
    \caption{Integrating over the multiverse of patches of size $R$, summing up the contributions to the pairwise-distance distribution of all patches, for three assumptions of (i) generalized pair distribution, (ii) pair distribution of a uniform, circular patch of buildings, and (iii) a parabola pair distribution, with varying $\alpha$. Here, we used a minimum patch size of $R=10\mathrm m$, which fixes a scale and leads to linear growth of the pairwise-distance distribution for small distances.}
    \label{fig:integration-multiverse-numerical}
\end{figure}

Let us consider the implications of this for the area of the patches. The area would scale as $A\propto R^2$ (or $R\propto\sqrt A$), indicating that the distribution of the area would follow
\begin{align}
    \tilde f(A) &= \frac{\mathrm dR}{\mathrm d A} f(R=\sqrt{A})\\
                &\propto \frac{1}{A^{(5-\alpha)/2}}.
\end{align}
In the data, we observe $\alpha=0.67$. That means that the area of the patches would have to be distributed according to a power law with exponent $\mu=-(5-\alpha)/2\approx -2.17$, which is well within what has been found empirically for cities \cite{nature_stanley}. Furthermore, our city radius inference analysis (cf.~Fig.~\ref{fig:cities-gpdd-fits-all-in-once}) indicates that $f(R)\propto1/R^{3.29}$, which leads to $\mu=-2.15$. This demonstrates that the results of these two separate analyses are consistent.
As previously demonstrated, the sublinear scaling exponent $\alpha$ is independent of the tail parameter $m$. We extend our analysis by using the parabola pair distribution model Eq.~\ref{eq:pdd-parabola}. We have 
\begin{align}
    n(r) &\propto \int\limits_0^\infty dR\ n(r,R) f(R)\\
     &= \int\limits_{r/2}^\infty dR\ f(R) R^4  \left(-\frac 3 {4R} \left(\frac r R \right)^2 + \frac {3r} {2R^2}\right)
     .
\end{align}
Here, the lower bound in the integral comes from the condition that $r\leq 2R$. Now, as above, we demand $f(R)\propto 1/R^{4-\alpha}$ to find
\begin{align}
    n(r) &\propto \int\limits_{r/2}^\infty dR\ R^\alpha \left(-\frac 3 {4R} \left(\frac r R \right)^2 + \frac {3r} {2R^2}\right)
     \\
     &= \int\limits_{r/2}^\infty dR\ \left(-\frac {3r^2} {4R^{3-\alpha}} + \frac{3r}{2R^{2-\alpha}}
     \right)\\
     &\propto r^\alpha.
\end{align}
We solve the respective integrals $n(r)$ for the circle, generalized pair distribution, and parabola pair distribution models numerically and find that the above derivation holds for all three (see Fig.~\ref{fig:integration-multiverse-numerical}).

\subsection{Pair distribution  of a self-similar modular hierarchical model of building locations}
\label{sec:pdd-fractal}
One limitation of the multiverse approach is that it is only applicable if the patches are truly independent or sufficiently separated so that the scale of the pair distribution number density $n(r)$ does not affect the outcome. However, it is plausible that patches may be in close proximity to each other. This is supported by findings in \cite{nature_stanley}. Additionally, they discovered that a collection of these patches forms a fractal, or self-similar structure. While it is possible to demonstrate that a fractal dimension of building location does not necessarily result in sub-linear growth of the pair distribution function, it is certainly possible to investigate the consequences of such self-similar patch location on it.

\begin{figure}
    \centering
    \includegraphics[width=\textwidth]{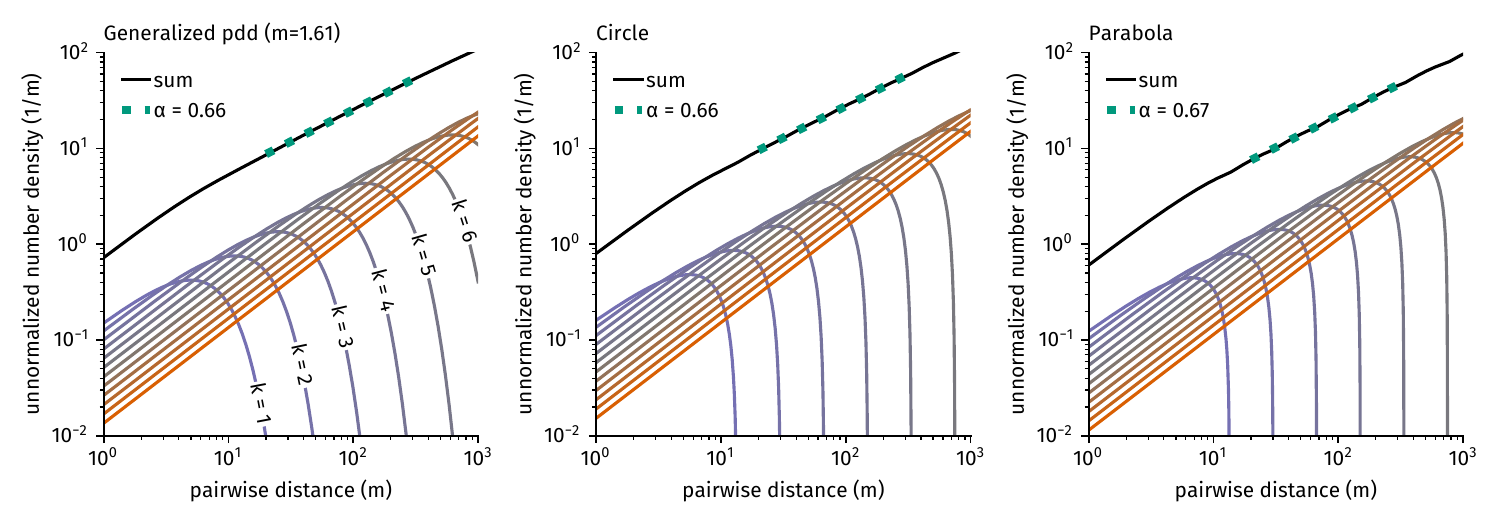}
    \includegraphics[width=.8\textwidth]{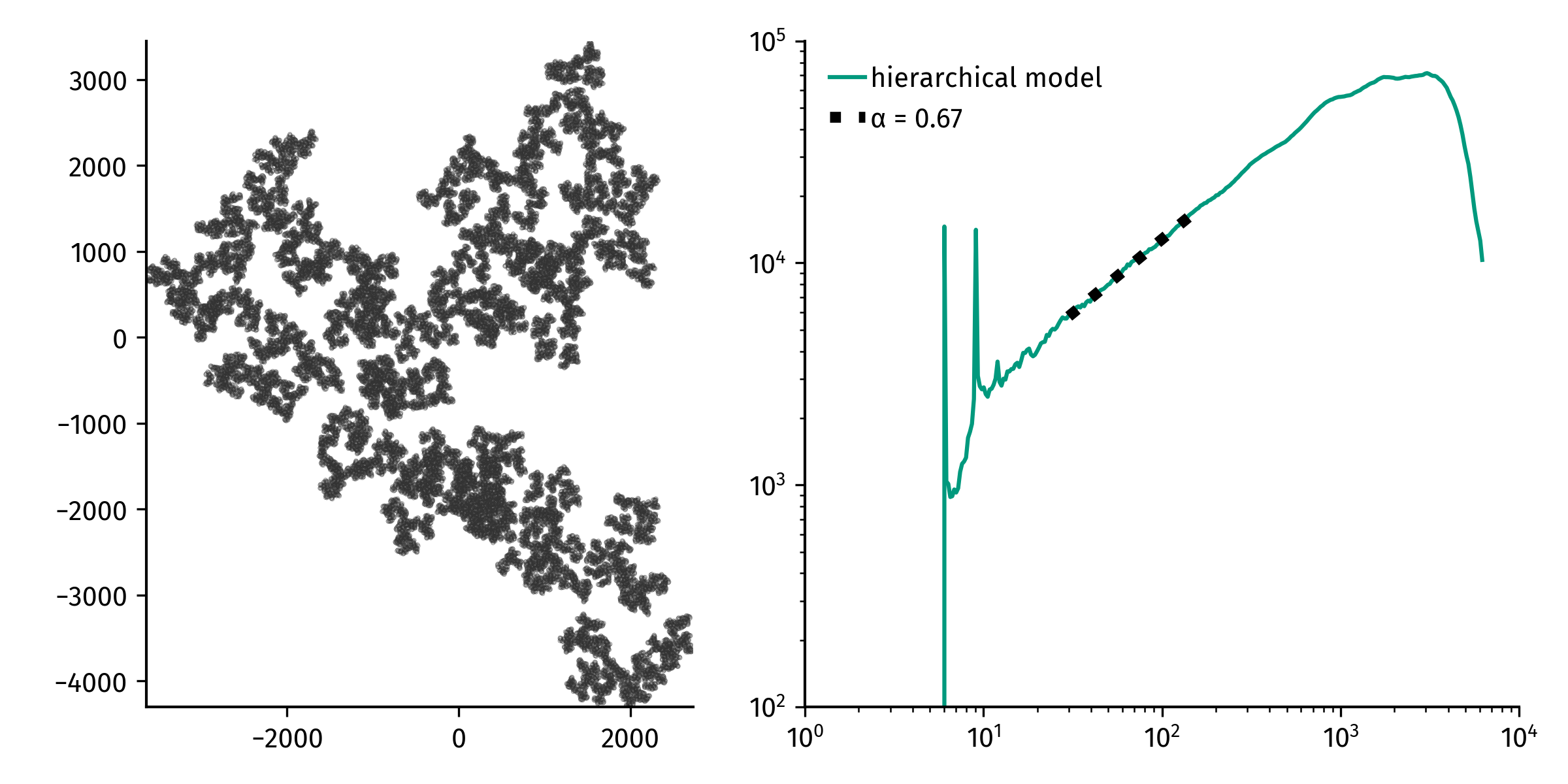}
    \caption{Upper two panels: Theoretical pairwise-distance distributions of the hierarchical building placement model with $N=4$ and $\theta=0.8$, in color the single contributions Eq.~\eqref{eq:hierachical-model-theory-single-layer} of scale $k$, in black the total joint distribution function over all scales Eq.~\eqref{eq:hierachical-model-theory-joint}, including a power-law fit showing the resulting scaling exponent $\alpha\approx 2/3$. 
    Lower two panels: A sample from the hierachical placement model with $L=8$ hierarchy layers and the corresponding resulting pairwise-distance distribution of the whole structure, showing the desired scaling behavior on the mesoscale.}
    \label{fig:hierarchical-model-tests}
\end{figure}

We assume a self-similar modular hierarchical structure comprising patches of buildings. We start with a single unit of $b$ buildings, each of which has a radius $z_1$. These are located within a patch of size $R_1$. Our objective is to regulate the building density in such a way that the number of buildings in a patch is given by $b = \theta (R_1^2/z_1^2)$. Here, $\theta$ represents the packing fraction. This implies that the patch radius is computed as follows: $R_1 = z_1\sqrt{b/\theta}$.

Now, consider that there are $b$ of these patches of radius $R_1$, located in a larger patch of higher order, which is (self-)similar to the basic patch. We posit that each of the lower-order patches has a radius of $z_2=R_1$, while the higher-order patch has a radius of $R_2=z_2\sqrt{b/\theta}$. 

Subsequently, we add $b-1$ similarly constructed patches to form an even larger patch. This entails constructing a self-similar structure of patches where the size of each patch of hierarchical order $k$ is given by
\begin{align}
    R_k = z\left(\frac{b}{\theta}\right)^{k/2}
\end{align}
and the size of each sub-patch of a patch is given by
\begin{align}
    z_k = z_1\left(\frac{b}{\theta}\right)^{(k-1)/2}.
\end{align}
Given a maximum number of $L$ orders (layers), the total number of houses is eventually $b^L$. In each layer $k$, there are $b^{L-k}$ patches of order $k$.

Now assume that for each patch $i$ of order $k$ and location $\bm s_{i,k}$, we distribute the location of its $b$ sub-patches randomly within this patch such that the pair distribution function of sub-patches leads to a pair distribution  $p(r, R_k)$ of scale $R_k$. To prevent the patches from overlapping excessively, a collision algorithm is employed for the $b$ sub-patches (preventing collisions between disks with radius $z_k$). We repeat this process recursively for each patch until the depth of the $N$-ary hierarchy tree reaches $L$. The leaves of the tree represent buildings. As these buildings will overlap, another collision algorithm is run until they no longer overlap.

We now estimate the joint pairwise-distance distribution of the entire structure. Consider a container patch that contains $b$ sub-patches. If we assume that the total the $B$ buildings within a sub-patch of scale $k-1$ are sufficiently concentrated within its center, the contribution of this patch of radius $R_k$ will be proportional to the pair distribution  $p(r,R_k)$, weighted with the total number of pairs of buildings within this container-patch, except the pairs of buildings within the same sub-patch. Consequently, the number of pairs therefore scales as $\propto B^2 b(b-1)$, or, neglecting the linear contribution, as $\propto (bB)^2$. For each patch in layer $k$, there are going to be $B=b^{k-1}$ buildings in a sub-patch. Therefore, the number of pairs of buildings in a patch of scale $k$ grows as $\propto b^{2k}$.

Each of these patches contributes to approximately $b^{2k}$ pairs to the joint pairwise-distance distribution. There are $b^{L-k}$ patches of order $k$ with radius $R_k$. Therefore, the total contribution to the joint pairwise-distance distribution of this scale is
\begin{align}
    n_k(r, R_k) &\propto b^k p(r, R_k) \\
                &= b^k p(r, z(b/\theta)^{k/2})
                \label{eq:hierachical-model-theory-single-layer}
\end{align}
where $p(r,R)$ depends on how sub-patches are distributed within patches.

The total pair distribution number density is given by
\begin{align}
    n(r) &= \sum_{k=1}^{L} b^k p(r, z(N/\theta)^{k/2}).
                \label{eq:hierachical-model-theory-joint}
\end{align}
An explicit form using the generalized pair distribution function model is
\begin{align}
    n(r) &\propto \sum_{k=1}^{L}
                        b^k
                        {\frac r {R_k^2}} 
                        \exp\left(-\left(\frac r {R_k}\right)^m\right)
                   .\\
         &=  {\frac r {z^2} }\sum_{k=1}^{L}
                         {\theta^k} 
                        \exp\left(-\left(\frac r z\right)^m \left(\frac \theta b\right)^{km/2}\right).
\end{align}
We compute this equation numerically for varying $\theta$, $b$, and different models for $p(r,R)$. Figure~\ref{fig:hierarchical-model-tests} illustrates examples for $b=4$ and $\theta=0.8$, which leads to a scaling of approximately $\propto r^{0.67}$. 

To derive the dependence of the exponent of the observed scaling law on the parameters, we use the saddle-point approximation. First, we approximate the sum over hierarchy layers with an integral over a constant hierarchy layer density
\begin{align}
    n(r) \propto r \int\limits_1^\infty dk
                        \exp\left[\underbrace{
                                        k\ln\theta
                                        -\left(\frac r z\right)^m \exp\left(\frac{mk}2 \ln\frac{\theta}b
                                        \right)
                            }_{=-w(k)}\right].
\end{align}
Relying on the saddle-point method, we can approximate the integral to find
\begin{align}
    n(r) \propto r  \frac{\exp\big[-w(k_0)\big]}{\sqrt{w''(k_0)}}
\end{align}
where $k_0$ is the minimum of $w(k)$. We compute the derivatives
\begin{align}
    w'(k) &= -\ln\theta + \left(\frac r z\right)^m \frac m 2 \ln\frac \theta b
                         \exp\left(\frac{mk}{2}\ln\frac\theta b\right)\\
    w''(k) &= \left(\frac r z\right)^m \left(\frac m 2 \ln\frac \theta N\right)^2
                         \exp\left(\frac{mk}{2}\ln\frac\theta b\right)
\end{align}
and with $w'(k_0)=0$ find the following equations for the minimum
\begin{align}
      \exp\left(\frac{mk_0}{2}\ln\frac\theta b\right) &=
            \frac {\ln\theta} {\ln(\theta/b)} \left(\frac z r\right)^m \frac 2 m\\
        k_0 &= \frac 2 m \frac 1 {\ln \theta/b}
                       \left[
                            \ln\left(\frac{\ln\theta}{\ln(\theta/b)}\right)
                        +   m\big( \ln z - \ln r\big) + \ln\left(\frac 2 m \right)
                       \right].\label{eq:k0}
\end{align}
Using the first of the two equations we note that for $w(k_0)$, the dependence on $r$ cancels out in the second term and so
\begin{align}
    -w(k_0(r)) = k_0(r)\ln\theta - W
\end{align}
where $W$ is an irrelevant constant. Second, we use the same equation to see that $w''(k_0)$ does not depend on $r$ and therefore does not concern us any further either. As a step in between, this means that 
\begin{align}
    n(r) \propto r \exp\big[k_0(r)\ln\theta\big].
\end{align}
Looking at Eq.~\eqref{eq:k0}, we see that the only $r$-dependent term gives the minimum a structure of
\begin{align}
    k_0 = K - \frac{2}{\ln(\theta/b)} \ln r,
\end{align}
and therefore we find 
\begin{align}
    n(r) &\propto r \exp\left[- \frac{2\ln\theta}{\ln(\theta/b)} \ln r\right]\\
         &= r^{1- \frac{2\ln\theta}{\ln(\theta/b)}},
\end{align}
i.e. the exponent of the sub-linear growth in the pair distribution function is given as
\begin{align}
    \alpha = 1- \frac{2\ln\theta}{\ln(\theta/b)}.
    \label{eq:alpha-approx}
\end{align}
We can compare this estimation with numerical results. Above we used $N=4$ and $\theta=0.8$ to find $\alpha\approx 0.67$. Eq.~\eqref{eq:alpha-approx} yields $\alpha=0.72$, an acceptable approximation. Notably, this result is also independent of the parameter $m$, which represents the explicit form of the pairwise distance distribution per patch. This explains why similar results are obtained despite the use of different geometries.

We can perform the same analysis for the parabola pair distribution model Eq.~\eqref{eq:pdd-parabola}. We have 
\begin{align}
    n(r) &\propto \sum_{k=1}^L \frac{b^k}{R_k^2} \left(-\frac 3 {4} \frac{r^2} {R_k} + \frac {3r} {2}\right)\\
         &= \frac{r}{z^2} \sum_{k=1}^L \theta^k \left(-\frac 3 {4} \frac{r} {R_k} + \frac {3} {2}\right).
\end{align}
Again, we use the saddle point method to approximate
\begin{align}
    n(r) &\propto r \frac{\exp\left[-w(k_0)\right]}{\sqrt{w''(k_0)}},
\end{align}
with $k_0$ being the minimum of the function
\begin{align}
    w(k) = -k\ln\theta - \ln\left(-\frac{3}{4}\frac{r}{z}\frac{\theta^{k/2}}{b^{k/2}} + \frac 3 2\right).
\end{align}
We arrive at
\begin{align}
    n(r)\propto r \frac{ \theta^{\frac{2 \ln{\left(\frac{4 z \ln{\left(\theta \right)}}{r \left(2 \ln{\left(\theta \right)} + \ln{\left(\frac{\theta}{b} \right)}\right)} \right)}}{\ln{\left(\frac{\theta}{b} \right)}}} 3 \sqrt{\pi} \ln{\left(\frac{\theta}{b} \right)}}{z^{2} \sqrt{\left(2 \ln{\left(\theta \right)} + \ln{\left(\frac{\theta}{b} \right)}\right) \ln{\left(\theta \right)}} \left(2 \ln{\left(\theta \right)} + \ln{\left(\frac{\theta}{b} \right)}\right)}.
\end{align}
Leaving aside the irrelevant prefactors, we have
\begin{align}
    \ln n(r) &\propto \ln r + \ln\theta \left(K - \frac{2\ln r}{\ln(\theta/b)}
    \right) \\
             &\propto \ln r \left(1-\frac{2\ln \theta}{\ln(\theta/b)}\right),
\end{align}
i.e.~the same result as with the generalized pair distribution.

\newpage

\section{Studying mobility with the pair distance function}
\subsection{Comparison with other Mobility models}
\subsubsection{Gravity}
\label{sec:mobility-gravity}
 %explain the parish exponential decay, two step, the city, non city why you need our model

The gravity model as originally introduced in \cite{zipf1946p1p2}, states that the number of moves $T$ from area $i$ to $j$ is proportional to,
\begin{align}
    T_{i \to j} \propto \frac{N_i N_j}{d_{i,j}}
\end{align}
with $N_i$ the population at location $i$ and $d_{i,j}$ the distance between locations $i$
and $j$.

 \begin{figure}
    \centering
    \includegraphics[width=\textwidth]{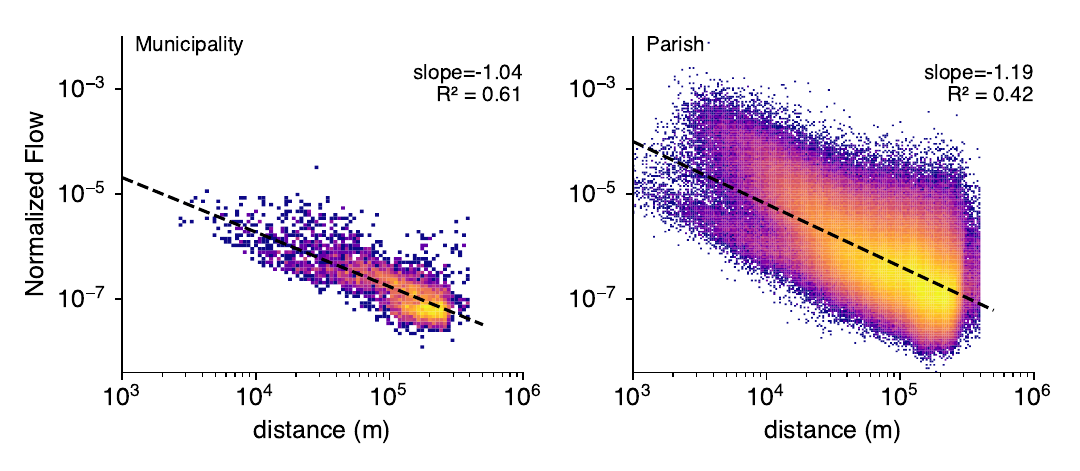}
    \caption{Gravity model as in \cite{zipf1946p1p2} for migration between (left) municipalities in Denmark, (right) parishes in Denmark. The normalized flow represented on the $y$-axis is the number of moves from $i$ to $j$ divided by the product of population between $i$ and $j$, $T_{i,j}/ (N_i N_j)$. }
    \label{fig:gravity_municipalites}
\end{figure}

\subsubsection{Radiation}
\label{sec:radiation}

In this section, we derive a continuous radiation model and demonstrate how it differs from our continuous gravity model. The radiation model of human mobility is usually represented by the following equation \cite{simini2012universal},

\begin{equation}
    P_{ij} \propto \frac{m_i m_j}{(m_i + s_{ij})(m_i + m_j + s_{ij})}
\end{equation}

with $P_{ij}$ the predicted number of people traveling from location $i$ to location $j$, $m_k$ the population at the origin location $k$, $s_{ij}$ represents the total population in a circle centered at $i$ with radius equal to the distance between $i$ and $j$, excluding the populations of $i$ and $j$.

Following \cite{simini2012universal}, the probability of having a single particle emitted from location $i$ to location $j$ is,

\begin{align}
\label{eq:radiation_origin}
    P(1| m_{i},m_{j},s_{i,j}) = \int_{0}^{\infty} \mathrm{d}z P_{m_i}(z) P_{s_i}(<z)P_{m_j)(>z)}
\end{align}

with,
\begin{equation}
    P_{m_i}(z) = \frac{\mathrm{d} P_{m_i}(<z) }{\mathrm{d} z} = m_i p(z)^{m_i-1}\frac{\mathrm{d} p(<z)}{\mathrm{d} z}
\end{equation}

if $m_i=1$, then it simplifies to $P_1(z)=\mathrm{d} P(<z)/\mathrm{d}z$.
Equation \ref{eq:radiation_origin} simplifies to,

\begin{align}
    P(1| 1,1,s_{i,j}) &= \int_{0}^{\infty} (1-P(<z)) P(<z)^{s_{i,j}} \mathrm{d} P(<z)\\
    &= \frac{1}{(1+s_{i,j})(2+s_{i,j})}
\end{align}

In the general case when $s_{i,j} \gg 1$, that is, when the number of houses in the circle of center $i$ and radius the distance $i$ and $j$ is greater than one, which is often the case when the distance is sufficiently large (see Extended Data Fig.~1),
\begin{equation}
    P_{ij} = \frac{1}{(1 + s_{ij})(1 + 1 + s_{ij})} \sim s_{ij}^{-2}.
\end{equation}

However, the number of addresses in the circle $s_{i,j}$ of center $i$ and radius $R(i,j)$ is closely related to the primitive of the pair distribution function,
\begin{align}
    s_{x,y} &= \int_{\Omega} dz p^{d}(z) \delta\{R(x,z)<R(x,y)\},\\
    s(r) &= \int_x dx \int_y dy \, s_{x,y} p^{d}(x) p^{d}(y) \delta(r-R(x,y)), \\
    s(r) &= \int_x dx \int_y dy \, s_{x,y} p^{d}(x) p^{d}(y) \delta\{r>R(x,y)\}
\end{align}

In the meantime, the cumulative distribution of pair distribution is (with $p$ the pair distribution function),

\begin{align}
    \Phi(r) &= \int_0^r p(s)ds, \\
    \Phi(r) &= \int_0^r ds \int_\Omega d  x\ \int_\Omega d  y\ ^d(  x) p^d(  y) \delta(s - R(  x,  y)), \\
    \Phi(r) &= \int_\Omega d  x\ \int_\Omega d  y\ p^d(  x) p^d(  y) \delta\{r>R(  x,  y)\}, \\
    \Phi(r) &= s(r).
\end{align}

If we come back to the derivation of the pair distribution, 
\begin{equation}
    f(x,y)\mathrm{d}^2x\mathrm{d}^2y = \pi(x,y) \varrho^d(x) \varrho^d(y) \mathrm{d}^2x\mathrm{d}^2y.
    \label{eq:exp_mov}
\end{equation}

In our continuous gravity law model, we assume that $\pi(x,y)=\pi(R(x,y))=\pi(r)$. On the other hand, the radiation model assumes that,
\begin{equation}
    \pi(x,y) = s(x,y)^{-2}= \Phi(x,y)^{-2}
\end{equation}

If we assume that on average, $s(x,y) \simeq \int_0^{|x-y|}\mathrm{d}r p(r) = \Phi(r)$. Then the two models will agree on average if,
\begin{equation}
\Phi(r)^2 = r
\end{equation}

The pair distribution function would be $p(r)=\pm r^{-0.5}$. While the empirical pair distribution (Fig.2a) exhibits more complex patterns than this, and the two models appear to be irreconcilable.

\subsection{The intrinsic distance cost as the inverse of distance}
\label{SI:log-utility}
Consider an individual situated at location $i$. The utility associated with moving to location $j$ can be expressed as follows,
\[
U(i,j) = V(j) - C(r(i,j))
\]

where $V(j)$ represents  the inherent values of location $j$ and $C(r(i,j))$ is a function of the distance $r(i,j))$ between $i$ and $j$. 

The probability of selecting location $j$ can be calculated using the multinomial logit model, as outlined in \cite{train2009discrete},

\[
f(i,j) = \frac{e^{ U(i,j) }}{\sum_k{e^{ U(i,k)}}}
\]

The probability to move of a distance $r$ is the sum over all locations $k$ for which $r(i,k)=r$. This probability becomes independent of $i$. Thus, the probability of moving of a distance $r$ aggregates over all locations $i$ as $f(r) = \sum_i f(i,r)f(i)$. For simplification, let us that $f(i)$ is constant, implying that the probability of movement from any given location is equal. Furthermore, let us assume that for all $i,j \in \Omega$, $f(i,r) = f(j,r)$. In this case $f(r) \propto N f(i,r)$.

Moreover, the number of locations $k$ satisfying $r(i,k)=r$ corresponds to the pair distribution function, $p(r)$, up to a constant, with this constant being the total number of locations.

\[
f(r) =  \frac{\sum_{i\in \Omega}\sum_{k| r(i,k)=r} e^{ U(i,k) }}{\sum_k{e^{ U(i,k)}}}
\]

Assuming a uniform value $V(k)$ across all locations, we derive:
\[
f(r) =  \frac{\sum_{i\in \Omega}\sum_{k| r(i,k)=r} e^{ V - C(r(i,k)) }}{\sum_k{e^{ U(i,k)}}} \\
\sim {p(r) e^{ V -C(r) }}
\]
with $p(r)$ the pair distribution function evaluated at $r$. 

Given the human tendency to perceive distances logarithmically, especially for larger magnitudes, we posit that the cost function $C(r) \propto \log(r)$. This hypothesis is supported by the observation that humans might perceive numerical differences in a logarithmic fashion, such as perceiving the difference between 1 and 2 similarly to the difference between 10 and 20 or even 100 and 200, as demonstrated in \cite{dehaene1997number}. Consequently, upon normalization by the pair distribution function, we derive the equation 

\[
\frac{f(i,r)}{p(i,r)} \sim e^{ V  - \log(r) }   = e^{ \log(V_2)  - \log(r) } =  \frac{V_2}{r} 
\]

where $V = \log(V_2)$. This leads to
\[
\frac{f(i,r)}{p(i,r)} \sim \frac{1}{r}
\]
which is equivalent to equation (\ref{eq:pi_r})
of the intrinsic distance cost, given $\pi(r)=\frac{1}{r}$.

\subsection{Reconciling Opportunity and Distance-Based Mobility Paradigms }
\label{SI:opportunity-distance}

As highlighted in \cite{barbosa2018human,noulas2012tale,mazzoli2019field}, in classical mobility studies, two divergent theories explain the movement of individuals. The first, inspired by Newton's law of gravity, suggests that mobility decreases as the physical distance between locations increases, often modeled as a power law of distance \cite{zipf1946p1p2}. The second theory argues that mobility is not directly related to distance but to intervening opportunities. It suggests that individuals prioritize locations based on the availability of closer opportunities rather than distance itself, leading to movements driven more by the opportunity distribution than by travel constraints \cite{stouffer1940intervening, simini2012universal}.

The gravitational model describes how the flow of individuals between two locations decreases as the distance between them increases, analogous to the decrease in gravitational pull with distance in Newton's theory. This decline in mobility is modeled using functions such as exponential or power laws \cite{lenormand2012universal,lenormand2016systematic,erlander1990gravity}. In this framework, the population sizes of the start and end locations act like masses, attracting travel in direct proportion to their size and inversely to the square of the distance between them. Known for its simplicity and ease of computation, the gravity model is widely used in various fields such as migration, intercity communication, and traffic flow analysis. It can also be extended by integrating socio-economic characteristics of locations to improve its accuracy and applicability \cite{patuelli2007network}.

Conversely, the intervening opportunity model emphasizes that mobility is mainly driven by the availability of intervening opportunities rather than distance \cite{schneider1959gravity}. It suggests that individuals are influenced more by the availability of nearby opportunities than by the distance to distant opportunities, with the spatial distribution of these opportunities dictating destination choices, making distance a less critical factor. The radiation model \cite{simini2012universal}, an extension of this theory, streamlines the analysis by assuming that the chosen opportunity is the optimal one. This model correlates opportunity density with population density and provides a mathematical formula for predicting trip endpoints. The model has also been extended to a continuous spatial framework, providing a nuanced understanding of how social factors influence movement patterns \cite{siminiContinuumApproach}.

Our framework provides a unifying approach to the intervening opportunity model and distance-based model
In the equation \ref{eq:pi=f/p},
\begin{align}
    f(r) \propto \pi(r)p(r).
\end{align}

The pair distribution function $p(r)$ captures the available opportunities to move, when we divide the moving distance distribution by the pair distribution, we normalize the empirical mobility traces by their intervening opportunities. Consequently, the remaining quantity is found to be dependent on the distance, with the result that $pi(r)=f(r)/p(r)\simeq 1$. This can be interpreted as a distance term analogous to the gravity model. Furthermore, our framework explain why intra-city movements are better fitted by a gravity model with an exponential decay, which is similar to the exponential decay of population density \cite{liang2013unraveling}.

\subsection{Local piece-wise power law}
\label{SI:mobility-piecewise}
In this section, we investigate whether the continuous gravity model still holds at the scale of cities. The observed moving distribution $f(r)$ can be decomposed into the observed moving distribution of each city $c_i$ in the set $\mathcal{C}$ of cities in Denmark,
\begin{equation}
    f(r) = \sum_{c_i \in \mathcal{C}} f_{c_i}(r),
\end{equation}
where $f_{c_i}(r)$ only counts moves originating from city $c_i$, i.e. $f_{c}(r)=\int_{x \in c}  \mathrm{d}x \int_0^\infty  \mathrm{d}y f(x,y)$.
For each city's observed moving distance the appropriate normalization is not the country-wide pair distribution but the relative pair distribution function, which counts pairs with at least one location in the city of interest. We define,
\begin{equation}
    p_c(r) = \int_{x \in c}  \mathrm{d}x \int_{y \in \Omega} \varrho^d(x) \varrho^d(y) \mathrm{d}x\mathrm{d}y.
\end{equation}
We can now compute the relative intrinsic distance cost, $\pi_c(r) = f_c(r)/p_c(r)$. In the main manuscript Figures.3f-g show that the relative intrinsic distance cost does not follow the continuous gravity law and leads to a piece-wise intrinsic distance cost, Figures \ref{fig:piecewise-1}-\ref{fig:piecewise-5} show the same process for many more cities. The piece-wise intrinsic distance cost is a piece-wise power law distribution of the form:
\begin{align}
\label{eq:piecewise-powerlaw}
    \pi_c(r) = 
\begin{cases} 
C_1 (R_{c}/r)^{\beta}, & 0 \leq r \leq R_{c} \\
C_2 (R_{c}/r)^{\alpha}, & r > R_{c}
\end{cases}
\end{align}

where $R_{c}$ is the mobility inferred city radius, $\alpha>1$, $\beta \in \mathbb{R}$, and  $x \in \mathbb{R}^{+}$. The three parameters $\alpha, \, \beta, \, r_{c}$ are estimated by maximum-likelihood and compared to a log-normal, Pareto, and exponential Pareto distribution \cite{maier2023maximum}. The likelihood ratios and p-values for all cities are reported in figures \ref{fig:piecewise-pvalues0}-\ref{fig:piecewise-pvalues2}. We obtain that on average, $\beta=0.60\pm 0.20$, $\alpha=2 \pm 0.21$, $R_{c}$ follow a heavy-tailed distribution between a power law and a log-normal distribution. 

\subsubsection{The remarkable case of islands}

\begin{figure}
    \centering
    \includegraphics[width=\textwidth]{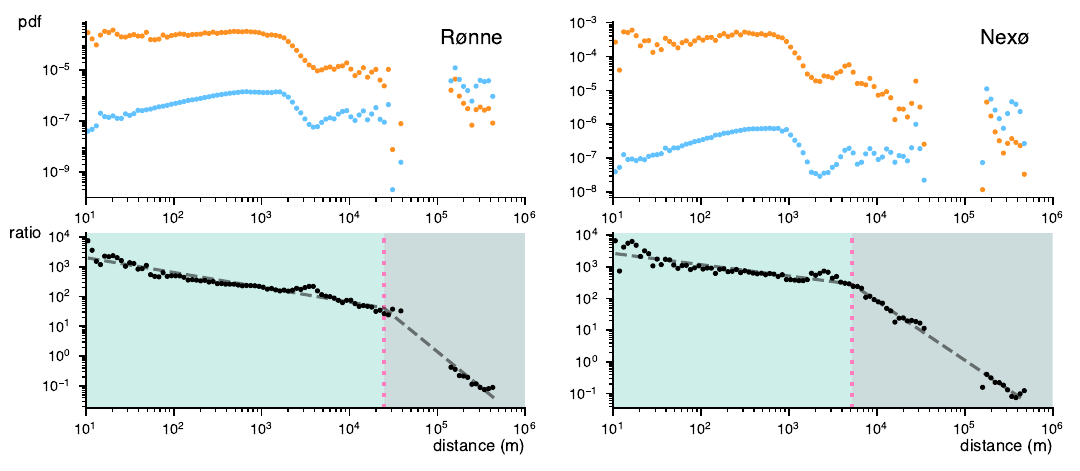}
     \caption{Piecewise intrinsic distance for the city of Rønne and Nexø both situated on an island.} 
    \label{fig:piecewise-islands}
\end{figure}

In the section, we highlight the remarkable point that the piece-wise two-step intrinsic distance cost for cities still holds remarkably well for islands (Fig.\ref{fig:piecewise-islands}). 
In the case of islands, the inter-city part of the piece-wise process (when the exponent is on average 2), it mostly across the sea, and therefore the relative pair distance distribution is null. However, if the distance is large enough to reach the nearest landmass, the pair distance distribution is positive again and the $2$ exponent can be interpolated between the island and the nearest landmass (see Fig.\ref{fig:piecewise-islands}).

\subsection{Simulation of piece-wise intrinsic distance cost}
\label{SI:piecewise-simulation}
In this section, we try to recover the intrinsic distance cost, $\pi(r)=1/r$ from the aggregation of the piece-wise process of equation \ref{eq:piecewise-powerlaw}. We can decompose the intrinsic distance cost as follows,  

\begin{align}
    \pi(r) = \frac{f(r)}{p(r)} =& \sum_{c \in \mathcal{C}} \frac{f_c}{p(r)} 
    = \sum_{c \in \mathcal{C}} \pi_{c}(r)\frac{p_c(r)}{p(r)}
\end{align}

The only term that is unknown in this equation is $p_c(r)$ the relative pair distance distribution function for each city. However, we have already shown that the pair distribution of cities follows a generalized gamma distribution within the city boundary. Beyond the city boundary, we have shown that the positions of cities follow an ideal gas (i.e. random position but no overlap), which implies that we can model the relative pair distribution function with a linear growth beyond the city radius. The relative pair distribution function of a city, $c$, is thus, 

\begin{align}
\label{eq:piecewise-pddc}
    \hat{p}_c(r) = 
\begin{cases} 
C_1 (r/R_c^2)\exp(-(r/R_c)^m) , & 0 \leq r \leq R_{c} \\
C_2 r, & r > R_{c}
\end{cases}
\end{align}
where $C_1$ and $C_2$ are two constants such that $\hat{p}_c$ is continuous. To obtain the global pair distribution we need to weight each relative pair distribution by its contribution, which is equal to the number of addresses. We assume that the number of addresses is proportional to the square of the radius and $p_c= R_c^2 \hat{p}_c$. For simplicity, we also assume that the general pair distribution function $p$ follows,

\begin{align}
\label{eq:piecewise-pdd-hypothesis}
    p(r) = 
\begin{cases} 
C_3 r^{2/3} , & 0 \leq r \leq  R_{\text{max}} \\
C_4 , & r > R_{\text{max}}
\end{cases}
\end{align}
where $R_{\text{max}}= \max\{R_c\, , c\in \mathcal{C}\}$ the largest city radius. The form of the pair distribution function closely follows the empirical one of Figure 2a. Therefore, we obtain the intrinsic distance cost when aggregating over the city scale,

\begin{align}
\label{eq:piecewise-cost-normalised}
 \pi_{c}(r)\frac{p_c(r)}{p(r)} = 
\begin{cases} 
C_5 (r^{-1/6}/R_c^2)\exp(-(r/R_c)^m) , & 0 \leq r \leq R_{c} \\
C_6 r^{-5/3} , &  R_{c} < r \leq R_{\text{max}} \\
C_7 r^{-1} , &   r > R_{\text{max}} 
\end{cases}
\end{align}

\begin{figure}
    \centering
    \includegraphics[width=\textwidth]{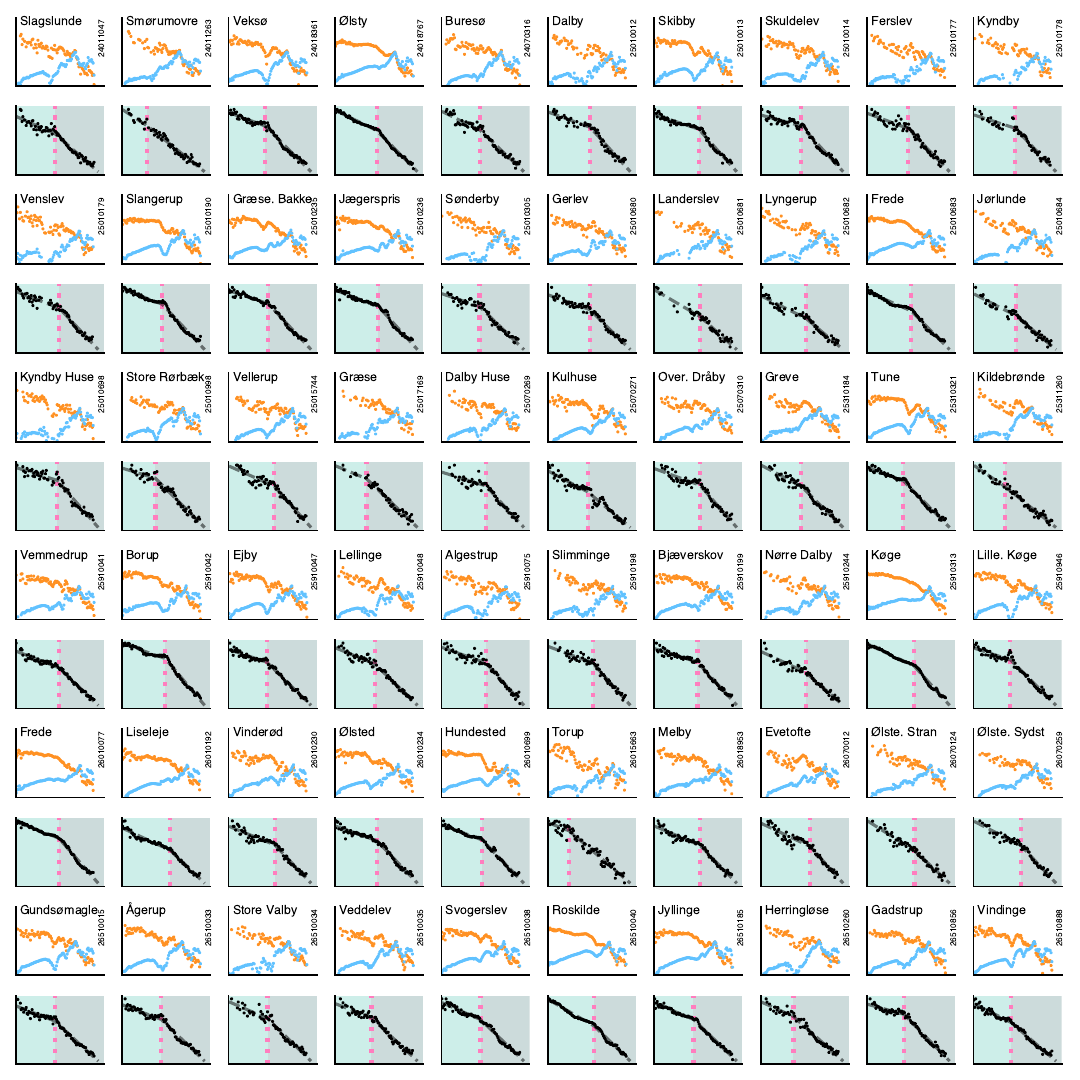}
        \caption{For each city in Denmark, the distribution of observed moving distance (orange) and the relative pair distribution function. Pairs are restricted to those containing at least one address from the city of interest. The second plot shows the intrinsic distance cost (black), as the ratio between the observed moving distance and the relative pair distribution function. The pink dashed line is the inferred mobility city radius, the light teal corresponds to intra-city moves, and the dark teal to inter-city moves. }
    \label{fig:piecewise-idc_simulation}
\end{figure}

\begin{figure}
    \centering
    \includegraphics[width=\textwidth]{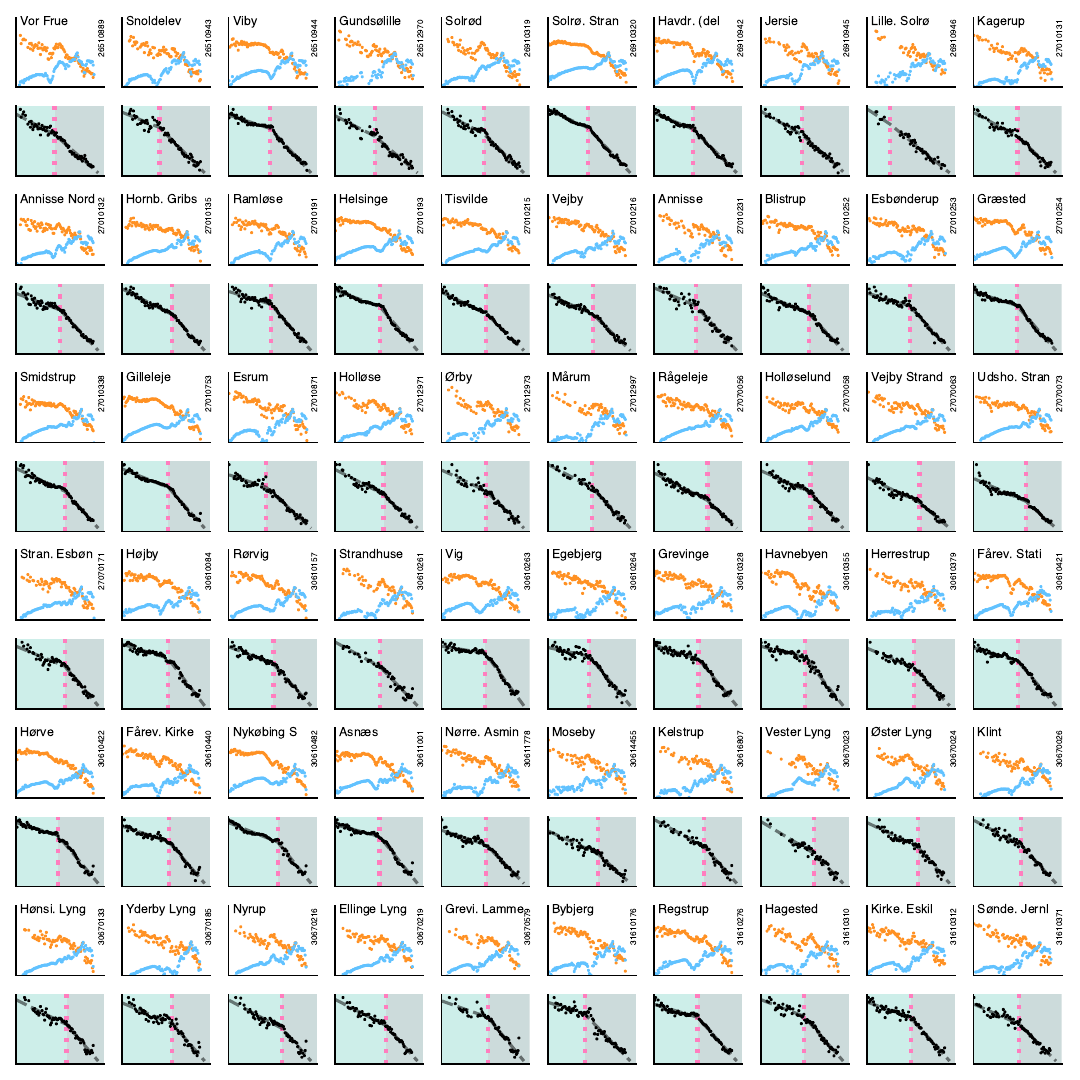}
   \caption{For each city in Denmark, the distribution of observed moving distance (orange) and the relative pair distribution function. Pairs are restricted to those containing at least one address from the city of interest. The second plot shows the intrinsic distance cost (black), as the ratio between the observed moving distance and the relative pair distribution function. The pink dashed line is the inferred mobility city radius, the light teal corresponds to intra-city moves, and the dark teal to inter-city moves. }
    \label{fig:piecewise-1}
\end{figure}

\begin{figure}
    \centering
    \includegraphics[width=\textwidth]{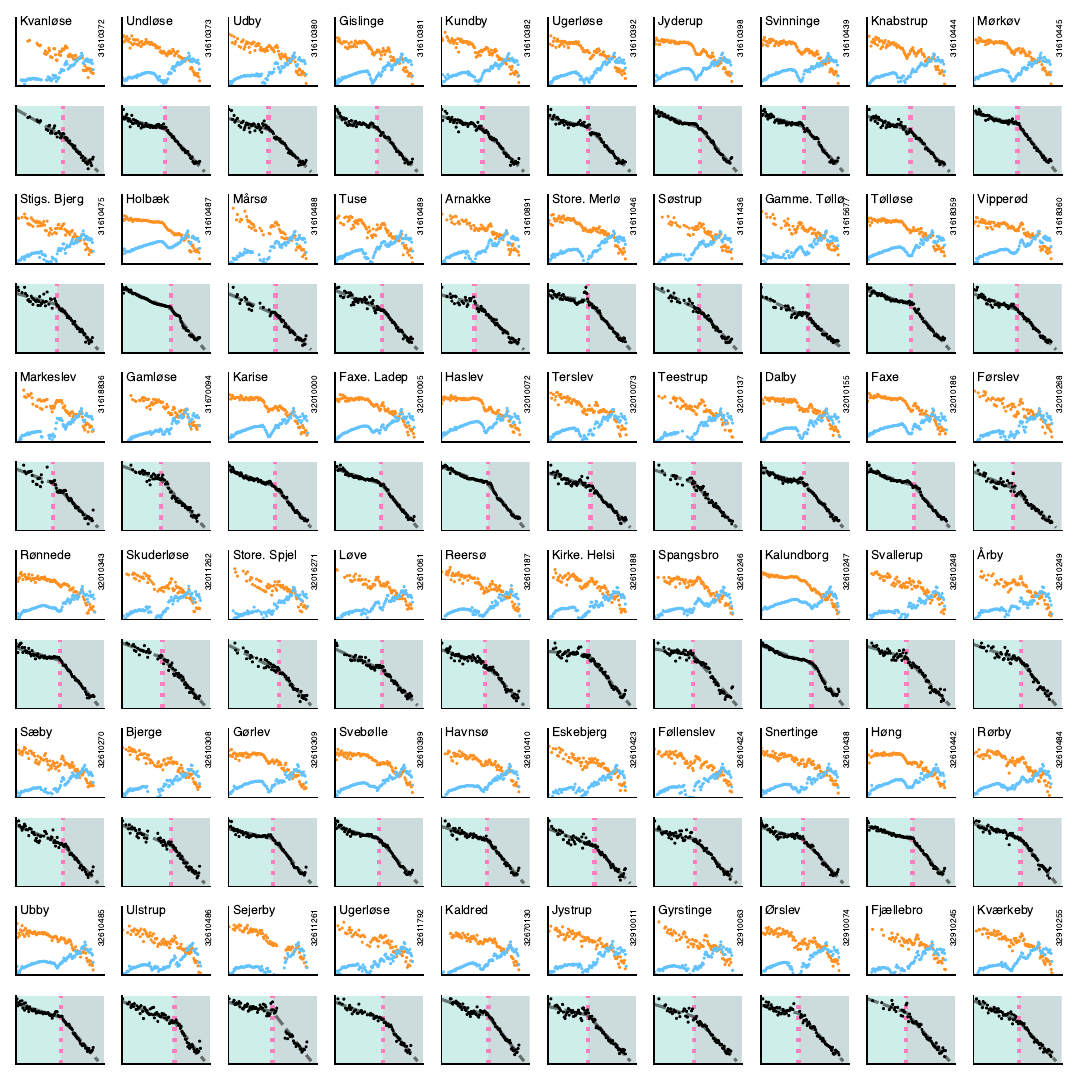}
   \caption{For each city in Denmark, the distribution of observed moving distance (orange) and the relative pair distribution function. Pairs are restricted to those containing at least one address from the city of interest. The second plot shows the intrinsic distance cost (black), as the ratio between the observed moving distance and the relative pair distribution function. The pink dashed line is the inferred mobility city radius, the light teal corresponds to intra-city moves, and the dark teal to inter-city moves. }
    \label{fig:piecewise-2}
\end{figure}

\begin{figure}
    \centering
    \includegraphics[width=\textwidth]{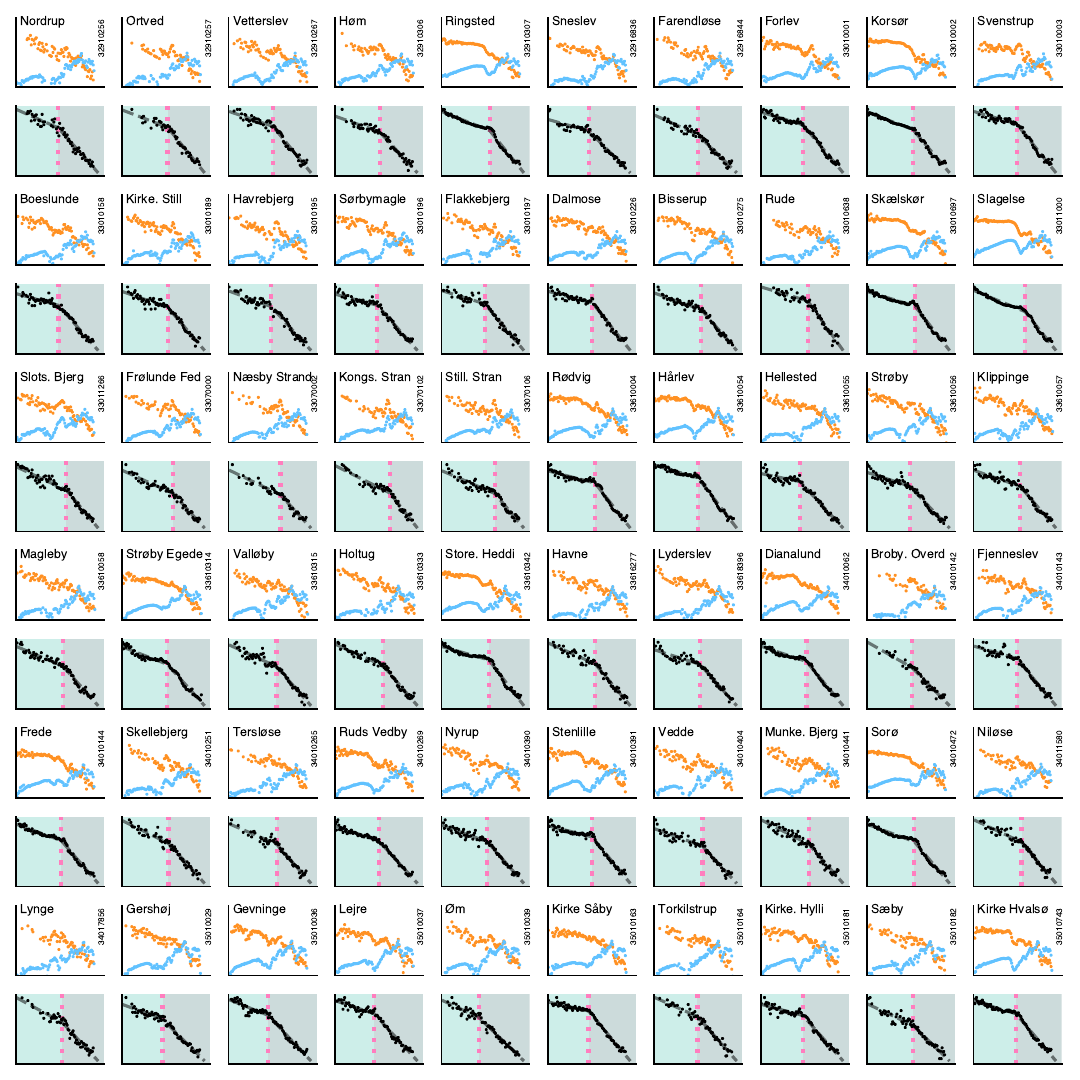}
\caption{For each city in Denmark, the distribution of observed moving distance (orange) and the relative pair distribution function. Pairs are restricted to those containing at least one address from the city of interest. The second plot shows the intrinsic distance cost (black), as the ratio between the observed moving distance and the relative pair distribution function. The pink dashed line is the inferred mobility city radius, the light teal corresponds to intra-city moves, and the dark teal to inter-city moves. }
    \label{fig:piecewise-3}
\end{figure}

\begin{figure}
    \centering
    \includegraphics[width=\textwidth]{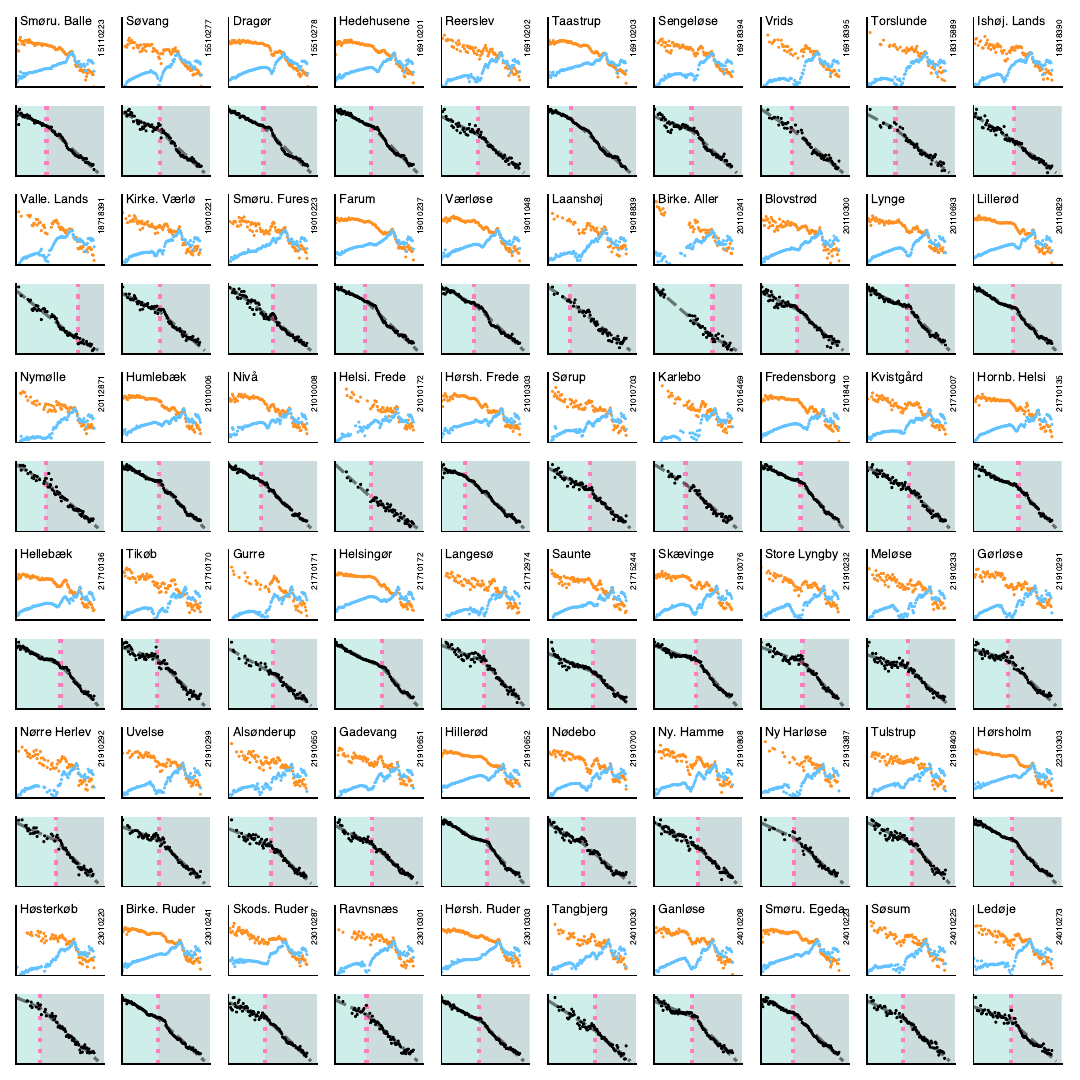}
    \caption{For each city in Denmark, the distribution of observed moving distance (orange) and the relative pair distribution function. Pairs are restricted to those containing at least one address from the city of interest. The second plot shows the intrinsic distance cost (black), as the ratio between the observed moving distance and the relative pair distribution function. The pink dashed line is the inferred mobility city radius, the light teal corresponds to intra-city moves, and the dark teal to inter-city moves. }
    \label{fig:piecewise-4}
\end{figure}

\begin{figure}
    \centering
    \includegraphics[width=\textwidth]{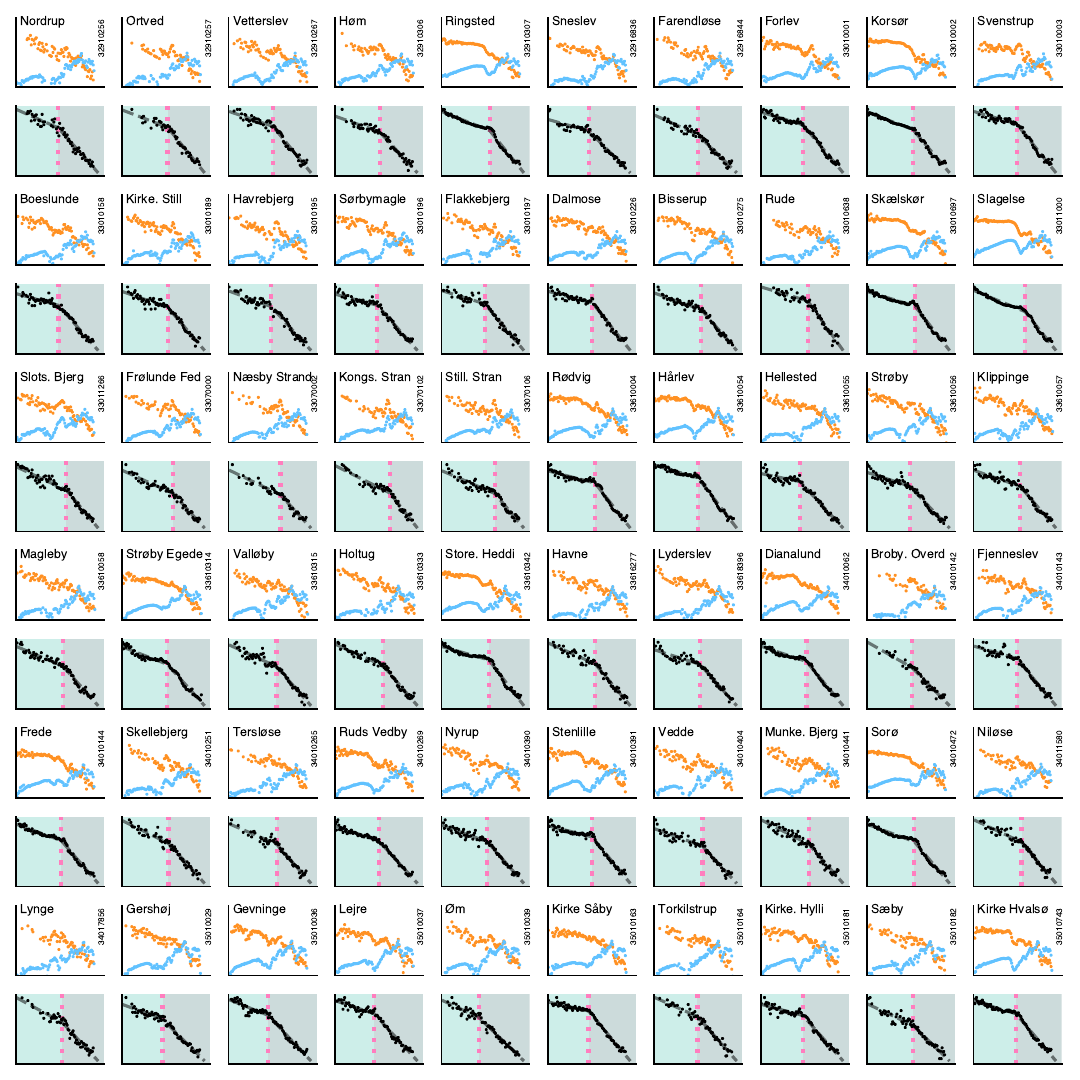}
\caption{For each city in Denmark, distribution of observed moving distance (orange) and the relative pair distribution function. Pairs are restricted to those containing at least one address from the city of interest. The second plot shows the intrinsic distance cost (black), as the ratio between the observed moving distance and the relative pair distribution function. The pink dashed line is the inferred mobility city radius, the light teal corresponds to intra-city moves, the dark teal to inter-city moves. } 
    \label{fig:piecewise-5}
\end{figure}

\begin{figure}
    \centering
    \includegraphics[width=\textwidth]{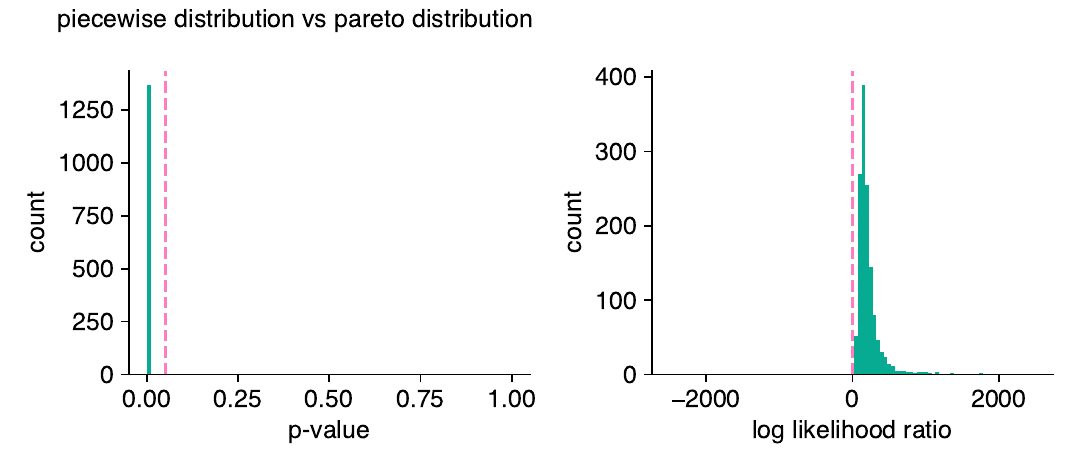}
     \caption{For each city of the piecewise fit, Figure.\ref{fig:piecewise-1}-\ref{fig:piecewise-5}, we compare the log-likelihood ratio between the piecewise power law distribution of equation \ref{eq:piecewise-powerlaw} and a Pareto distribution. The left plot shows the distribution of p values with a pink line at $p=0.05$. The right plot shows the distribution of the log-likelihood ratio, with a pink line at the decision boundary $R=0$. A positive log-likelihood ratio indicates that the piecewise models describe the data better, the p-value give the significance of the result.} 
    \label{fig:piecewise-pvalues0}
\end{figure}

\begin{figure}
    \centering
    \includegraphics[width=\textwidth]{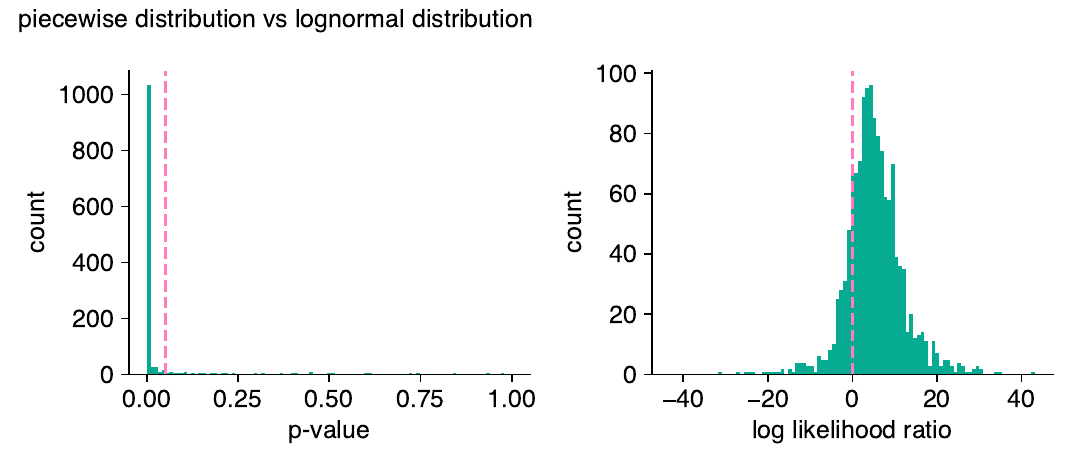}
     \caption{For each city of the piecewise fit, Figure.\ref{fig:piecewise-1}-\ref{fig:piecewise-5}, we compare the log-likelihood ratio between the piecewise power law distribution of equation \ref{eq:piecewise-powerlaw} and a lognormal distribution. The left plot shows the distribution of p values with a pink line at $p=0.05$. The right plot shows the distribution of the log-likelihood ratio, with a pink line at the decision boundary $R=0$. A positive log-likelihood ratio indicates that the piecewise models describe the data better, the p-value give the significance of the result.} 
    \label{fig:piecewise-pvalues1}
\end{figure}

\begin{figure}
    \centering
    \includegraphics[width=\textwidth]{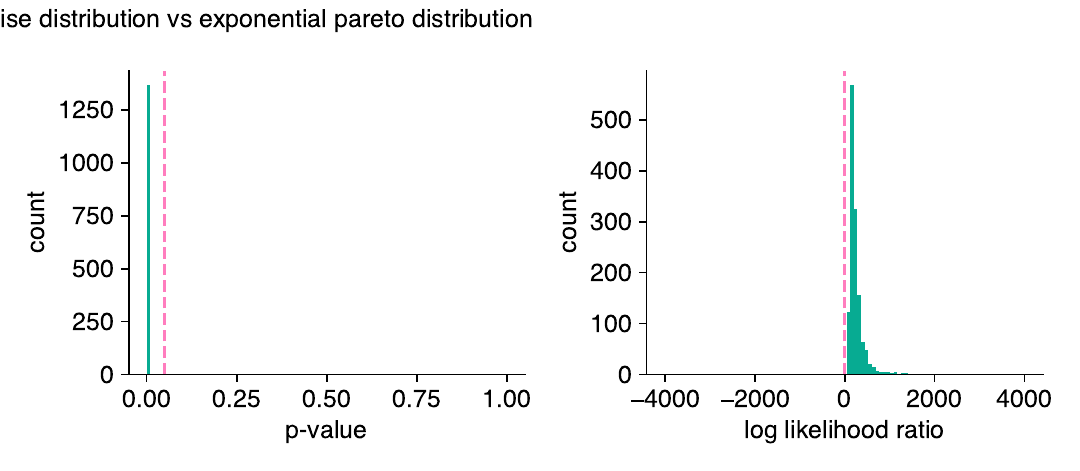}
     \caption{For each city of the piecewise fit, Figure.\ref{fig:piecewise-1}-\ref{fig:piecewise-5}, we compare the log-likelihood ratio between the piecewise power law distribution of equation \ref{eq:piecewise-powerlaw} and a piecewise exponential-Pareto distribution. The left plot shows the distribution of p values with a pink line at $p=0.05$. The right plot shows the distribution of the log-likelihood ratio, with a pink line at the decision boundary $R=0$. A positive log-likelihood ratio indicates that the piecewise models describe the data better, the p-value give the significance of the result.} 
    \label{fig:piecewise-pvalues2}
\end{figure}

Remarkably, the inverse of distance scaling of the intrinsic distance cost for $r\geq R_{\max}$emerges as a sole consequence of the random distribution of city locations. The scaling $r\leq R_{\max}$ results from the aggregation over the scale of cities, akin to our patch model SI.~\ref{sec:patch-model} or \cite{alessandretti2020scales}. We obtain that for $r\leq R_{\max}$,

\begin{align}
\label{eq:piecewise_idc_integral}
    \pi(r) = \int_{0}^{r} C_5 (r^{-1/6}/R_c^2)\exp(-(r/R_c)^m) g(R) \mathrm{d}R + \int_{r}^{\infty} C_6 r^{-5/3}  g(R)\mathrm{d}R 
\end{align}

where $g$ is the distribution of the city radius. From the main text Fig.3, $g$ can be modeled as a log-normal or a power-Pareto distribution. We solve the equation \ref{eq:piecewise_idc_integral} numerically by simulating over the city intrinsic distance cost over an artificial geography of Denmark (eq.\ref{eq:piecewise-pddc},\ref{eq:piecewise-pdd-hypothesis}). We recover the intrinsic distance cost $\pi(r)= 1/r$ (Fig.\ref{fig:piecewise-idc_simulation}).

\newpage
\section{The definition of city}

\label{sec:HDBSCAN}
The precise demarcation of a city — where a city ends and its boundaries — is a pressing issue in academic literature \cite{arcaute2015constructing}. Depending on the definition adopted, different results can be obtained \cite{bettencourt2013origins}. To ensure that our results are not just a byproduct of our chosen city definition and that they could potentially be applied to other definitions, we evaluated the robustness of the city boundaries.

First, we adopted the city definition provided by Danmarks Statistik \cite{dst_befolkning_valg}. According to this definition, Denmark consists of 1,473 cities.

To ensure more robust results, we compare the city definition to cities defined by density-based clustering techniques. These methods distinguish densely clustered data points that represent urban or urbanized areas from sparse or noisy regions that typically correspond to rural or sparsely populated areas. DBSCAN is particularly well-suited for this task because of its innate ability to delineate clusters of different geometries, which is critical for accommodating the non-uniform shapes of urban regions. This algorithm sorts data points into core points, boundary points, and noise based on surrounding data density. However, one of the main challenges is to determine the optimal values for $\epsilon$ (neighborhood search radius) and minimum data points, as these parameters can determine the level of detail of the identified urban zones. \cite{ester1996density} .

For data covering cities with different population densities, HDBSCAN stands out. Building on the foundation laid by DBSCAN, HDBSCAN employs a hierarchical tactic that makes the fixed $\epsilon$ value redundant. By examining the density hierarchy, HDBSCAN can differentiate between densely populated cities and smaller towns within the same data collection. This granularity, coupled with the flexible clustering results, provides a detailed view of the boundaries of cities \cite{campello2013density}.

\begin{figure}
    \centering
    \includegraphics[width=\textwidth]{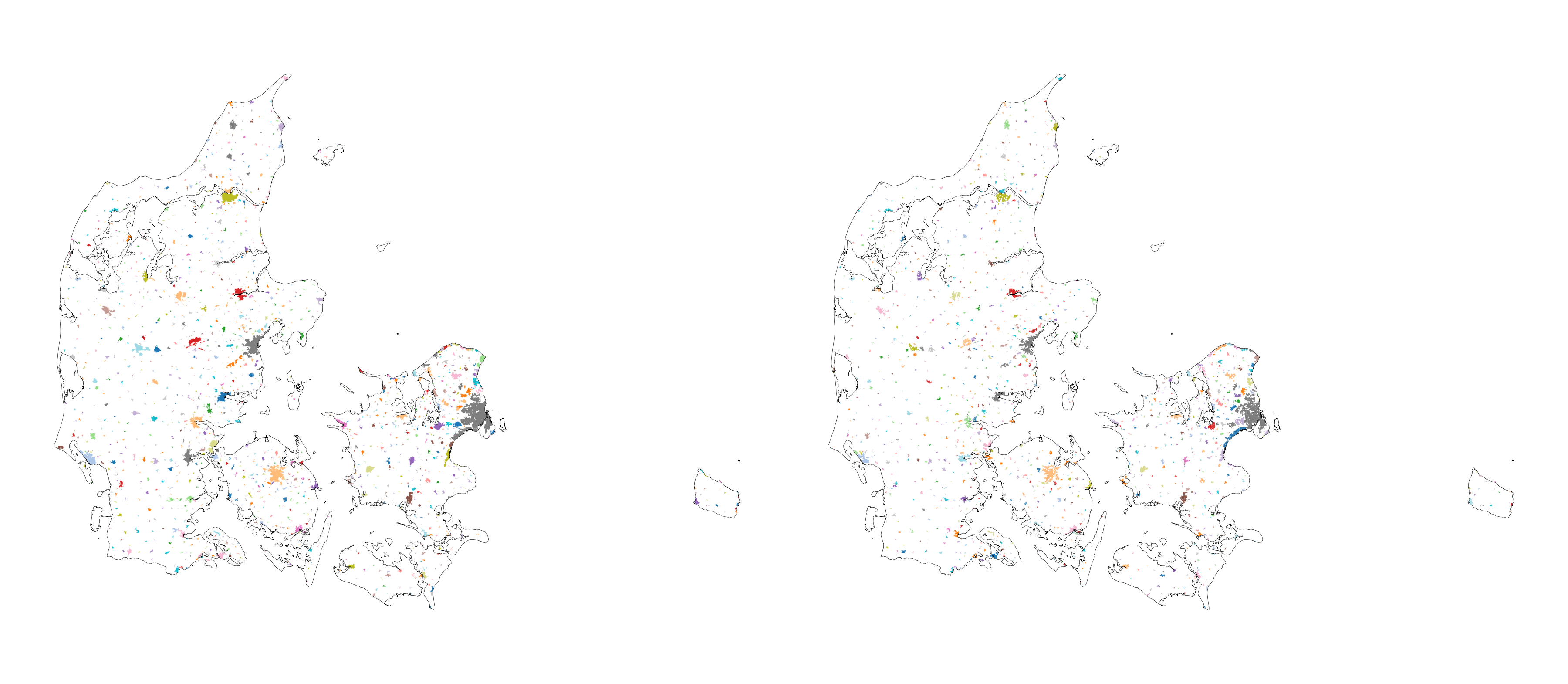}
    \caption{Comparison between the Danmarks Statistik definition of cities (on the left) and the cluster obtained from HDBSCAN (on the right). The colors are ordered according to the size of the cluster/city.}
    \label{fig:dk_cities_cluster}
\end{figure}

Copenhagen, however, presents a unique scenario. Its urban influence extends well beyond its administrative boundaries, as evidenced by phenomena such as commuting patterns. As a result, neighboring cities have been merged with Copenhagen to form the Copenhagen metropolitan area, or Hovedstadsområdet, as defined by DST. Notably, the clustering results from HDBSCAN reflected this merging, indicating the combined urban sprawl of the Copenhagen region.

Normalized Mutual Information (NMI) and Adjusted Mutual Information (AMI) quantify the similarity between two clusterings. Both metrics measure the information shared between the true labels and the labels assigned by a clustering algorithm.  Therefore, we can use them to evaluate the performance of clustering algorithms in the absence of ground truth labels. The NMI is defined as the mutual information between two clusterings divided by the geometric mean of their entropies \cite{strehl2002cluster}. However, a limitation of the NMI is that it ignores the random grouping of clusters, i.e. random cluster assignments can produce a non-zero NMI value. The AMI, on the other hand, corrects for this limitation by adjusting the score to account for chance, ensuring that random cluster assignments result in an AMI score close to zero \cite{vinh2010information}. Therefore, the AMI provides a more accurate representation of the similarity between two clusterings in our case where the number of clusters is not fixed.

The table \ref{tab:nmi_ami_values_CPH} shows the value of the indices for the two algorithms. An NMI/AMI score of 0.9 indicates that the two clusterings share a significant amount of information. Such a high score typically indicates that the two clusterings are almost identical, with only a few data points potentially clustered differently. Therefore, the city definition from Danmarks Statistics is substantially similar to the one we obtain using density-based clustering techniques. Furthermore, table \ref{tab:nmi_ami_values_HOV} shows that merging Copenhagen into the Hovedstadsområdet is a better definition of a city according to the density-based clustering techniques.

\begin{table}[ht]
\centering
\begin{tabular}{@{}lcc@{}}
\toprule
Algorithm & NMI & AMI \\
\midrule
DBSCAN    & 0.88 & 0.87 \\
HDBSCAN   & 0.91 & 0.89 \\
\bottomrule
\end{tabular}
\caption{NMI and AMI values between empirical city clustering (Copenhagen merged) and algorithmic clustering (DBSCAN and HDBSCAN).}
\label{tab:nmi_ami_values_HOV}
\end{table}
\begin{table}[ht]
\centering
\begin{tabular}{@{}lcc@{}}
\toprule
Algorithm & NMI & AMI \\
\midrule
DBSCAN    & 0.71 & 0.67 \\
HDBSCAN   & 0.67 & 0.65 \\
\bottomrule
\end{tabular}
\caption{NMI and AMI values between the empirical cities (Copenhagen not merged) clustering and the algorithmic clustering (DBSCAN and HDBSCAN)}
\label{tab:nmi_ami_values_CPH}
\end{table}

\begin{figure}
    \centering
    \includegraphics[width=0.7\textwidth]{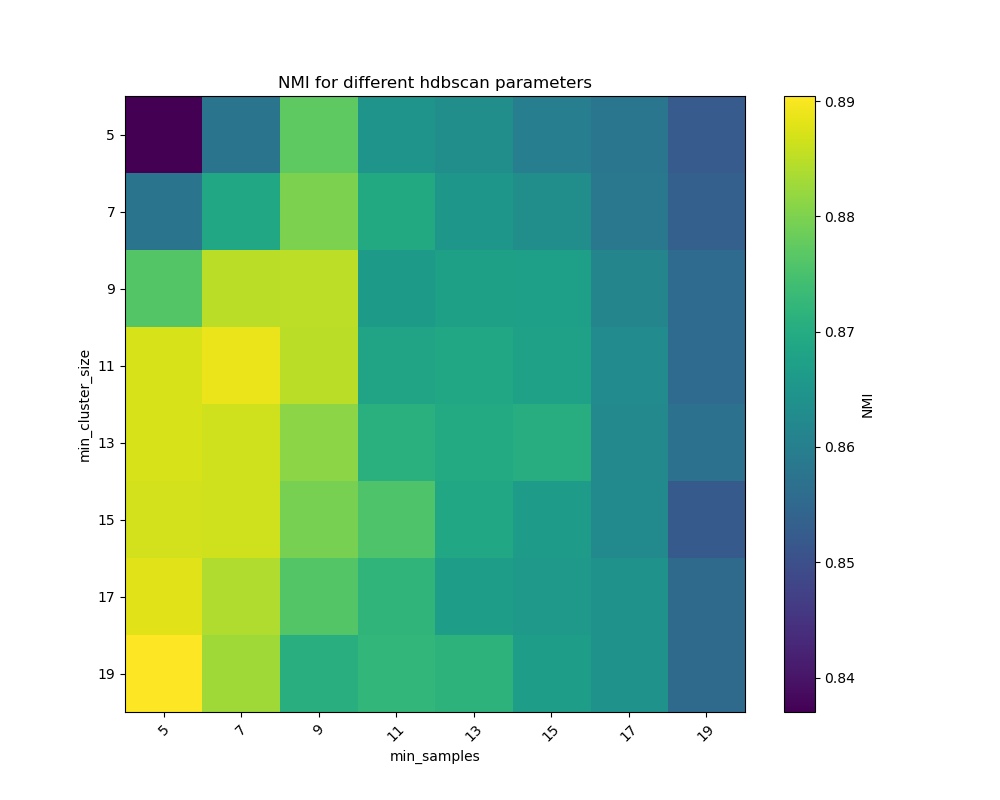}
    \caption{Heatmap of the NMI between the clustering of addresses locations and their official city and the values for different combinations of HDSCAN parameters.}
    \label{fig:NMI_HDBSCAN}
\end{figure}

The parameters for HDSCAN were chosen to optimize the NMI between the clustering and the real city labels. Figure \ref{fig:NMI_HDBSCAN} shows the values for different values of \textit{minimum samples} and \textit{minimum cluster size}, the rest of the parameters follow the default values of this implementation \cite{campello2013density}.

Figure \ref{fig:dk_cities_cluster} shows all addresses colored by the cluster they belong to. The background color is the official city border. 

\section{Maximum likelihood estimation of power laws}
\label{SI:power-law}
Fitting the statistical power-law model, to examine the distance distribution of human movement, we fit a truncated power law with the form,

\begin{align}
    p(x) \propto x^{-\alpha} e^{-\lambda x}.
\end{align}

where $\alpha$ is a constant parameter of the distribution (known as the scaling parameter or exponent), $x$ is the travel distance ($x > x_{min}>0$), and $\lambda$ is the parameter of the exponential distribution. $x_{min}$ represents the shortest distance above which the power law scaling relationship begins. The scaling parameter $\alpha$ must be estimated before finding the optimal values of $x_{min}$. The methods of \cite{clauset2009power} find $x_{min}$ by generating a power-law fit starting from each unique value in the data set, then selecting the one that gives the minimum Kolmogorov-Smirnov distance between the data and the fit. For any given value of $x_{min}$, we estimated the scaling parameter using maximum likelihood estimation. The goodness of fit for the truncated power law distribution was considered in comparison to the fit of other distributions (e.g., power law, exponential, and lognormal).

Although some of the power laws present a more complicated picture, we then used the methods developed in this \cite{maier2023maximum}, which extend \cite{clauset2009power} procedures for a wider class of distributions.

\begin{figure}
    \centering
    \includegraphics[width=0.7\textwidth]{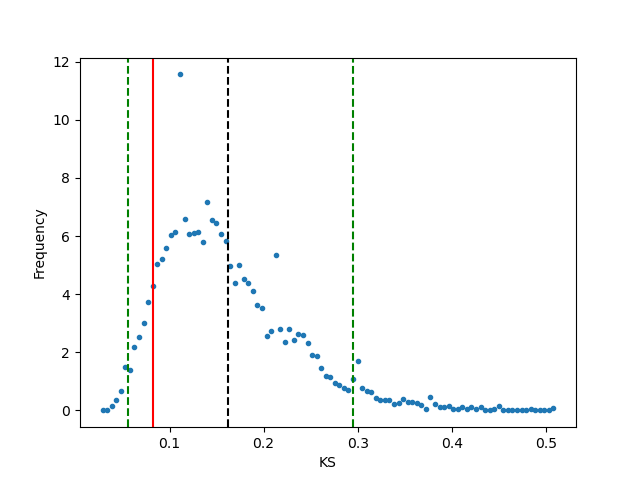}
    \caption{Kolmogorov-Smirnov (KS) test for Houston data, the blue dots represent the distribution of the test statistics over the sampled data. The mean is represented by a dashed black line. The 5th and 95th percentiles are represented by dashed green lines, indicating the range within which 90\% of the observed values fall.}
    \label{fig:houston-ks}
\end{figure}

\begin{figure}
    \centering
    \includegraphics[width=0.7\textwidth]{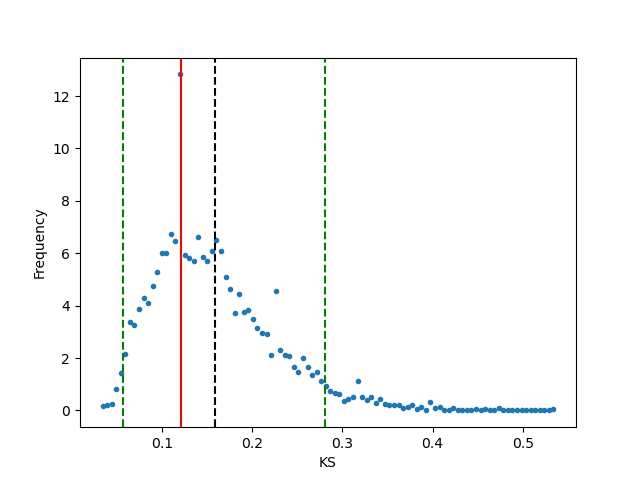}
    \caption{Kolmogorov-Smirnov test for Singapore data, the blue dots represent the distribution of the test statistics over the sampled data. The mean is represented by a dashed black line. The 5th and 95th percentiles are represented by dashed green lines, indicating the range within which 90\% of the observed values fall.}
    \label{fig:singapore-ks}
\end{figure}

\begin{figure}
    \centering
    \includegraphics[width=0.6\textwidth]{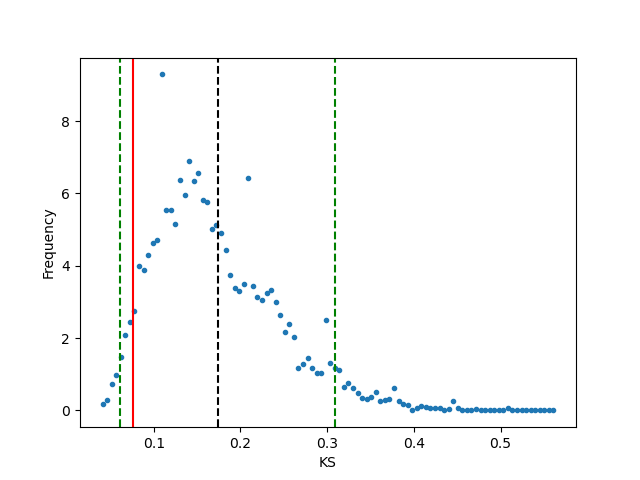}
    \caption{Kolmogorov-Smirnov test for San Francisco data, the blue dots represent the distribution of the test statistics over the sampled data. The mean is represented by a dashed black line. The 5th and 95th percentiles are represented by dashed green lines, indicating the range within which 90\% of the observed values fall.}
    \label{fig:sanfrancisco-ks}
\end{figure}

\begin{figure}
    \centering
    \includegraphics[width=0.6\textwidth]{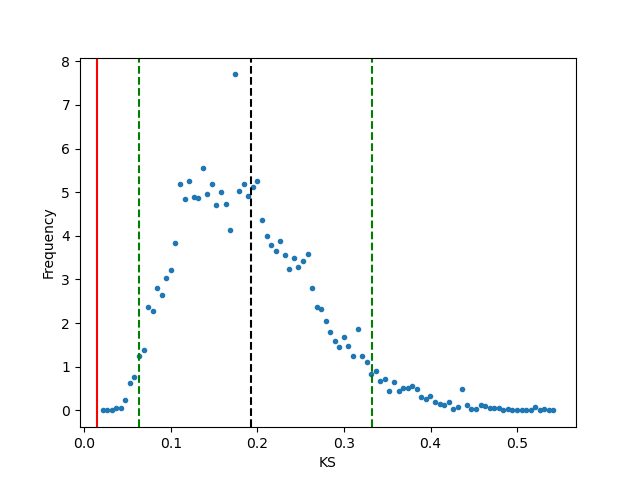}
    \caption{Kolmogorov-Smirnov test for France data, the blue dots represent the distribution of the test statistics over the sampled data. The mean is represented by a dashed black line. The 5th and 95th percentiles are represented by dashed green lines, indicating the range within which 90\% of the observed values fall.}
    \label{fig:fr-ks}
\end{figure}

\newpage

\bibliographystyle{unsrt}
\bibliography{bibliography,bibliography-SI}

\end{document}